\documentclass[a4paper,twoside]{article}
\pdfoutput=1

\newif\iftr 
\trtrue     

\usepackage{color}
\usepackage{xcolor}
\definecolor{darkblue}{rgb}{0,0.08,0.45}
\usepackage{ifpdf}  
\ifpdf
    \usepackage[pdftex]{graphicx}
    \usepackage[colorlinks,linkcolor=darkblue,citecolor=darkblue,urlcolor=darkblue]{hyperref}
    \DeclareGraphicsExtensions{.pdf}
\else
    \usepackage{graphicx}
    \usepackage[hypertex]{hyperref}  
\fi

\usepackage[disable]{todonotes} 
\newcommand{\bob}[1]{\todo[color=olive!40,inline]{Bob: #1}}

\usepackage{
url       %
,fancyhdr 
,lastpage 
,enumerate 
,amsmath  
,amsfonts 
,amssymb  
, enumitem 
}
\usepackage[hang]{footmisc} 
\usepackage[noindent, arraystretch, fullpage]{setlengths} 
\usepackage{
own         
}

\usepackage{amsthm}

\newtheorem*{test*}{Test}

\graphicspath{{images/}}

\newcommand*{\metaauthori}{Bob Briscoe}
\newcommand*{\metaauthorii}{Asad Sajjad Ahmed}
\newcommand*{\metashorttitle}{Classic ECN AQM Fall-back}
\newcommand*{\metatitle}{TCP Prague Fall-back on Detection of a Classic ECN AQM}
\newcommand*{\metano}{TR-BB-2019-002}
\newcommand*{\metakeywords}{Data Communications, Networks, Internet,
Control, Performance, Latency, Responsiveness, Dynamics, Algorithm, Standards}

\newcommand*{\metamaili}{\href{mailto:research@bobbriscoe.net}{research@bobbriscoe.net}}
\newcommand*{\metamailii}{\href{mailto:me@asadsa.com}{me@asadsa.com}}
\newcommand*{\metaaddress}{}

\newcommand*{\metaversion}{03}
\newcommand*{\metadate}{19 Feb 2021}

\hypersetup{                       
     pdfauthor = {\metaauthori and \metaauthorii
     },
     pdftitle = {\metashorttitle},
     pdfsubject = {},
     pdfkeywords = {\metakeywords}
}%

\title{\metatitle}%
\author{\metaauthori%
\thanks{\metamaili, %
\metaaddress}%
\ %
\and \metaauthorii%
\thanks{\metamailii}%
}
\date{\metadate}%

\fancypagestyle{first}{%
\fancyhead[LO,RE]{}%
\fancyhead[LE,RO]{}%
\fancyhead[C]{}%
}%

\begin{document}
\bibliographystyle{alpha}%


\maketitle%
\thispagestyle{first}

\begin{abstract}
{\small\noindent%
The IETF's Prague L4S Requirements~\cite{Briscoe15f:ecn-l4s-id_ID} expect an
L4S congestion control to somehow detect if there is a classic ECN~\cite{IETF_RFC3168:ECN_IP_TCP} AQM at
the bottleneck and fall back to Reno-friendly behaviour (as it would on a
loss). 
This paper addresses that requirement in depth. A solution has been
implemented in Linux and extensively tested, which distinguishes L4S from Classic AQMs primarily by their delay variation. This paper describes version 2 of the
design of that solution, giving extensive rationale and pseudocode. It
briefly summarizes a comprehensive testbed evaluation of the solution,
referring to the full details online.
The v2 algorithm very rarely falsely detects a Classic AQM as L4S. It also
rarely detects an L4S AQM as Classic in the majority of scenarios, but not at
low link rates and large RTTs.

This report is a work in progress. It suggests ideas for improving on the
approach. It also outlines new ideas that could solve the problem in
complementary or alternative ways.
}      
\end{abstract}
\tableofcontents
\newpage

\section{The Coexistence Problem}\label{ecn-fallbacktr_problem}

As the name implies, the Low Latency Low Loss Scalable throughput (L4S) architecture~\cite{Briscoe16a:l4s-arch_ID} is intended to enable incremental deployment of scalable congestion controls, which in turn are intended to provide very low latency and loss. 

Since 1988, when TCP congestion control was first developed, it has been known that it would hit a scaling problem. Footnote 6 of Jacobson \& Karels \cite{Jacobson88b:Cong_avoid_ctrl} said ``We are concerned that the congestion control noise sensitivity is quadratic in \(w\) but it will take at least another generation of network evolution to reach window sizes where this will be significant.'' The footnote went on to say, ``If experience shows this sensitivity to be a liability, a trivial modification to the algorithm makes it linear in \(w\).''

By the end of the 1990s that scaling problem had become very apparent~\cite{IETF_RFC3649:HSTCP}. A ``trivial modification to the algorithm'' would indeed make it linear, which is the definition of a scalable congestion control. However, the problem was not how to modify the algorithm, it was how to deploy it. Such a linear congestion control would not coexist with all the traffic on the Internet that had evolved in coexistence with the original TCP algorithm.

A scalable congestion control induces frequent congestion signals, and the frequency remains invariant as flow rate scales over the years. Therefore, modern scalable congestion controls, such as DCTCP~\cite{Alizadeh10:DCTCP_brief}, use Explicit Congestion Notification (ECN) rather than loss to signal congestion. They use the same ECN codepoints as the original ECN standard~\cite{IETF_RFC3168:ECN_IP_TCP}, but they induce much more frequent ECN marks~\cite{Briscoe15f:ecn-l4s-id_ID} than Classic (Reno-friendly) congestion controls.

\begin{figure}[h]
  \centering
  \includegraphics[width=0.7\linewidth]{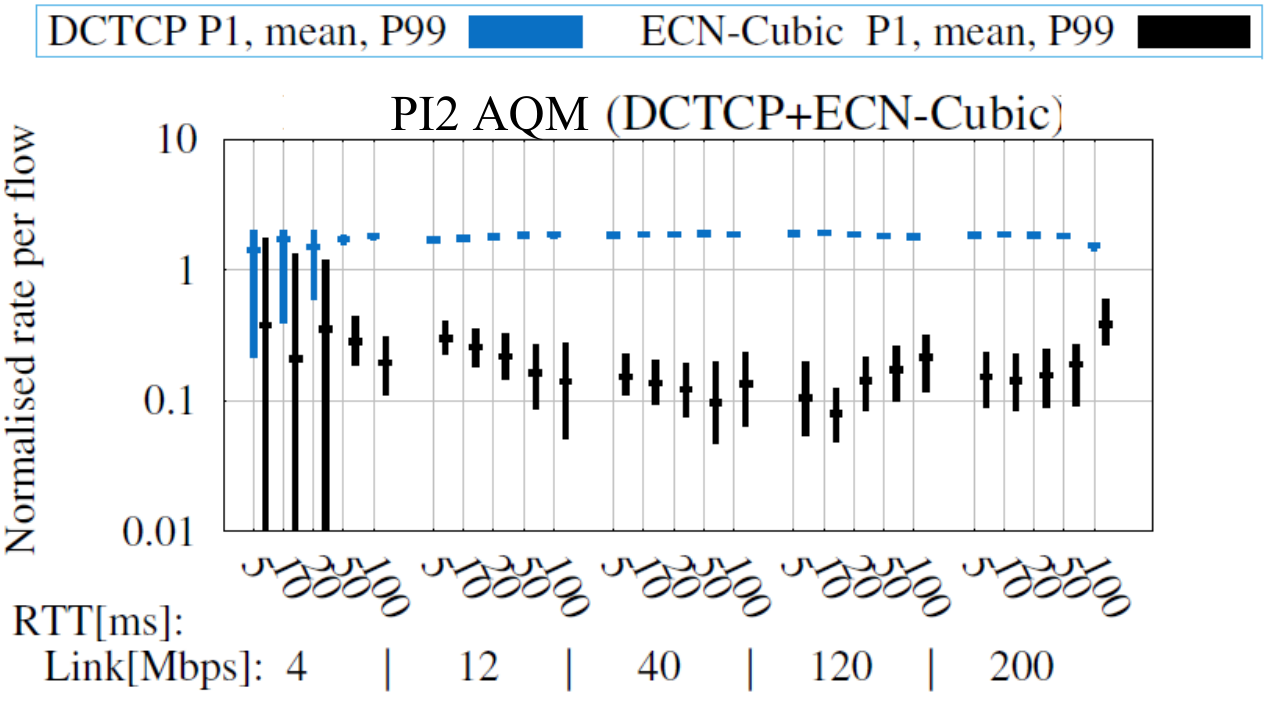}
  \caption{Quantification of the Rate Imbalance Problem between L4S and Classic ECN flows in a Classic ECN Queue. Flow rates are of long-running flows after a period of stabilization and normalized relative to precisely equal flow rates, 1 DCTCP vs. 1 ECN-Cubic, equal base RTTs, single queue PI\(^2\) AQM; default parameters (see \S\,\ref{ecn-fallbacktr_eval_fairness} for full experiment details.)}\label{fig:ecn-fallbacktr_rate_flow}
\end{figure}

Thus, if a scalable congestion control finds itself sharing a queue with a congestion control that conforms to the `Classic' definition of ECN as equivalent to loss~\cite{IETF_RFC3168:ECN_IP_TCP}, the Classic flow will imagine that there is heavy congestion and back off its flow rate. It will not actually starve itself, but
it will reduce to a rate that can be 4--16 times less on average than any competing L4S flow
(\autoref{fig:ecn-fallbacktr_rate_flow}).

Coexistence is only a problem if the bottleneck is a shared queue that supports Classic ECN marking. It is not known whether any shared queue Classic ECN AQMs are operational on the Internet, and so far no evidence has been forthcoming from the search for one. However, for L4S to proceed through standardization, it has to be assumed that such AQMs do exist or that they might exist. Therefore scalable congestion controls ought to respond appropriately if they detect a classic ECN AQM. In 2015, this was codified as one of the `Prague L4S' requirements for scalable congestion controls, which have since been adopted by the IETF~\cite{Briscoe15f:ecn-l4s-id_ID}. That requirement sets the problem that motivates the present report.

Other aspects of the L4S architecture already address coexistence in all the other possible cases, viz:
\begin{itemize}
	\item If the bottleneck does not support any form of ECN (as is the case
	for the vast majority of buffers on the Internet), the only sign of
	congestion will be loss. It is easy to ensure that scalable congestion
	controls respond to loss in a Reno-friendly way, and all known scalable
	congestion controls do so (this is top of the list of ``Prague L4S
	requirements''~\cite{Briscoe15f:ecn-l4s-id_ID}).
	
	\item If the bottleneck applies per-microflow\footnote{A microflow is a data-flow 
	between two application end-points that is identified by the 5-tuple of source and 
	destination addresses \& ports plus the protocol. Per-flow queuing can be 
	configured for other definitions of `flow', but per-microflow is the most common.} 
	scheduling, it enforces
	coexistence without the need for the present algorithm. There are two
	sub-cases:
	\begin{itemize}
		\item If there is a Classic AQM in each per-flow queue (understood to be
		a common deployment with FQ\_CoDel~\cite{Hoeiland18:fq-codel_RFC} and
		COBALT~\cite{Palmei19:Cobalt}, which is used in CAKE), then the detection
		algorithm in the present paper ought to detect it and switch to Classic
		behaviour, which could provide performance benefits;
		
		\item If the AQM in each per-flow queue supports L4S by detecting the L4S
		ECN identifier, then the full benefits of L4S will be available without
		the present algorithm, which will remain quiescent;
	\end{itemize}
	
	\item If the bottleneck supports the DualQ Coupled
	AQM~\cite{Briscoe15e:DualQ-Coupled-AQM_ID}, that will ensure that L4S and
	classic flows coexist and the full benefits of L4S will be available
	without the present algorithm, which will again remain quiescent.
\end{itemize}

\section{Modular Approach}\label{ecn-fallbacktr_modular}

\subsection{Modularity Requirements}\label{ecn-fallbacktr_modular_reqs}

An operator might want to determine the type of AQM on the path
\textbf{in-band}, that is within live traffic, or \textbf{out-of-band} using
test traffic.

An aim of in-band testing is to detect the type of AQM fast
enough%
\footnote{Perhaps within a dozen or so round trips, given the original
	paper on TCP Cubic\cite{cubic} described convergence as
	`short' when it took one or two hundred rounds (assuming flows even last that
	long).} %
that a flow can start with the L4S behaviour but if necessary change
over to Classic behaviour in the early stages of convergence.

However, a server operator might not want to trigger any change in behaviour.
For instance, a CDN operator might want to use in-band or out-of-band testing
purely to check the likelihood that the problem even exists over the paths they
serve.

Therefore, a complete algorithm needs to work in two separable stages:
\textbf{detection} then if necessary \textbf{fall-back} to Classic behaviour.
More generally, fall-back ought to be called \textbf{changeover} because, as
detection continues, the algorithm might need to reverse the change (either
because the first change was premature, or because the bottleneck has changed).

Detection might purely measure the traffic's characteristics (\textbf{passive}),
or it might alter the way traffic is sent (\textbf{active}), e.g.\ altering
send-timing, the sizes of certain packets, their markings, or adding extra
probes. Active techniques tend to provide more certainty at the expense of
altering live traffic. So passive detection is preferable at first, but if
experimentation finds it is insufficient, active techniques could be kept in
reserve as a double-check just before a changeover actually occurs, perhaps only
in more challenging scenarios, e.g.\ high RTT, low link rate or when a bursty
radio link appears to be present on the path.

Out-of-band testing is not applied to live traffic, so it can typically resort
to active techniques straight away (unless it could potentially harm other live
traffic).

The main body of this report is divided into the same modular sections; on
passive detection (\S\,\ref{ecn-fallbacktr_detection}), active detection (both
out-of-band in \S\,\ref{ecn-fallbacktr_OOB_active} and in-band in
\S\,\ref{ecn-fallbacktr_active}) and change-over
(\S,\ref{ecn-fallbacktr_changeover}). Different modules can then be mixed and
matched to produce different solutions, one of which (so far) is evaluated in
\S\,\ref{ecn-fallbacktr_evaluation}.

\subsection{Modular Code Structure}\label{ecn-fallbacktr_structure}

Beyond solving the coexistence problem, the following principles are proposed to structure the code for multiple purposes:
\begin{enumerate}
	\item Rather than fall-back being a binary switch between modes, it should be
	a gradual changeover, the more certain it is that the AQM supports classic ECN;
	\item Nonetheless, at either end of the spectrum of (un)certainty, there
	should be ranges where the CC behaves on the one hand purely scalably and on
	the other purely classically;
	\item Minimal additional persistent TCP state;
	\item The code should be structured with detection separate from changeover of
	behaviour, so that detection can eventually apply to more than one CC, while
	changeover is likely to be CC-specific;
	\item However, until the concept is proven, it will be OK initially to
	implement the whole algorithm within the TCP Prague CC module, and only
	rationalize it once mature.
\end{enumerate}

To simplify pseudocode, a float called \texttt{c} controls how much the CC
should behave as classic, from 0 (scalable) to 1 (classic). In practice this
might be an integer variable in the range 0 to
\texttt{CLASSIC\_ECN}. This will be driven from the variable \texttt{classic\_ecn}, which is defined as the outputof the detection algorithm.

\begin{figure}[h]
  \centering
  \includegraphics[width=0.35\linewidth]{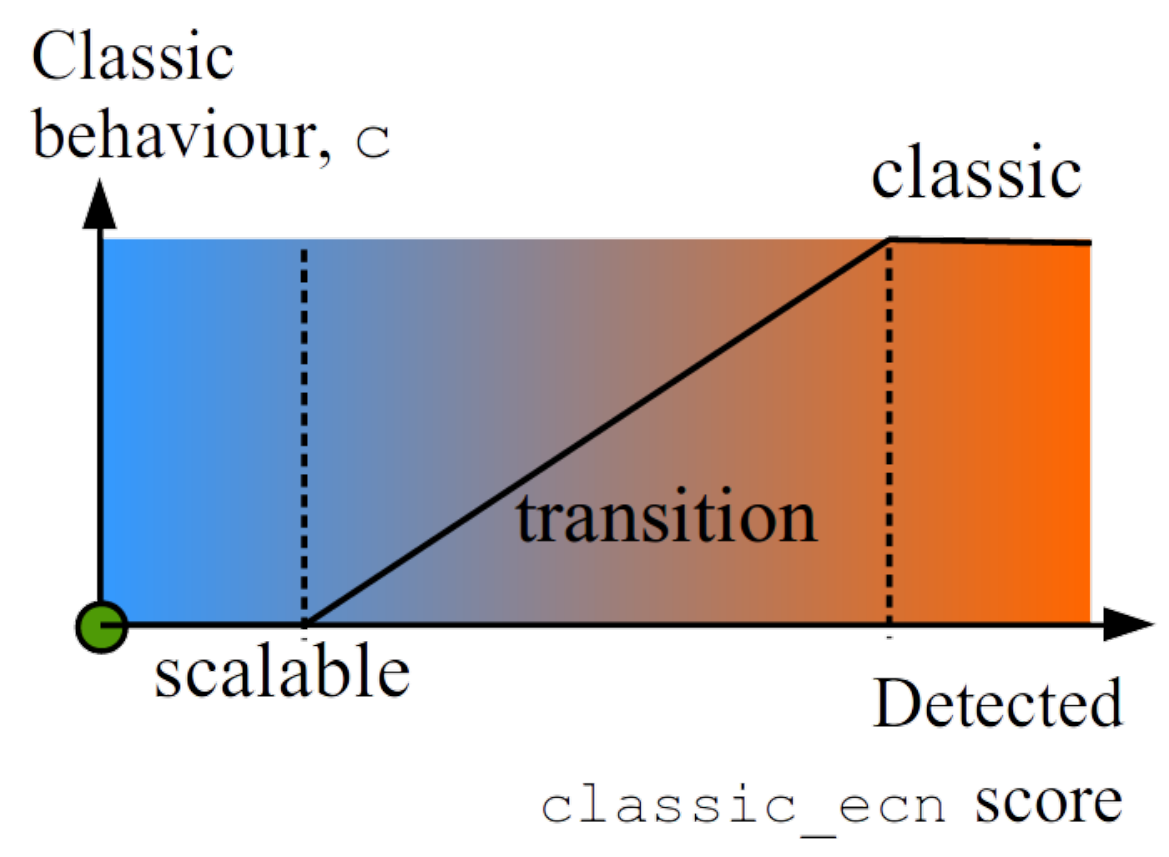}
  \caption{Score-based transition, rather than modal switch, between Scalable and Classic behaviour.}\label{fig:ecn-fallbacktr_c-trans}
\end{figure}

The \texttt{classic\_ecn} indicator can continue beyond either end of this
range. as depicted in \autoref{fig:ecn-fallbacktr_c-trans} and tabulated below, in order to implement a degree of stickiness where
the CC algorithm behaves purely as L4S or purely classically.

\begin{tabular}{rcccll}
\hline
\texttt{-L\_STICKY}         & \(\le\) & \texttt{classic\_ecn} &\(\le\) & \texttt{0}                            & Pure L4S behaviour\\
\texttt{0}                            & \(\le\) & \texttt{classic\_ecn} & \(\le\) & \texttt{CLASSIC\_ECN} & Transition between L4S and classic\\
\texttt{CLASSIC\_ECN}  & \(\le\) & \texttt{classic\_ecn} & \(\le\) & \texttt{C\_STICKY}         & Pure classic behaviour\\
\hline
\end{tabular}

It is up to the detection algo, not the CC algo, to maintain the
\texttt{classic\_ecn} variable. However, any particular CC algorithm can override
the default parameters of the detection algorithm. For instance, a CC module
could alter how 'sticky' the hysteresis is at either end by overriding the
default \texttt{L\_STICKY} and/or \texttt{C\_STICKY} parameters.
 
\section{Passive Detection of Classic ECN AQMs}\label{ecn-fallbacktr_detection}

\subsection{Candidate Metrics}\label{ecn-fallbacktr_metrics}

The following metrics are likely to be relevant when detecting a classic ECN AQM:
\begin{itemize}
	\item The onset of CE marking;
	\item Application-limited (no buffering at the sender);
	\item Receive window limited (\texttt{rwnd} \(<\) \texttt{cwnd});
	\item The moving mean deviation of the RTT (\texttt{fbk\_mdev})\footnote{An alternative to
	standard deviation that is a little easier to compute and no less valid as a
	variability metric~\cite[Appx A]{Jacobson88b:Cong_avoid_ctrl}} of the RTT.
	\item The difference between the smoothed RTT (\texttt{fbk\_srtt}) and a 
	minimum RTT (\texttt{rtt\_min}), with suitable
	safeguards against a false minimum and against step changes;
	\item The distribution of the spacing between ECN marks
\end{itemize}

Note that the variable names \texttt{fbk\_mdev} and \texttt{fbk\_srtt} are prefixed with \texttt{fbk\_} to emphasize that they are specific to the fall-back algorithm, and not necessarily the same as the \texttt{mdev} and \texttt{srtt} variables that Linux TCP already maintains for its retransmit timer (see \S\,\ref{ecn-fallbacktr_pseudocode_srtt}).

\subsubsection{Dependence on Presence of CE marking}\label{ecn-fallbacktr_CE_presence}

Obviously no transition to classic should occur unless there has been a CE
mark. Any transition should be suppressed for a number of RTTs after the onset of CE
marking, both to allow the connection to stabilize and because aggressive
competition for bandwidth is not a great concern with short flows. Indeed when
rate `fairness' is considered over time, it is more fair if long-running flows are
less aggressive than short flows~\cite{Guo02:Size-based_scheduling,
Yang02:Sized-basedTCP, Zielger04:TCP_Vienna, Moktan12:FavoringShort,
Bai15:PIAS, Briscoe19c:FQ_v_e2e}---though a delay is sufficiently motivated by the need for a
stabilization period rather than any desire to use a different form of
fairness.

\begin{description}
	\item[How long for instability to end?] A classical slow-start ends on the
	first CE (unless it had already ended due to a loss). In the next few rounds,
	all the flows suffer a period of instability as they recover from the
	transient overshoot of the new flow. The random nature of this period leaves
	them all at different shares of capacity. Then they might take a few hundred
	further round trips to converge on stable shares. It is likely that any flow-start
	approach developed for shallow threshold ECN might have a less clear cut
	transition between a flow-start phase and a period of convergence. Given this
	likely heterogeneity in approaches to flow-start, it is not feasible to
	quantify how long any single flow should wait after the first CE before
	starting to detect a classic ECN AQM, because the period of instability
	depends more on the behaviour of other flows than on its own behaviour.

	Therefore it is proposed to start maintaining metrics as soon as the first
	feedback of a CE mark arrives, rather than attempting to wait for the
	subsequent instability to subside. As will be seen next, even if the RTT
	metrics start during this period of instability, they can be given plenty of time
	to stabilize before they alter the CC behaviour, while they remain scalable in 
	the `sticky' region.

	\item[How long before inappropriate convergence becomes significant?]
	In the original paper on TCP Cubic\cite{cubic}, convergence was still described 
	as 'short' when it took one or
	two hundred rounds (assuming flows even last that long). Therefore, relative to
	overall convergence time, it will be insignificant if a flow takes a couple of
	dozen rounds to work out whether it should be converging to an L4S or to a
	classic target.
\end{description}

Rather than take an absolute number of rounds before the CC behaviour starts
to transition, it would be better to depend on how strongly the other metrics are
indicating that a transition is necessary. For instance, the higher RTT
variance is, the fewer rounds would
need to elapse before allowing a transition to start.

If CE marking stops for a protracted period, it will be likely that a non-ECN
link has become the bottleneck. Then the choice between classic and scalable
ECN behaviour would be moot and the default loss response would be sufficient.
If CE marking were to pick up again later, it would be best to ignore (i.e. not
measure) any period with more than one loss but no  CE marking, then restart the 
detection algorithm if a CE mark ever appears again.

\subsubsection{Dependence on being Self-Limited}\label{ecn-fallbacktr_self-limited}

If a TCP Prague flow is app-limited or receive-window limited (i.e.
self-limited), there is no great need to fall back to classic behaviour on
receipt of an ECN mark.

The presence of CE marks while the flow is not
trying to fill the pipe (its send buffer is empty\footnote{With the
introduction of pacing, it is no longer correct to measure whether a flow
is network limited by whether it is fully using the congestion window. It
is necessary to check whether the send buffer is empty instead.})
probably implies that a greedy
flow or other short flows are sharing the link. Then (assuming cwnd
validation is being used) the flow will not be increasing cwnd as much as
competing traffic could be. In that case, a large classic response to a
CE-mark could under-utilize the link until cwnd returned. So a small scalable
response would be more appropriate.

Also, any tendency towards classic behaviour due to RTT variability (see below) will be 
more due to other flows. So 
the classic ECN variable ought to reduce by a certain amount per RTT while a
sender is self-limited. However, not being self-limited alone is not a reason
to increase the variable---for that there has to be a positive sign of a
classic ECN AQM, such as RTT variabiity.

Similarly, while the sender is idle, any previous detection of a classic ECN
bottleneck could become stale. However, during an idle period there are no
events to trigger any actions, so an adjustment to the classic ECN variable will have to be
made at the restart of activity based on TCP's idle timer.

\subsubsection{Dependence on RTT Variability}\label{ecn-fallbacktr_RTT_deviation}

A large degree of RTT variability is the surest way to detect a
classic-ECN bottleneck. So, if accompanied by CE marking it is likely to imply
a classic ECN AQM at the bottleneck. For the Internet, 'large variability' can be
quantified as more than about 1.5\,ms of variability, given the target L4S delay
will generally be 1\,ms or less while the lowest target delay to which classic
AQMs are recommended to be configured is about 5\,ms. So any classic queue
could vary from zero to slightly above that.

Pseudocode for dependence of classic ECN fall-back on RTT variability will be given in
\S\,\ref{ecn-fallbacktr_passive_detection}. But first, the following two
subsections will discuss possible false positives and false negatives.

\paragraph{Non-queuing causes of RTT variability with an L4S bottleneck}\label{ecn-fallbacktr_false_positives}\leavevmode\\

RTT variability can have other causes than queuing:
\begin{itemize}
	\item A reroute.
	\item Variability in Interrupt handling, processor scheduling and batched
	processing by the endpoints and by nodes on the path. 
\end{itemize}

\paragraph{Reroute:} A moving average of RTT and deviation of the RTT from
this average does not filter out step changes in the base RTT
(e.g.\ due to a reroute), which could cause the moving average to be
temporarily 'incorrect' so that the mean deviation from this incorrect average
would temporarily expand (see \autoref{fig:ecn-fallbacktr_reroute}). The pseudocode in
\autoref{ecn-fallbacktr_alt-srtt} is intended to fill that gap.

\paragraph{Other Non-Queue Variability:} The passive detection algorithm assumes that the
combined result of all these variations will be small compared to variability
of a classic ECN queue. This assumption has turned out to be sufficient in testing so far. 
But it will need to be tested in a wider range of scenarios and parameters
altered accordingly (\S\,\ref{ecn-fallbacktr_detection_params}). 
If necessary active detection will need to be added, which is designed to 
complement passive detection where this assumption breaks down.

\paragraph{Low RTT variability with a classic ECN bottleneck}\label{ecn-fallbacktr_false_negatives}\leavevmode\\

RTT variability will not distinguish a classic ECN bottleneck in the following cases:
\begin{itemize}
	\item A high degree of flow multiplexing at a shared-queue bottleneck with a
	classic ECN AQM. The averaging effect of large numbers of uncorrelated
	sawteeth causes the mean deviation of the RTT of \(N\) flows sharing a buffer
	to be about \(1/\sqrt{N}\) of that of 1 flow, derived straightforwardly from
	the Central Limit Theorem~\cite{Appenzeller04:Sizing_buffers}.
	\item ...any others?
\end{itemize}

Few networks are designed so that sharing at a link serving a large number of
individual flows is controlled by the end-points, let alone with a classic ECN
AQM at this link as well. Nonetheless, we will consider three cases where this
could possibly occur:
\begin{description}
	\item [Commercial ISP's access link with a shared-queue classic ECN AQM:]
	Invariably, the operator designs the network so that the bottleneck is in the
	access link allocated to each customer, the capacity of which is isolated from
	other customers using a scheduler. For the mean deviation of a flow to appear
	to be 5\(\times\) lower (e.g.\ \(800\,\mu\)s rather than 4\,ms), at least 25 classic
	flows would have to be multiplexed together, by the Central Limit formula
	above. Such a scenario can occur within a single customer's access. However,
	for the fall-back algorithms of each flow to be fooled into thinking the
	bottleneck was L4S, that many classic flows would all have to run continually
	with no disruption from other flows. We have to accept that the algorithm
	could give a false negative in such a scenario, which is unlikely but
	possible.
	
	\item [Commercial ISP's core or peering link with a shared-queue classic ECN
	AQM:] The bottleneck can sometimes shift to a core link or more likely a
	peering point, where there per-customer scheduling is unlikely to be deployed and 
	flow multiplexing will be high enough to keep RTT very
	smooth. Usually this occurs during some sort of anomalous conditions, e.g.\ a
	provisioning mistake, a core link failure or a DDoS attack. If it does, the
	concern is that L4S flows could out-compete classic flows. Nonetheless, the
	scope for a high degree of flow rate inequality is very limited, as explained
	in \autoref{ecn-fallbacktr_common}.
	
	\item [Campus network access link:] A corporate or University network is
	rarely designed with an individual bottleneck for each user. Rather, each user
	typically has high speed connectivity to the campus (e.g.\ 1Gb/s Ethernet) and
	all stations using the Internet at any one time bottleneck at the campus
	access link(s) from the Internet. In such an access link, L4S flows will not
	coexist well with classic flows (as in
	\autoref{fig:ecn-fallbacktr_rate_flow}). It is not known whether any campus
	networks use classic ECN AQM in their access link, but they might do. Until
	the operator of such an AQM can deploy an L4S AQM, an unsatisfactory
	work-round would be to reconfigure the AQM to treat ECT(1) as Not-ECT so that
	it uses drop not CE as a signal for L4S flows. However, this would
	disincentivize L4S deployment in the affected campus networks. The 
	alternative of some campus networks just allowing the unfairness 
	would also be an option (applications already open multiple flows to
	achieve a similar advantage in the access to existing campus networks).
\end{description}

\subsubsection{Dependence on Minimum RTT}\label{ecn-fallbacktr_RTT_min}

Minimum RTT metrics are known to be problematic, especially where the buffer
is already filled by other traffic before a flow arrives. Therefore, it may be
preferable not to use this metric at all, and rely solely on RTT variability.

Nonetheless, it would do no harm to use a min RTT metric, as long as the
outcome was asymmetric. In other words, a large difference between
\texttt{fbk\_srtt} and \texttt{srtt\_min} would make classic fall-back more likely,
while a small difference would not make classic fall-back less likely. This is 
the approach taken in the pseudocode below.

\begin{figure}[h]
  \centering
  \includegraphics[width=0.45\linewidth]{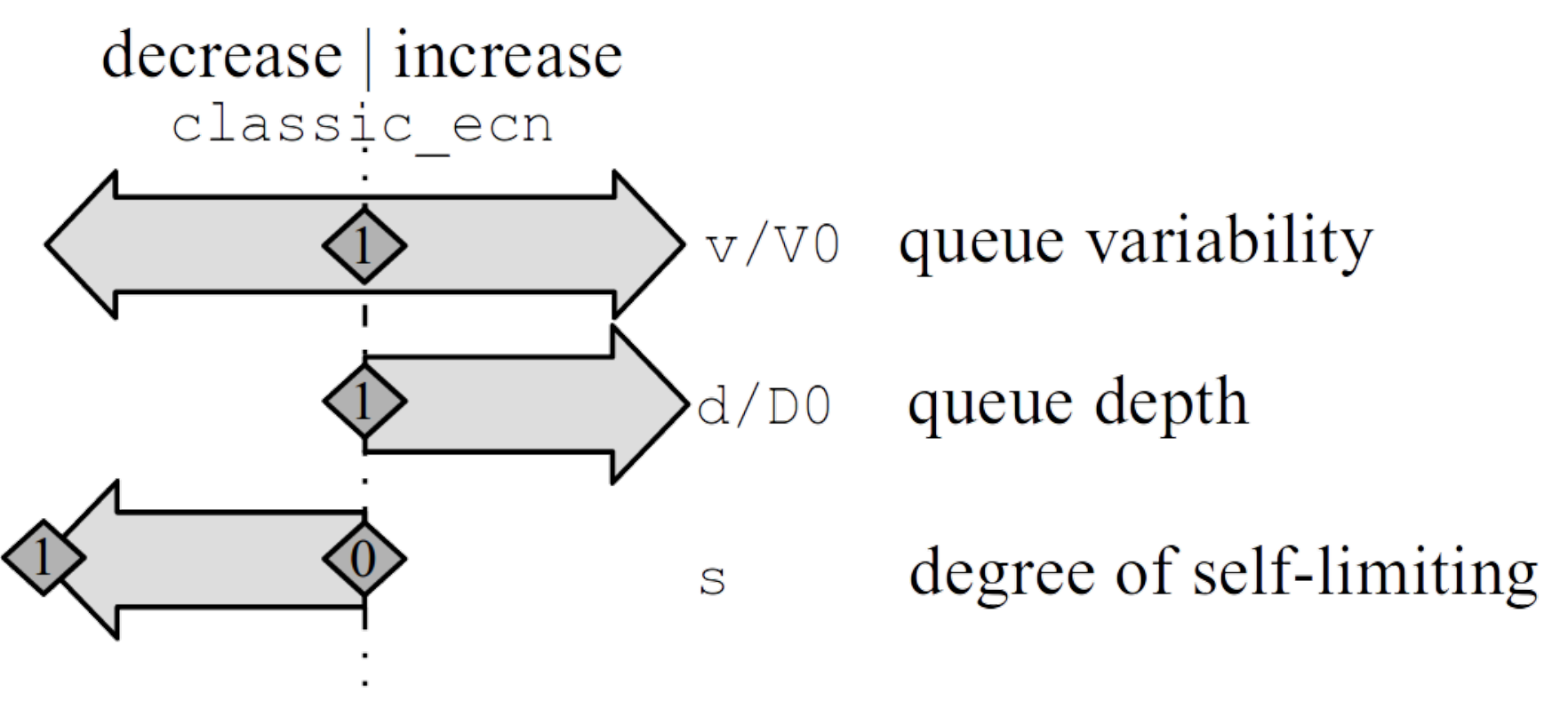}
  \caption{Visualization of the three main metrics, and how their effect on the \texttt{classic\_ecn} score will be constrained}\label{fig:ecn-fallbacktr_metric-push-pull}
\end{figure}

\paragraph{Subsection Summary:} \autoref{fig:ecn-fallbacktr_metric-push-pull} depicts the three main variables that will be used to drive Classic ECN AQM detection. It shows that queue variability should be able to increase or decrease the \texttt{classic\_ecn} score symmetrically. In contrast queue depth will be limited to only increasing the score, due to the well-known problem of measuring a false \texttt{rtt\_min}. The degree to which a flow is self-limiting is also asymmetric, but in this case it can only decrease the score.

\subsubsection{Dependence on Spacing Between ECN Marks}\label{ecn-fallbacktr_inter-mark}

This metric was only first thought of after v2 of the algorithm had been implemented and evaluated. At present it is just an idea, but it seems the most promising approach. It might prove to be the only metric that is needed, which would provide a really simple solution. Otherwise, it would complement the other metrics above.

\begin{figure}[h]
	\centering
	\includegraphics[width=0.7\linewidth]{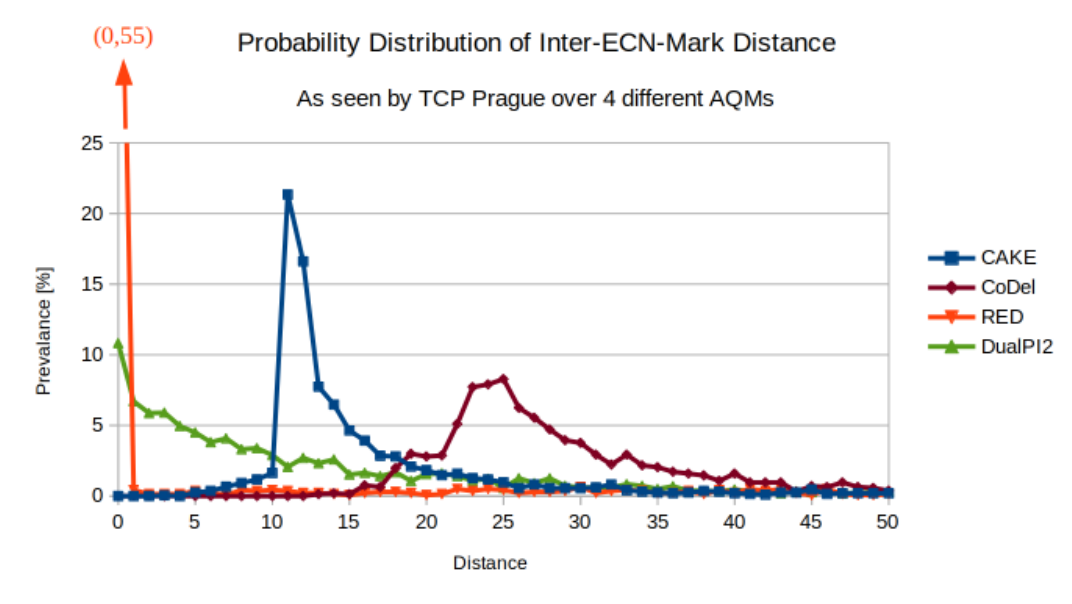}
	\caption{Distributions of spacing between marks for four different AQMs}\label{fig:ecn-fallbacktr_ce-space-distrib}
\end{figure}
The idea is to record the number of non-marked packets between one ECN mark to the next, and monitor the distribution of different spacings. At its simplest it might be sufficient to monitor the prevalence of just one selected spacing (e.g.\ 1, meaning a mark every other packet) relative to the total number of marks. The initial experiments shown in \autoref{fig:ecn-fallbacktr_ce-space-distrib} shoe that all the low spacings (except zero) are much higher for the L4S (DualPI2) AQM than the others, and there is fairly sound reasoning for that, as explained next.

By definition, all Classic AQMs filter out short term fluctuations in the queue. Most, if not all, known Classic AQMs also attempt to even out the spacings between the markings to avoid clustering---a process sometimes called derandomization. For instance:
`\begin{itemize}
	\item After RED calculates the marking probability, to prevent marks clustering it converts the marking probability into a uniform random variable (see \cite[\S\,7]{Floyd93:RED}).
	\item CoDel~\cite{Nichols18:codel_RFC} and any AQM derived from it (e.g.\ COBALT~\cite{Palmei19:Cobalt}) uses a control law that inherently introduces a measured space between markings.
	\item PIE~\cite{Pan13:PIE} and derivatives such as DOCSIS PIE~\cite[Appx.~M]{CableLabs:DOCSIS3.1} use a derandomization algorithm similar to that of RED.
\end{itemize}

In contrast, by definition, an L4S AQM is specifically precluded from smoothing its markings~\cite[\S\,5.2]{Briscoe15f:ecn-l4s-id_ID}, which is the job of the end system in the L4S architecture~\cite[\S\,4.4]{Briscoe15f:ecn-l4s-id_ID}. Therefore, typically the distribution of the spacings will be geometric. A geometric distribution has the counter-intuitive property that the greater the spacing, the less likely it will occur (much like the DualPI2 distribution in \autoref{fig:ecn-fallbacktr_ce-space-distrib}). So the most likely spacing is always 0, the next most likely 1, and so on.

A higher prevalence of zero spacing is a special case, and not a useful distinguishing feature, because any AQM might occasionally mark all packets for a while when it overshoots. 

\autoref{fig:ecn-fallbacktr_ce-space-distrib} is only an initial experiment based on single runs of 20s duration including flow startup with one link rate and one RTT (40\,Mb/s, 20\,ms) for two long-running flows started simultaneously; one Prague and one Cubic ECN.

It is very likely that the picture is less clear-cut in the Internet at large, although that will be hard to determine given no-one has yet captured a FIFO Classic ECN AQM in the wild, so we do not have one to dissect in the lab. Below some known complicating factors are discussed, some of which turn out not to be of concern, others need further investigation, and some are truly unknown and likely to remain so:
\begin{itemize}
	\item Different L4S Distributions:
	\begin{itemize}
		\item In a DualQ Coupled AQM, when there is a mix of L4S and Classic traffic, the L4S traffic is marked by coupling from an internal probability (termed \(p^{\prime}\)) pronounced p-prime, that also drives the Classic AQM. This would seem to mean that the distribution of spacing of L4S marks could be similar to that from the AQM in the Classic queue. However, on the Classic side, \(p^{\prime}\) has to be squared then derandomized. While on the L4S side \(p^{\prime}\) is merely factored up. So it would make no sense to derive the coupled L4S probability from the derandomized Classic probability, which would involve an unnecessary square root operation. This argument is very likely to apply whatever algorithm is used for the Classic AQM.
		\item L4S AQMs often use a simple step threshold. So marking will not be driven from a random variable, and therefore it will not follow a geometric distribution. A pure step threshold is likely to cause its own distinctive distribution of spacing, with long runs of solid marking followed by long runs of none. However, different patterns are likely to result from different levels of short flows in the background.
	\end{itemize}
	\item Different Classic Distributions (assuming FIFO Classic ECN AQMs do exist):
	\begin{itemize}
		\item Even if a Classic ECN AQM derandomizes marking in a FIFO, it only derandomizes the spacing between marks in the aggregate, not in each flow. If there are multiple flows the packets from each flow tend not to take perfect turns as they arrive, so the spacing between markings in each flow becomes more random again (it un-derandomizes)~\cite[\S\,5.6]{Briscoe15d:PIE_rvw}. This could start to look more like the distribution of spacing from an L4S AQM.
		\item It is possible that some Classic ECN AQMs will not include derandomization code, for instance in high-speed switches to simplify the hardware process.
		\item Large numbers of AQM designs have been proposed in research literature and in patents. If any were implemented, they might not all have included derandomization.
	\end{itemize}
\end{itemize}

If further experimentation proves that the spacing between marks still has merit as a metric, it will need to be designed into an algorithm that increases the \texttt{classic\_ecn} variable the more likely the AQM is Classic. And it will need to take account of periods when the flow might be application limited, receive-window-limited or idle. However, as already explained, the algorithm described below was designed implemented and tested before using an inter-mark spacing metric had been thought of.

\subsection{Passive Detection Pseudocode }\label{ecn-fallbacktr_passive_detection}

The following pseudocode pulls together all the passive detection ideas in the preceding sections.
\begin{verbatim}
/* Parameters */
#define V 0.5          // Weight of queue *V*ariability metric
#define D 0.5          // Weight of mean queue *D*epth metric
#define S 0.25         // Weight of *S*elf-limiting metric
#define C_FRAC_IDLE 2  // Multiplicative reduction in classic_ecn each idle timeout
#define CLASSIC_ECN 1 // Max of transition range for classic_ecn score
#define L_STICKY 16*V  // L4S stickiness incl. min rounds from CE onset to transition
#define C_STICKY 16*V  // Classic stickiness
#define V0 750         //  Reference queue *V*ariability [us]
#define D0 2000        // Reference queue *D*epth [us]

/* Stored variables */
classic_ecn; // Signed integer. The more +ve, the more likely it's a classic ECN AQM
rtt_min;    // Min RTT (using Kathleen Nichols's windowed min tracker in Linux)
fbk_srtt;           // The smoothed RTT (see later pseudocode)
fbk_mdev;           // The mean deviation of the RTT (see later pseudocode)
s;           // Proportion of the latest RTT that was self- (app- or rwnd-) limited

// Temporary variables to improve readability
v = fbk_mdev;
d = fbk_srtt - rtt_min;           // The likely mean depth of the queue. 
delta_;

/* The following statements are intended to be triggered by the stated events */

{   // On connection initialization
    classic_ecn = -L_STICKY;
}

{   // On CE feedback, enable delta_ calc'n if classic_ecn is clamped at its minimum 
    classic_ecn += (classic_ecn <= -L_STICKY);
}

{    // On expiry of idle timer
    if (classic_ecn > 0) {
        classic_ecn = classic_ecn/C_FRAC_IDLE;
        re-arm_idle_timer();
    }
}

{   // Per RTT
    if (classic_ecn > -L_STICKY) {    // Suppress delta_ calc'n if classic_ecn at min
        delta_ = V*lg(v/V0) + D*lg(max(d/D0, 1)) - S*s;
        classic_ecn = min(max(classic_ecn  + delta_, -L_STICKY), C_STICKY);
    } else {
        ect_tracers = 0;    // Unsuppress ect_tracers (for active detection in Section 5)
    }
}

{   // Per ACK
    // Update fbk_mdev and  fbk_srtt (see later pseudocode)
}
\end{verbatim}

\paragraph{Passive Detection Pseudocode Walk-Through}

While \texttt{classic\_ecn} sits at \texttt{-L\_STICKY}, calculation of
\texttt{delta\_}, the change in \texttt{classic\_ecn}, is suppressed to save
unnecessary processing. Maintenance of the variables used in this calculation
could also be suppressed (not shown).

At connection initialization, maintenance of the \texttt{classic\_ecn}
variable starts off in the above quiescent state. Feedback of a CE mark awakens it
by incrementing \texttt{classic\_ecn} by its minimum integer granularity (1).

Every RTT, as long as \texttt{classic\_ecn} is not in its quiescent state, the
per-RTT change in \texttt{classic\_ecn} is calculated. This is the core of the
passive classic ECN detection algorithm. To aid readability, a
temporary variable (\texttt{delta\_}) is assigned to this intermediate
calculation.

The change in \texttt{classic\_ecn} consists of three terms, each weighted
relative to each other by the three parameters \texttt{V}, \texttt{D} and
\texttt{S}: 
\begin{description} 
	\item[RTT Variability, \texttt{v} (\S\,\ref{ecn-fallbacktr_RTT_deviation}):]
	The metric \texttt{lg(v/V0)} is used, where lg() is an approximate (fast)
	base-2 log (see Appendix \ref{ecn-fallbacktr_ilog}) and \texttt{V0} is a reference mean-deviation parameter (default
	\(750\,\mu\)s). It is increasingly hard to achieve smaller deviations, so it is
	necessary to use a log function in order to ensure that a mean deviation of,
	say, \(23\,\mu\)s moves the classic ECN variable as much downwards as a mean
	deviation of 24\,ms moves it upwards (respectively 32 times smaller and 32 times
	larger than \texttt{V0} = \(750\,\mu\)s).
	
	\item[Likely Mean Queue Depth, \texttt{d} (\S\,\ref{ecn-fallbacktr_RTT_min}:]
	The metric \texttt{lg(max(d/D0), 1)} uses the log of the ratio of \texttt{d}
	over the reference queue depth \texttt{D0} for the same reason as the previous
	bullet. However, the max() with respect to 1 ensures it ignores queue
	depths below the threshold, which could be suspect, as explained in
	\S\,\ref{ecn-fallbacktr_RTT_min}.
	
	\item[Self-Limitation, \texttt{s} (\S\,\ref{ecn-fallbacktr_self-limited}):]
	Here the fraction of the RTT that was self-limited can be used directly. 
\end{description}

The last two variables can each only push \texttt{classic\_ecn} in one
direction (see \autoref{fig:ecn-fallbacktr_metric-push-pull}). Mean queue depth can only push it upwards (more classic), while
self-limitation can only push it downwards (less classic). If mean queue depth
is below the reference queue depth \texttt{D0}, or there is no self-limitation
in a round, the \texttt{classic\_ecn} indicator remains unchanged.

If, on balance, the calculations to detect classic ECN AQM are positive, 
delta\_ increases \texttt{classic\_ecn} towards its maximum
(\texttt{C\_STICKY}). But if they are negative, delta\_ decreases \texttt{classic\_ecn} 
towards its minimum (\texttt{-L\_STICKY}) where further calculations
will be suppressed, at least until calculations are reawakened by the next CE
mark.

During an idle period, \texttt{classic\_ecn} is exponentially reduced by
default to 1/2 of its previous value on every expiry of the idle timer, but
only if it is positive. Thus, while idling, a connection that had detected a
classic ECN AQM will gradually drift to the L4S end of the transition, but
towards the cusp of the transition. Then, if it continues to detect a classic
ECN AQM once it restarts, It will immediately transition to classic ECN mode
again, while it is restarting.

\subsection{RTT Smoothing}\label{ecn-fallbacktr_srtt}

The smoothed RTT and the mean deviation from that smoothed RTT are the 
primary metrics used by the passive classic ECN AQM detection algorithm. They 
both use exponentially weighted moving averages. 

All implementations of TCP already maintain two such variables; the
EWMA of the RTT and an EWMA of the mean deviation of RTT samples from
that smoothed RTT. In Linux, they are called \texttt{srtt} and
\texttt{mdev} and both are updated on every ACK. It was originally hoped
to reuse these
variables for classic AQM detection. However, they are unsuitable for two
reasons:

\begin{itemize}
	\item TCP maintains \texttt{srtt} and \texttt{mdev} for calculation of
	its retransmission time out (RTO). For this it needs a worst-case measure
	of the duration between sending a packet and seeing an ACK. So, whenever
	TCP receives an ACK, it measures the RTT for RTO purposes (\texttt{mrtt})
	from when it sent the \emph{oldest} newly acknowledged packet. This
	includes the duration of the delay introduced by the receiver's delayed
	ACK mechanism, making these metrics unusable as a smoothed measure of the
	actual RTT. The presumption that RTT is used for RTO calculation is also
	built into the time the receiver is expected to stamp into TCP's
	time-stamp option, making timestamps unusable for measuring actual round
	trip delay as well.
	
	\item In Linux, TCP smooths these variables (\texttt{srtt} and \texttt{mdev}) over very 
	few ACKs (8 and 4 respectively) whereas, for classic ECN AQM detection,
	ideally RTT needs to be smoothed over at least one sawtooth of the flow's own
	window, in order to pick up the full depth and variability of the bottleneck
	queue.
\end{itemize}

Terminology: After \(g\) iterations, a step change in an EWMA's input
will have changed the moving average by about 63\% of the step, or
precisely \(1-1/e\), where \(e\) is the base of the natural logarithm.
This is what the phrase smoothed `over' a certain number of iterations
means. It is another way of saying that the smoothing gain of the EWMA is
\(1/g\). For instance, saying that TCP smooths \texttt{srtt} over 8 ACKs,
is alternative way of saying the smoothing gain is 1/8. The smoothing
gain of an EWMA is the fraction of each newly measured value that is
added to the average on every update (and the fraction of the old average
that is subtracted).

In the original discussion of RTO estimation in Jacobson and
Karels~\cite[Appx A]{Jacobson88b:Cong_avoid_ctrl} the EWMAs for
\texttt{srtt} and \texttt{mdev} were recommended to be smoothed over
respectively a little greater and a little less than the congestion
window, measured in segments. However, the Linux code has never related
these parameters to \texttt{cwnd} and they remain hard coded as they were
when typical values of \texttt{cwnd} were hundreds of times lower than
they are today.\footnote{Linux TCP uses a third gain value of 1/32 in
the case where \texttt{mrtt} is less than the smoothed average AND its
distance from the average has increased. A comment in the code points to
the Eifel algorithm as a possible rationale, but another comment
sarcastically says that the code implements the opposite of what was
intended, without saying why it has not been fixed.}

Linux already maintains a more precise RTT metric on each ACK. It stores
the times at which it sends every packet and it measures a variable we
shall call \texttt{acc\_mrtt} from the time at which it sent the packet that
elicited the ACK to when it receives the ACK.

It is not costly to maintain an extra pair of EWMAs based on this more
precise \texttt{acc\_mrtt} variable. However, it is necessary to detect queue variations over the timescale of TCP's sawteeth, which requires smoothing over a very large number of ACKs; far more than the 4 or 8 currently used in Linux and other OSs for RTO calculation.

\autoref{ecn-fallbacktr_srtt_adapt} addresses the question of how many ACKs to smooth over. The theoretical number of ACKs in a Classic sawtooth is \((\texttt{ssthresh})^2\). However, the appendix explains that this can be considered as a rare upper limit. In practice the sawteeth of Classic congestion controls rarely reach this size, especially in the presence of other traffic, such as short flows.

In experiments with a range of link rates between 4\,Mb/s and 200\,Mb/s
and RTTs between 5\,ms and 100\,ms, the formula that
resulted in a good compromise between precision and speed of response
was: \[\mathtt{fbk\_g\_srtt} = 2 * (\mathtt{ssthresh})^{3/2}.\]

And the mean deviation is smoothed twice as slowly as the RTT
itself, i.e.: \[\mathtt{fbk\_g\_mdev} = \mathtt{fbk\_g\_srtt} * 2.\]
\bob{Currently, the Linux code uses g\_mdev = g\_srtt * 2, but I would like to try g\_mdev = g\_srtt and even g\_mdev = g\_srtt / 2}

\subsection{RTT Smoothing Pseudocode}\label{ecn-fallbacktr_pseudocode_srtt}

\begin{verbatim}
// fbk_g_srtt = U2 * (ssthresh)^(U1)
#define U1 (3/2)
#define U2 2
#define FBK_G_RATIO 2    // fbk_g_mdev = fbk_g_srtt *  FBK_G_RATIO

/* Stored variables */
fbk_srtt;                // Smoothed RTT
fbk_mdev;                // Mean deviation
fbk_g_srtt;              // srtt gain (initialized dependent on ssthresh)
fbk_g_mdev;              // mdev gain (initialized dependent on ssthresh)

/* Temporary variables */
u32 acc_mrtt;            // Measured RTT from newest unack'd packet
s64 error_;

{   // At start of connection, either after first RTT measurement or from dst cache
    fbk_srtt = acc_mrtt;
    fbk_mdev = 1;        // No need for conservative init, unlike for RTO
}

{   // Per ACK
    // Update EWMAs
    error_ = acc_mrtt - fbk_srtt;
    fbk_srtt += error_  /  fbk_g_srtt;
    fbk_mdev += (abs(error_) - fbk_mdev) / fbk_g_mdev;
}

{   // Per ssthresh change, including when initialized
    fbk_g_srtt = U2 * ssthresh^U1
    fbk_g_mdev = fbk_g_srtt * FBK_G_RATIO
}
\end{verbatim}

\paragraph{RTT Smoothing Pseudocode Walk-Through}

The EWMAs in the `Per ACK' block are straightforward. The mean deviation
is defined as the EWMA of the non-negative error, so it is calculated
using the \texttt{abs()} function, which returns the absolute
(non-negative) value of its argument.

In the `Per-\texttt{ssthresh} change' block, the gains used for the EWMA
are adjusted to maintain their relationship with \texttt{ssthresh} as
just explained in \S\,\ref{ecn-fallbacktr_srtt}.

Appendix \ref{ecn-fallbacktr_srtt_adapt} gives more detailed pseudocode
based on that above, to address the questions of EWMA precision and
upscaling when using integer arithmetic in the kernel.

\subsection{Questioning Assumptions used for Passive Detection}\label{ecn-fallbacktr_detection_open}

\subsubsection{Clocking Interval for \texttt{classic\_ecn}}\label{ecn-fallbacktr_clocking}

So far, it has been assumed that \texttt{classic\_ecn} should be recalculated
once per RTT, for all the detection metrics except idling time. The question
of which event is appropriate to update this variable needs to be addressed explicitly.

Four potential intervals on which to clock changes are:
\begin{itemize}
	\item round trips
	\item absolute time intervals
	\item after a certain amount of sent packets, or even sent bytes
	\item after a certain amount of feedback (ACK counting).
\end{itemize}

\begin{description}
	\item[Stabilization and convergence:] It makes sense to count how long to wait
	for a connection to stabilize and converge in round trips, because each flow
	adjusts iteratively on a round trip timescale.
	
	\item[RTT variability:] There is an argument that changes to
	\texttt{classic\_ecn} due to RTT variability should be clocked on a count of
	the ACKs received, e.g. every 32 or every 64 ACKs. This is because the
	precision of round trip smoothing and measurements of mean deviation depends
	on how many ACKs have contributed to the average. However, the variability of
	the queue itself alters dependent on evolution of each flow's congestion
	window, which adjusts on a round trip timescale. Therefore dependence of the
	value of \texttt{classic\_ecn} on RTT variability metrics should be clocked
	against round trips.
	
	\item[Self-limitation:] Self-limitation is measured as a proportion, so it
	does not matter whether it is a proportion of a round, a proportion of a
	certain time period, or a proportion of any other metric. Given other metrics
	should be clocked on round trips, it makes sense to clock self-limitation
	calculations on the same events.
	
	\item[Idling:] There is no basis to argue that changes to \texttt{classic\_ecn}
	due to idling should be clocked on any particular metric. Nonetheless, the
	only metric that continues to clock during an idle period is time, so this is
	the only practical metric to use.
\end{description}

Counting either in sent packets or absolute time would be easy to implement,
but neither seem to have any logical backing, for any of the metrics.

\subsubsection{Clocking Interval for RTT EWMAs}\label{ecn-fallbacktr_clocking_ewmas}

\autoref{ecn-fallbacktr_srtt_adapt} addresses the question of how many
ACKs to smooth the RTT EWMAs over. However, it does not question whether
these EWMAs should be or need to be updated on every ACK, particularly
given the gain has to be so tiny, which implies that less frequent
updates with a larger gain would be no less precise but less costly in
processing terms.

It would be possible to measure the RTT of only a sample of ACKs, and
increase the gain accordingly. However, this would involve extra
complexity, and the actual additional processing cost is only 4 adds, 2
bit-shifts and a compare per ACK (see Appendix
\ref{ecn-fallbacktr_srtt_adapt}), which is not particularly problematic.
Sampling has its own complexity because the sampling ratio would have to
be adaptive, in order that it could revert to 100\% when the number of
ACKs per round trip was small. It would also require an extra state
variable to be maintained per ACK, in order to record when the next
sample was due.

Therefore, on balance it has been decided to maintain the EWMAs of RTT
and mean deviation per ACK. But there is nothing to stop other
implementers using sampling.

\subsection{Parameter settings}\label{ecn-fallbacktr_detection_params}

The parameters currently given in the passive detection pseudocode
(\S\S\,\ref{ecn-fallbacktr_passive_detection} \&
\ref{ecn-fallbacktr_pseudocode_srtt}) are heuristics arrived at as a
result of a large number of calibration experiments over a testbed with
link rates of 4--200\,Mb/s and RTTs of 5--100\,ms. Higher link rates
could have been included, but the most challenging part of the range is
the low end. \S\,\ref{ecn-fallbacktr_evaluation} summarizes the results
of evaluations using the parameters resulting from these calibration
experiments.

The current set of parameter values has solely been informed by
experiments with CoDel~\cite{Nichols18:codel_RFC} as the Classic AQM 
and DualPI2~\cite{Briscoe15e:DualQ-Coupled-AQM_ID} as the L4S AQM
(both with default settings). CoDel was chosen given it is a good 
candidate for the worst-case for the fall-back algorithm to detect, for two 
reasons:
\begin{itemize}
\item CoDel's default setting of \texttt{target} is very low (5\,ms). This pegs its
queue delay low relative to other Classic AQMs and tends to cause
under-utilization for all but the lowest RTTs. Both these factors lead to
very little, if any, queuing delay. This makes CoDel more difficult to
distinguish from an L4S AQM, and therefore likely to be a worst-case from
a parameter-setting viewpoint.
\item CoDel's design is highly specific to 
Classic congestion controls, probably exhibiting the slowest response of 
all AQMs to the high levels of ECN signalling induced by scalable 
congestion controls~\cite{Palmei19:Cobalt}. 
\end{itemize}
As the algorithm is evaluated with other
AQMs, it may be necessary to tune the parameters further.
\section{Out-of-Band Active Detection `of Classic ECN AQMs}\label{ecn-fallbacktr_OOB_active}

Out-of-band testing presents few constraints on the traffic patterns that can be
used. So it is possible to use unsubtle approaches like running two flows in
parallel, which is described here.

A server could be set up with ECN enabled so that, when a test client accesses
it, it serves a script that gets the client to open two parallel long-running
flows. Then it could serve one with a Classic CC (that sets ECT(0)) and one with
a scaleable CC (that sets ECT(1)).

If neither flow induces any ECN marks, it can be presumed the path does not
contain a Classic ECN AQM.

If either flow induces some ECN marks, the server could measure the relative
flow rates and round trip times of the two flows.
\autoref{tab:ecn-fallbacktr_OOB_active} shows the AQM that can be inferred for
various cases.

\begin{table}[h]
\centering
\begin{tabular}{|c|c|l|}
	\hline
	Rate          & RTT & Inferred AQM \\
	\hline
	\(L>C\)        & \(L=C\) & Classic ECN AQM (FIFO)\\
	\(L=C\)        & \(L=C\) & Classic ECN AQM (FQ)\\
	\(L=C\)        & \(L<C\) & FQ-L4S AQM\\
	\(L\approx C\) & \(L<C\) & DualQ Coupled AQM\\
	\hline
\end{tabular}
\caption{Out-of-band testing with two parallel flows, showing those cases with
	some ECN marks apparent. L=L4S, C=Classic.}\label{tab:ecn-fallbacktr_OOB_active}
\end{table}

The power of this approach is that is can identify whether a Classic ECN AQM is
in a FIFO (which is the type being sought), or in an FQ scheduler (which is
not).

In the case of the DualQ Coupled AQM, the relative rates will depend on the
configuration, so they are only shown as '\(\approx\)'. Other combinations of
outcomes are theoretically possible, e.g.\ \(L>C\) for RTT, but the test would
have to be abandoned in such cases, because no known AQM would cause such an
outcome.

\section{In-Band Active Detection of Classic ECN AQMs}\label{ecn-fallbacktr_active}

\subsection{In-Band Active Detection: ACK Problems in TCP}\label{ecn-fallbacktr_active_problem}

When active detection is used in-band, by definition it alters the traffic. So
care has to be taken not to intrude too much into live sessions.

The most likely testing pattern would be to start with passive testing, then
only introduce a little active testing before changing the CC behaviour,
particularly if the results were not clear cut. This is the approach expected
for Solution No.1 (\S\,\ref{ecn-fallbacktr_active_solution1}). Alternatively,
active testing could be used on a small sample of flows, which is the
most likely approach to be taken for Solution No.2
(\S\,\ref{ecn-fallbacktr_active_solution2}).

We start out on the assumption that an L4S sender will set some packets within a
single flow to ECT(0) even though it should set them to ECT(1). In Solution No.1
the sender checks for differences in RTT. In Solution No.2 it checks for
differences in marking.

One can imagine a number of na\"ive active measures that a sender could take:
\begin{itemize}
    \item It could duplicate a small proportion of ECT1 packets and set them as ECT0.

    \item it could set a small proportion of packets to ECT0 instead of ECT1;
\end{itemize}

We shall call these 'ECT tracer' packets,
because they trace whether the ECT field causes a packet to be classified into
a different queue. However, if TCP is being used, and if the receiver was using delayed ACKs (most do),
it would confound these na\"ive approaches:
\begin{itemize}
	\item in the first case, even if both duplicates were acknowledged (the first
	to arrive might not be), the sender would not be able to tell from the
	acknowledgement(s) which duplicate had arrived first\footnote{Unless AccECN
	TCP feedback with the TCP Option was implemented and it successfully traversed
	the path, but that is too unlikely to rely on}.
	    
	\item In the second case, some ECT0 packets would not trigger an ACK so their
	delay could not be measured. Also, if the bottleneck were an L4S DualQ Coupled
	AQM, any queuing delay suffered by the ECT0 packets would hold back the
	connection, and some might be delayed enough relative to ECT1 packets to make
	TCP believe they had been lost, causing the sender to spuriously retransmit
	and spuriously reduce its congestion window.
\end{itemize}

\subsection{In-Band Active Detection: Solution No.\,1}\label{ecn-fallbacktr_active_solution1}
\subsubsection{Active Solution No.\,1: Approach}\label{ecn-fallbacktr_active_solution1_approach}

A better strategy would be:
\begin{itemize}
	\item for the sender to make the receiver override its delayed ACK mechanism
	by ensuring that at least part of both tracer packets duplicates bytes already
	sent. This is because standard TCP congestion control
	\cite{IETF_RFC5681:TCP_algorithms, IETF_RFC2581:TCP_algorithms} recommends
	that a receiver sends an immediate ACK in response to duplicate data to
	expedite the fast retransmit process, and this recommendation has stood since
	the first Internet host requirements in 1989~\cite{IETF_RFC1122:HostReqs}.
		
	\item for either tracer packet to push forward the acknowledgement counter, so
	that the sender can tell which probably arrived first (there can be no
	certainty, because ACKs can be reordered). \end{itemize}

The best sender-only strategy so far conceived would be as follows (also
illustrated in \autoref{fig:ecn-fallbacktr_cc-changeover}):
\begin{enumerate}
	\item If, during passive testing, the \texttt{classic\_ecn} indicator is approaching the transition
	range from below, i.e.\ negative but close to zero, for a small proportion of
	segments send instead the following three smaller packets, all back-to-back:
    \begin{itemize}
		\item a larger front segment marked ECT1;
		 
		\item a smaller middle segment marked ECT0, duplicating at least the last 2\,B
		of the first segment;
		 
		\item a rear segment marked ECT1 of the same size as the second but only
		duplicating the last byte of the first segment;
    \end{itemize}
	\item If the ACK for the middle tracer arrives after that for the rear tracer,
	the AQM is likely to be L4S (unless some other mechanism happens to have
	coincidentally re-ordered the packet stream at this point);
\end{enumerate}

\begin{figure}
  \centering
  \includegraphics[width=0.49\linewidth]{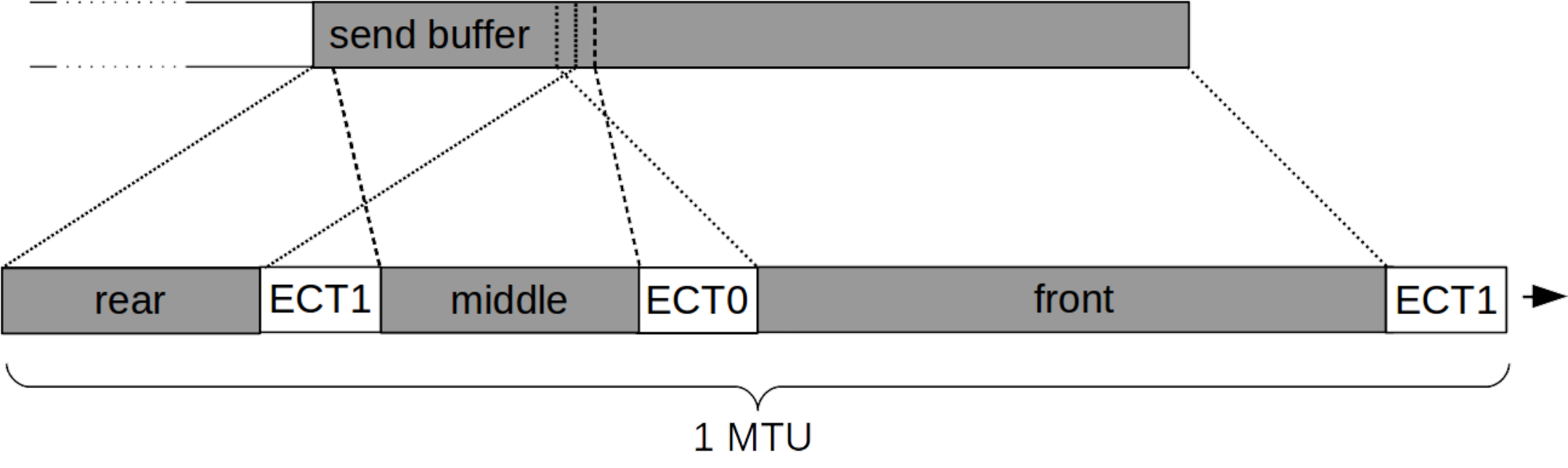}
  \caption{Tracer packets to detect separate treatment of ECT1 packets}\label{fig:ecn-fallbacktr_ect-classifier-tracer}
\end{figure}

Note that the ECT0-marked packet only includes redundant bytes, so if it is
delayed (or dropped) by a classic queue, it does not degrade the L4S service.

The combined size of all three packets should be no greater than 1\,MTU so
that, if packet pacing is enabled, all three packets will remain back-to-back
without having to alter pacing (also the two back-to-back ECT1 packets will
cause no more of a burst in an L4S queue than a single packet would).

The front packet is larger to reduce the risk that detection of L4S AQMs will
sometimes fail. Being larger, it is more likely to still be dequeuing when the
rear packet arrives at the bottleneck. Otherwise, if there was a DualQ Coupled
AQM at the bottleneck, and if there was no other classic traffic queued ahead
of the middle tracer, it could start dequeuing after the front packet had
dequeued, but before the rear tracer arrived.

Nonetheless, in order to minimize the possibility that the small tracer
packets are treated differently by middleboxes, they should be larger than the
size \(S\) of the largest packet that might be considered 'small' by common
acceleration devices (\(S = 98\)\,B would probably be sufficient).




\bob{ToDo: Write up a symmetric facility at the classic end of the spectrum.
And write-up pseudocode in the following section, including a way to take
multiple active measurements and act on their combined outcome.}

\subsubsection{Active Solution No\.1: Pseudocode}\label{ecn-fallbacktr_active_solution1_pseudocode}

The following pseudocode implements the active detection ideas in
\S\,\ref{ecn-fallbacktr_active_solution1_approach}. It uses some of the macros and variables
defined in the passive detection pseudocode above.

\begin{verbatim}
// Parameters
#define TRACER_NUM 4    // Number of sets of 3 active tracers to send
#define REAR_SIZE 98    // Min size of middle and rear tracers [B]

ect_tracers = 0;  // Unsigned int storing remaining tracers (-ve means disarm sending)
tracer_nxt = 0;   // Point in the sequence space after the most recently sent tracer
                  // special (tracer_nxt == 0) disables checking for tracer ACKs

// Functions
send_tracer(start, size, ecn);    // Sends ECT tracer seg from 'start' in send buffer

// The following statements are intended to be inserted at the stated events

{   \\ Per RTT
    if (classic_ecn >= -L_STICKY/TRACER_NUM && !ect_tracers) {
        ect_tracers = TRACER_NUM;
    }

    if (ect_tracers < 0)    // The tracer armed 1RTT ago has been sent
        ect_tracers *= -1;    // Arm sending of the next tracer
}
    
{   // Prior to sending a packet
    if (ect_tracers > 0 && snd_q >= smss) {
        front_size = smss - 2 * (sizeof_tcp_ip_headers() + REAR_SIZE);
        send_tracer(snd.nxt,                 front_size, ECT1);    // Front tracer
        send_tracer(snd.nxt - 2,             REAR_SIZE,  ECT0);    // Middle tracer
        send_tracer(snd.nxt - REAR_SIZE + 1, REAR_SIZE,  ECT1);    // Rear tracer
        tracer_nxt = snd.nxt;
        if (--ect_tracers) {
            ect_tracers *= -1;    // Negate to disarm sending of the next tracer
        } elif (classic_ecn >= -L_STICKY/TRACER_NUM) {
           ect_tracers -= TRACER_NUM + 1;    // Suppress further tracers
        }
    }
}

{   // On receipt of seg (pure ACK or data) 
    if (tracer_nxt) {
        if (rcv.nxt == tracer_nxt && seg.sack == tracer_nxt - 1)
            // middle arrived after rear, so probably L4S bottleneck
            classic_ecn = max(classic_ecn - L_STICKY/TRACER_NUM, -L_STICKY);
        if (ect_tracers == -TRACER_NUM - 1)    // Further ECT tracers have been suppressed
            tracer_nxt = FALSE;    // Suppress further ACK checking
    }
}
\end{verbatim}

\paragraph{Interaction between Active Testing and the \texttt{classic\_ecn} Indicator:}

Greater RTT variability during passive testing might imply either a classic bottleneck or an L4S
bottleneck combined with variability from another link (e.g.\ non-L4S WiFi).
Whereas low variability is more likely to imply an L4S bottleneck. Therefore
if the result of an active test is L4S, it pushes the \texttt{classic\_ecn}
indicator towards the L4S end, counteracting the pposite trend due to
variability. Whereas if the result of an active test is classic, it does not
need to alter \texttt{classic\_ecn}; it can leave variability to do that.

Active detection is more decisive, but it alters the normal transmission
pattern. So to avoid unnecessarily altering the sending pattern, passive
measurement alone is used first to determine whether active measurement is
worthwhile.

If active measurement proves necessary, the plan then is to send a small
number (default TRACER\_NUM = 4) of sets of three tracer packets. If any set
of tracers detects that an L4S AQM is likely, it moves the
\texttt{classic\_ecn} indicator towards the L4S end of the spectrum by an
amount \texttt{L\_STICKY/TRACER\_NUM}.

Thus if all 4 tests detect L4S, \texttt{classic\_ecn} reduces by
\texttt{L\_STICKY}. The tests start at \texttt{classic\_ecn >= -L\_STICKY/4},
so if all 4 tests detect L4S, it will return to its floor value of
\texttt{-L\_STICKY}, and the CC will never have behaved as anything other than
pure L4S. Bear in mind that the \texttt{classic\_ecn} indicator will still be
altered by the passive detection algorithm as well.

If, on the other hand, no set of tracers detects L4S, the active tests will
not alter the \texttt{classic\_ecn} indicator at all. Then, if the bottleneck
is classic, continuing passive tests will detect the higher RTT variability
and continue to push the \texttt{classic\_ecn} indicator towards the classic
end of the spectrum.

Between these two extremes, if not all the active tests detect L4S, the
\texttt{classic\_ecn} indicator will be pushed down less and stop short of its
floor. Then if RTT variability continues, passive detection will more rapidly
return it to the \texttt{-L\_STICKY/4} threshold where active tests resume.

\paragraph{State Variables}

To detect which tracer packet arrived first, it is necessary to store an
indication of which feedback to check. Therefore no more than one set of
tracers is sent per round trip, to minimize the per-connection state needed.
This also spaces out the tracer tests, so that the small amount of redundant
data each one sends hardly causes any inefficiency\footnote{With typical MTU
and header sizes, a set of 3 tracer packets consumes 1\,MTU, but sends about
12\% less TCP data that would normally be in a full MTU. If there were say 16
packets per round, this inefficiency would be reduced to 12\%/16 = 0.75\%}.

Two additional state variables are needed for each connection:
\begin{itemize}
	\item \texttt{ect\_tracers}: This state variable stores the number of ECT
	tracers outstanding. Zero is not really a special value; it just has the
	expected meaning---that no tracers are outstanding.
	
	\item \texttt{tracer\_nxt}: After a set of three tracer packets have been
	sent, \texttt{tracer\_nxt} stores the next byte in the sequence space. Then
	later the matching ACKs for the tracers can be found. If the ACK never
	arrives, there is just no outcome to the test.
\end{itemize}

Negative values of \texttt{ect\_tracers} are special; they store the number of
outstanding sets of tracers but disarm them for a round trip (so that the
feedback from the last one has time to return).\bob{There could be a race condition
here, where the variable will be needed for the next tracer before it has been used
to pick up the feedback from the last one.}

The negative value of \texttt{ect\_tracers} one lower than
\texttt{-TRACER\_NUM} (-5 by default) is a further special value that
suppresses all further tracers.

Tracers are not suppressed as long as the outcome after 4 tracers has reduced
the \texttt{classic\_ecn} indicator below the threshold at which active tests
are triggered (\texttt{-L\_STICKY/4}). Then, if the indicator rises to this
threshold again, another set of tracers can be triggered. But, if the
indicator has not reduced after the 4 tracer tests (i.e.\ all 4 tracer tests
pass without reordering), all further active tests are suppressed so that
continuing passive measurements are allowed to push the indicator upwards
towards the classic ceiling (causing the CC to transition to classic
behaviour).

If RTT variability reduces (e.g.\ because the bottleneck moves from a classic
to an L4S AQM) such that the passive tests on their own pull the indicator
down to the L4S floor, active tests suppression is removed by setting
\texttt{ect\_tracers = 0}.

\paragraph{Per Packet Processing Efficiency}

The special values of the variables \texttt{ect\_tracers} and
\texttt{tracer\_nxt} are used to suppress the more complex conditions that
would otherwise have to be checked per packet, respectively: whether each
packet to be sent should be replaced by a set of tracers; and whether each ACK
is feedback from a tracer.

For efficient implementation, rather than checking a flag variable on millions
of packets just to send or receive a few packets differently, it might be
better to somehow suppress regular packet sending, send the required number of
tracer packets manually, then resume sending. This will need to be
investigated during implementation.

\subsection{In-Band Active Detection: Solution No.\,2}\label{ecn-fallbacktr_active_solution2}
\subsubsection{Active Solution No.\,2: Approach}\label{ecn-fallbacktr_active_solution2_approach}

To date, the draft text of the experimental IETF RFC to specify the L4S
markings says:
\begin{verbatim}
   For backward compatibility in uncontrolled environments, a network
   node that implements the L4S treatment MUST also implement an AQM
   treatment for the Classic service as defined in Section 1.2.  This
   Classic AQM treatment need not mark ECT(0) packets, but if it does,
   it will do so under the same conditions as it would drop Not-ECT
   packets [RFC3168].
\end{verbatim}

It has been suggested that this could be amended to preclude any Classic AQM
from marking ECT(0) packets if it is coupled with an L4S AQM. Then, an L4S
source could detect a Classic ECN AQM by probing with ECT(0) packets to see if
they were marked. In the following, this will be termed \textbf{exclusive}
marking.

With a flow-queuing (FQ) scheduler, an L4S and a Classic AQM can both be applied
within each queue~\cite[\S\,5.2.7]{Hoeiland18:fq-codel_RFC} by applying
immediate ECT(1) marking at a shallower threshold. It has also been suggested
that it would not be necessary for the whole FQ system to choose between L4S and
Classic ECN. Instead, within each per-flow queue (FQ), the queue could continue
to support Classic ECT(0) marking, except it could be required not to mark
ECT(0) packets if the same queue had `recently' seen an ECT(1) marking. In the
following, this will be termed \textbf{FQ-exclusive} marking.

`Recently' could mean in the lifetime of the per-flow bucket, or after a
timeout. If the timeout option were chosen, it would require at least a 2-bit
counter per queue. The timeout could be implemented by initializing each counter
to zero, then setting it to 3 on every ECT(1) packet. Then a timer shared over
all queues could decrement all the counters at a regular interval, \(T\). ECT(0)
marking would be disabled in any queue with a non-zero counter, so the timeout
would last for at least \(2T\) and at most \(3T\).

The whole `Solution No.2' approach would still work even if some FQ AQMs
supported both L4S and Classic AQMs but did not implement FQ-exclusive marking
(important given such AQMs are already deployed). This is because the queue for
a flow with decent proportion of ECT(1) packets would tend to sit below the
shallow ECT(1) threshold and never reach the deeper point where ECT(0) marking
started.

A notable benefit of the exclusive marking idea is that it only needs to be
temporary, for instance during the early stages of the experimental phase of
L4S. If it proves not to be useful, perhaps because some better approach is
developed, the standards can be updated and Classic ECN AQMs can be allowed to
mark ECT0 packets alongside L4S ECT1 marking.

The following two test strategies have been proposed to exploit exclusive
marking, if it were standardized: 
\begin{description} 
	\item[ECT0 probes:] One approach would be to intersperse small ECT(0) probe
	packets within an L4S ECT(1) flow, while attempting to induce enough congestion
	for some of these probe packets to be likely to be dropped or marked. Then if
	any ECT(0) packets were marked it would be highly likely that there was a
	Classic ECN AQM at the bottleneck.

	\item[Late onset ECT1 stripe:] In this approach, a long-running flow would
	start with all ECT0 packets. Then after some congestion had been induced (CE
	marking or loss), a small proportion of packets (e.g.\ 5\%) would be marked
	ECT1 while the majority would continue as ECT0.
\end{description}

More details of each strategy including pros and cons are discussed below. It is
uncertain whether either would be considered acceptable for in-band testing;
ECT0 probes might be too slow, and the late onset ECT1 stripe might not be
sufficiently benign. However, other strategies might be developed.

\paragraph{ECT0 Probes:} This approach relies on being able to tell whether
specific packets have been ECN-marked. Therefore, if it were used over a TCP
connection, a technique similar to that in
\S\,\ref{ecn-fallbacktr_active_solution1_approach} would be needed to avoid the
sender being confused by delayed ACKs.

Although a marking on any ECT0 probe would prove that a Classic ECN AQM was
likely, absence of such a marking would not prove absence of a Classic ECN AQM.
If the ratio of ECT0 probes to ECT1 packets was \(1:P\), then the test would
have to continue until the number of ECT1-marked packets had exceeded some
multiple of \(P\), e.g. \(m*P\) (say  \(m=10\), to improve the chances. This
would only be reliable if the flow and its ECT1 marking was steady and regular.
Actually the number of ECT1 packets would have exceed \(m*r*P\), where \(r\) is
the ratio of the largest to the smallest packet size used by some AQMs to
biasing marking towards larger packets. Although deprecated by IETF RFC 7141,
some common AQMs adopt this practice, for instance DOCSIS
PIE~\cite[\S\,4.6]{White17:PIE_RFC}.

\paragraph{Late onset ECT1 stripe:} This approach would offer typical Classic
queue delay even if the bottleneck AQM supported L4S. So it would have to be
used sparingly on a small sample of flows, perhaps to characterize the paths to
different destinations (or for out-of-band tests). Given L4S connections are
designed to fall back to Classic ECN or no ECN at all where it is only partially
deployed, occasional flows without L4S support should be unremarkable,
particularly during the experimental phase of L4S.

\autoref{tab:ecn-fallbacktr_late_ect1} shows the various patterns of marking
that might result from this test, and what could then be concluded about the
bottleneck in the right-hand column.

\begin{table}[h]
	\newcommand{\drp}{\textrm{drop}}
	\newcommand{\mrk}{\textrm{mark}}
	\centering
	\(
	\begin{array}{|crrc|crrcrrc|l|}
	    \hline
	    \multicolumn{4}{|c|}{\textrm{All ECT0}}
	    		 &\multicolumn{3}{|c}{\textrm{95\% ECT0}}
	    				    &&\multicolumn{3}{c|}{\textrm{5\% ECT1}}
	    									 & \\
	    \cline{2-3}\cline{6-7}\cline{9-10}
	   &\drp&\mrk&&&\drp&\mrk&&\drp&\mrk      && \textrm{The implied type of bottleneck}\\
		\hline
		&>0 &  0 &&& >0 &  0 && >0 &  0       && \textrm{Tail-drop or Non-ECN AQM (FQ or FIFO)}\\
		&   & >0 &&&    & >0 &&    & >0       && \textrm{Classic ECN AQM (FQ or FIFO)}\\
		&   & >0 &&&    & >0 &&    & \approx1 && \textrm{FQ-L4S AQM (non-exclusive)}\\
		&   & >0 &&& >0 &  0 &&    & >0       && \textrm{FQ-exclusive L4S AQM}\\
		&>0 &  0 &&& >0 &  0 &&    & >0       && \textrm{DualQ L4S AQM}\\
		\hline
	\end{array}
	\)
	\caption{Possible patterns of marking resulting from `Late Onset ECT1 Stripe'
	testing. An empty cell means `don't care'.}\label{tab:ecn-fallbacktr_late_ect1}
\end{table}

A run would have to be aborted under the following conditions\footnote{It has
	been checked that these cater for every possible condition of the truth table}:
\begin{itemize}
	\item If no congestion indication (drop or marking) can be induced on an ECT0
	packet, either before ECT1 packets are introduced or after, the run has to time
	out;
	
	\item Whether or not there are congestion indications on any ECT0 packets, a
	run has to eventually time out if there is never a congestion indication on an
	ECT1 packet;
	
	\item If any ECT0 packet has been marked, a run has to eventually time out if
	no ECT1 packet is ever marked;
	
	\item If no ECT0 packet is marked before ECT1 packets are introduced, but an
	ECT0 packet is marked after, the run has to abort.
\end{itemize}
\section{CC Behaviour Changeover Algorithm}\label{ecn-fallbacktr_changeover}

\subsection{TCP Prague-Based Example}\label{ecn-fallbacktr_ex_dctcp}

\begin{figure}[h]
  \centering
  \includegraphics[width=0.45\linewidth]{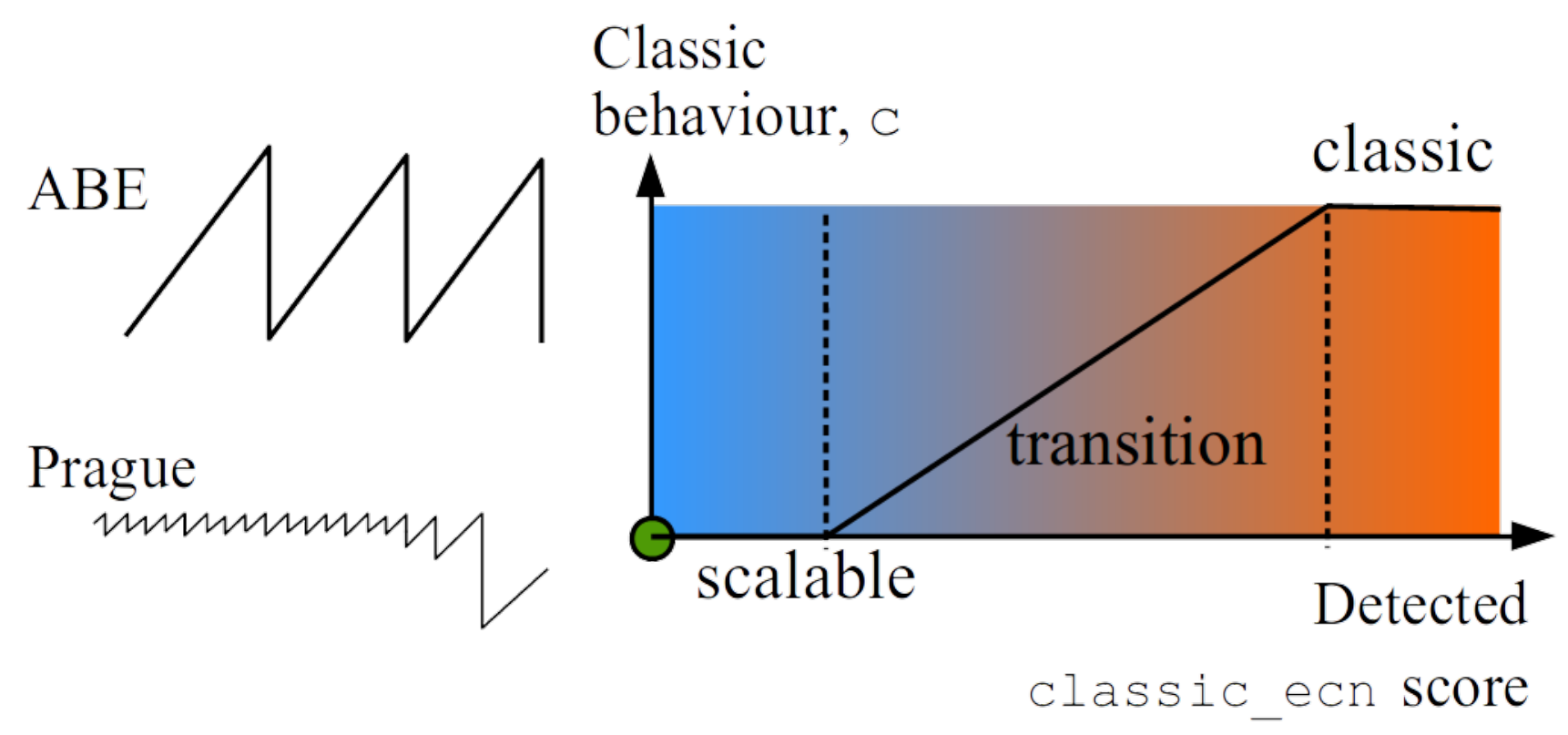}
  \caption{Visualization of the Transition between Scalable and Classic behaviour.}\label{fig:ecn-fallbacktr_c-trans-behaviour}
\end{figure}

The proposed changeover algorithm transitions its response to ECN from
scalable to ABE-Reno as \texttt{c} transitions from 0 to 1, as visualized in \autoref{fig:ecn-fallbacktr_c-trans-behaviour}.

Alternative Backoff with ECN (ABE) is an Experimental RFC~\cite{Khademi18:ABE}
that suggests it is preferable for the reduction in cwnd to be less severe in
response to an ECN signal than to a loss. The logic is that loss is more
likely to emanate from a deep buffer, whereas any ECN signals are likely to be
emanating from a modern AQM which will be configured with shallow target
queuing delay. Therefore, it is reasonable to reduce less in response to ECN
in order to improve utilization. A downside with ABE is that it will lead to
ECN flows competing more aggressively with non-ECN flows, but the difference
is not so great that non-ECN flows would be severely disadvantaged.

It is easiest for TCP Prague to fall back to Reno or ABE-Reno (though falling back to a less
lame congestion control such as Cubic or BBRv2 would be preferable). On an ECN
signal, the ABE RFC recommends a reduction to \(\mathrm{beta_{ecn}}\) of the
original cwnd, where for Reno \(\mathrm{beta_{ecn}}\) is in the range 0.7 to
0.85. The pseudocode below uses \(\mathrm{beta_{ecn}} =
0.7\). If ABE were disabled, for Reno it would be appropriate to transition
using \(\mathrm{beta_{ecn}} = \mathrm{beta_{loss}} =0.5\), but this detail is
not shown in the pseudocode.

Note that the maximum of Prague's variable reduction is 0.5, whereas
ABE's fixed reduction is less than 0.5 (it is 0.3 in our pseudocode).
Therefore, although Prague's reduction will usually be smaller than
ABE's, it can also be larger during periods of high congestion marking.

The example pseudocode below modifies Prague's congestion window reduction, by making
it a function of \texttt{alpha} and \texttt{c}, where:
\begin{itemize}
	\item \texttt{alpha} is already used in DCTCP and TCP Prague to hold an
	EWMA of the congestion level. 
	
	\item  \texttt{c} is the detected \texttt{classic\_ecn} score clamped between 0 and 
	1 (in floating point pseudocode)
\end{itemize}

Two types of simple algorithm are conceivable. 
They are compared in the two alternative statements to calculate
\texttt{reduction} following the original statement used by DCTCP in the
pseudocode below:

\begin{verbatim}
#define BETA_ABE 0.7                     // ABE: Alternative Backoff with ECN [RFC8511]
#define ALPHA_ABE 2*(1-BETA_ABE)         // 0.6

// For pseodocode clarity, c is a float covering the classic ECN transition (Section 3)
c = min(max(classic_ecn / CLASSIC_ECN}, 0), 1);

// original DCTCP reduction within prague_ssthresh()
reduction = cwnd * alpha / 2;

// reduction alternative #1
reduction = cwnd * (alpha + c * (ALPHA_ABE - alpha)) / 2;

// reduction alternative #2
reduction = cwnd * max(alpha, c * ALPHA_ABE) / 2;
\end{verbatim}

The macro \texttt{ALPHA\_ABE} is just the value that, when halved, would limit
the multiplicative reduction of cwnd to \texttt{BETA\_ABE}, by the formula:
\texttt{reduction = cwnd * BETA\_ABE = cwnd * (1 - ALPHA\_ABE / 2)}

\begin{figure}
  \centering
  \includegraphics[width=0.49\linewidth]{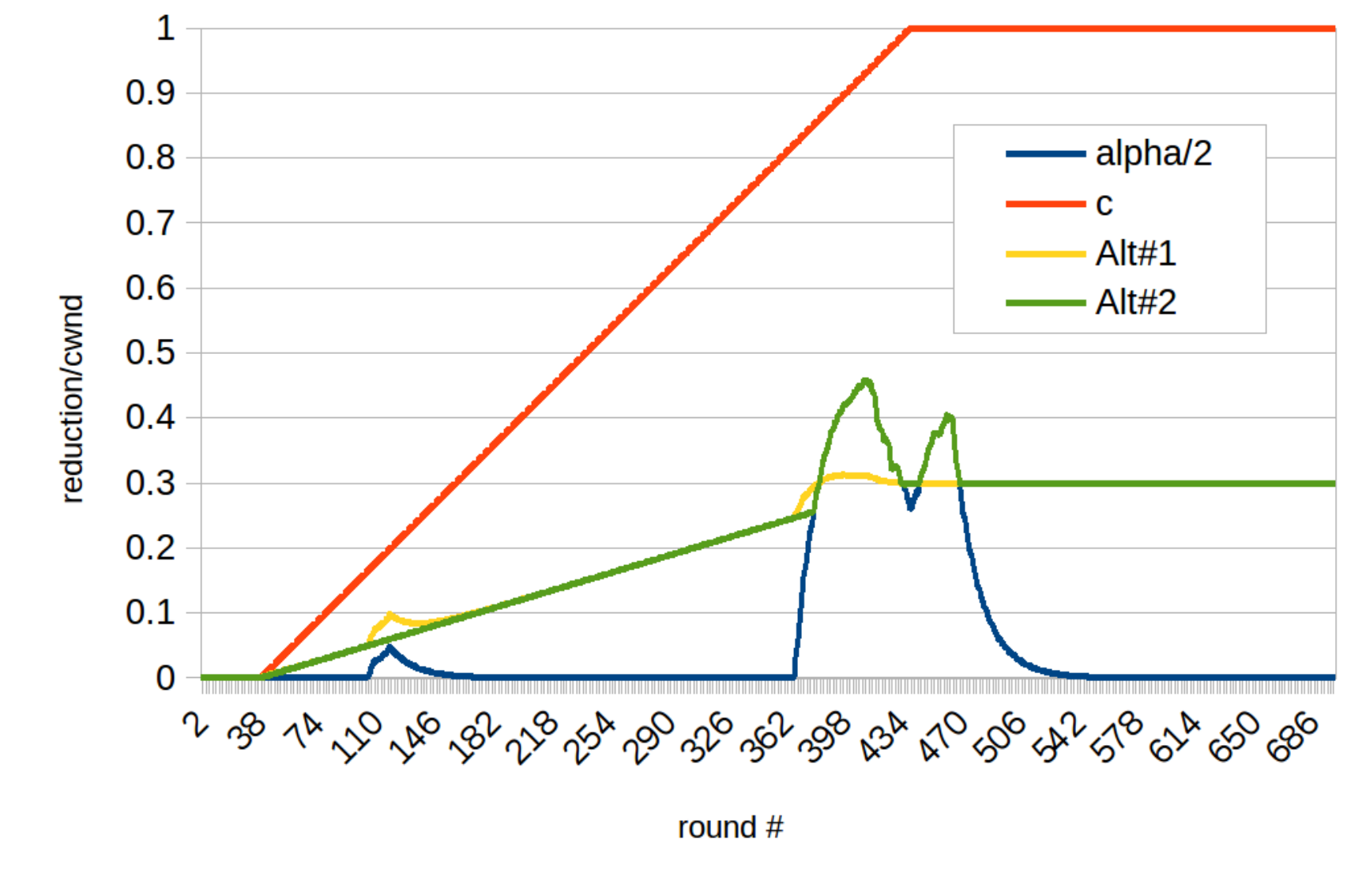}
  \includegraphics[width=0.49\linewidth]{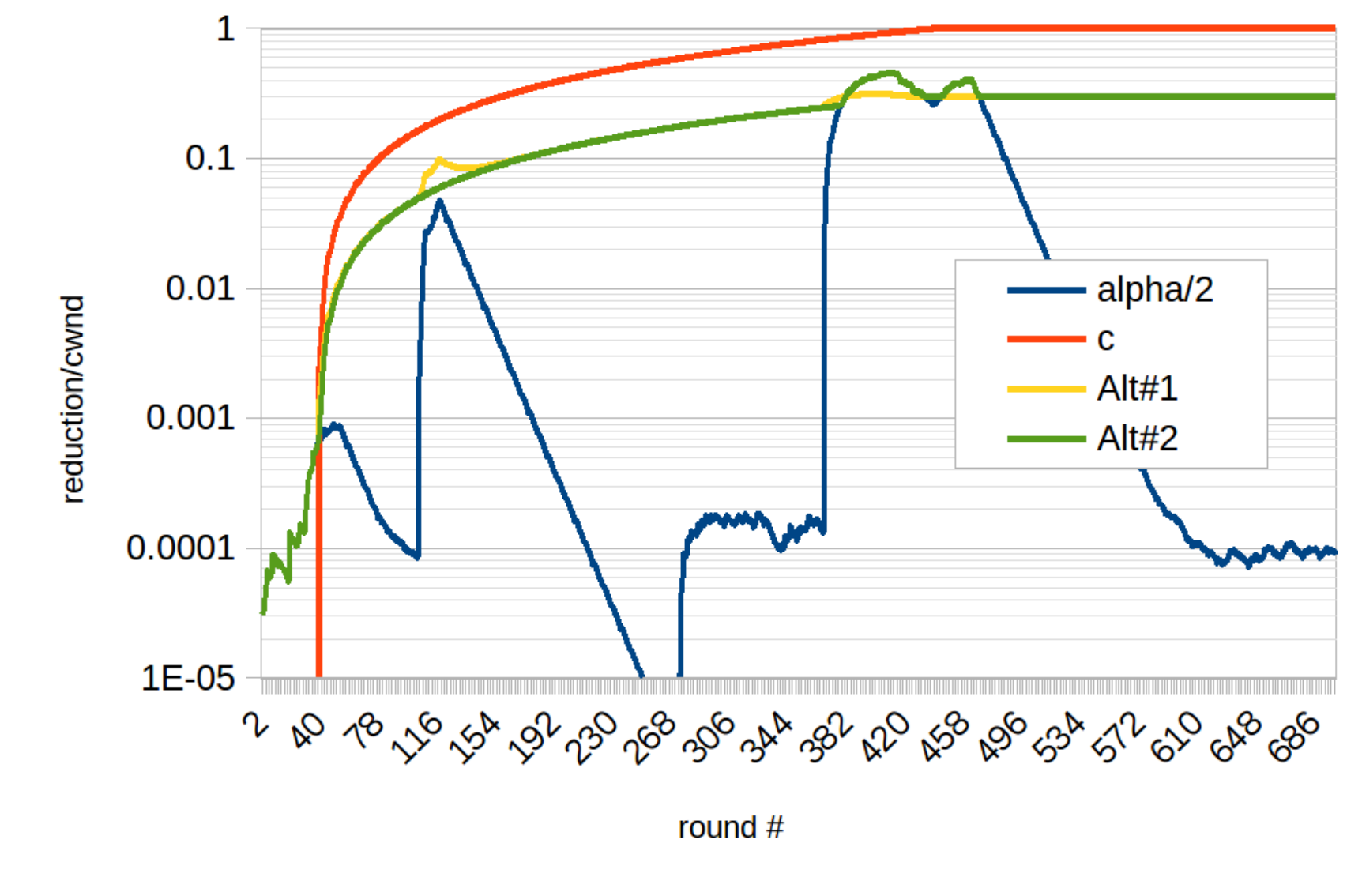}
\caption{Comparison of Two Alternative CC Behaviour Changeover Algorithms
while checking a sweep across the full range of \texttt{c} and
extreme values of \texttt{alpha}; linear (left) or log-scale
(right)}\label{fig:ecn-fallbacktr_cc-changeover}
\end{figure}

To quickly check all possible combinations,
\autoref{fig:ecn-fallbacktr_cc-changeover} shows an example time series
of \texttt{alpha} with extremely low and high values as it interacts with
a sweep of all the values of \texttt{c}, including plateaus at 0 and 1. It shows
the CC reduction on a linear
and log scale for the two alternative changeover algorithms.

Alt\#1 gives the reduction a pro-rata contribution from each of alpha and
\texttt{c}, dependent on the value of \texttt{c}. Alt\#2 takes the simple
maximum of \texttt{alpha} and the value of \texttt{c} scaled down by
\texttt{ALPHA\_ABE}.

As flow rates scale, the typical value of \texttt{alpha} becomes very small,
so it is deceptive to focus on the rounds when \texttt{alpha} is high.
Nonetheless, in current networks \texttt{alpha} can approach 100\%. With
either alternative, when \texttt{alpha} is small, the log plots show that
\texttt{c} dominates over most of its range. However, on the left of the log
plot it can be seen that \texttt{alpha} dominates when \texttt{c} is close to
zero.

Alt\#1 behaves more in the spirit of a transition, because it takes pro-rata
contributions from each approach. Whereas Alt\#2 is more like a binary
switch-over. However, the difference is very subtle and unlikely to be
noticeable by end-users.

Both alternatives can lead to a reduction greater than \texttt{APHA\_ABE} (at
about round \#400 in \autoref{fig:ecn-fallbacktr_cc-changeover}). This effect
is less severe with Alt\#1, but it not necessarily a bad thing to reduce cwnd
by more than \texttt{APHA\_ABE} when congestion is high.

\paragraph{Subsection Summary:} Ultimately, the two alternatives are similar 
enough that the choice between
them can be made on simplicity grounds, in which case Alt\#2 is slightly
preferable.

\subsection{Transition of ECT marking?}\label{ecn-fallbacktr_ect_marking}

When the CC transitions from scalable to classic, should the marking of
packets transition from ECT1 to ECT0?

Let us consider a bottleneck with each type of AQM in turn:
\begin{description}
	\item[Classic AQM:] The only concern here is the sender's CC behaviour, not
	its packet markings. If the CC does not transition to classic behaviour, it
	might outcompete classic flows (if the bottleneck is not FQ). But, it makes no
	difference whether the sender marks the packets ECT0 or ECT1. Because
	'classic' means RFC 3168, and RFC 3168 requires an AQM to treat ECT0 and ECT1
	identically.
	
	\item[L4S AQM:] Here both the packet markings and the CC behaviour need to
	comply with the L4S spec.~\cite{Briscoe15f:ecn-l4s-id_ID} in order to achieve
	any L4S performance benefit. If packets are not marked ECT1, they will never
	be classified into an L4S queue.
\end{description}

Therefore, it is not a good idea for an L4S-capable CC to transition packet
markings to ECT0, even if it transitions to classic CC behaviour (because it
detects a classic ECN bottleneck). It does no harm to anyone by marking its
packets ECT1. But if it uses ECT0, then if the bottleneck moves to one that
supports L4S~\cite{Briscoe15e:DualQ-Coupled-AQM_ID}, its packets will be
classified into the classic queue and it will never detect the lower delay
variability that would trigger its transition back to L4S.

Note that a classic ECN-capable CC does not harm other flows in an L4S
queue\footnote{In contrast to a non-ECN-capable classic CC, which overruns the
shallow ECN threshold until it detects tail drop}; it just unnecessarily
under-utilizes capacity on its own and competes lamely with L4S flows.

\paragraph{Subsection Summary:} A sender that is L4S-capable should always set its
packets to ECT1, irrespective of whether it has transitioned to classic CC
behaviour.

\section{Evaluation}\label{ecn-fallbacktr_evaluation}

A large number of evaluation results are available
online\footnote{\protect\label{note:eval_url}At \href{https://l4s.net/ecn-fbk/}{l4s.net/ecn-fbk/} or via the Evaluations section
of the L4S landing page at \href{https://riteproject.eu/dctth/\#eval}{riteproject.eu/dctth/\#eval}.},
with more being added continually. This section gives a brief summary of
the results so far. It solely evaluates the passive detection algorithm,
and none of the tests yet exercise detection of self-limiting or idling.

\subsection{Experimental Conditions} 
\begin{description}
	\item[Topology:] Dumbell. Experiments were conducted using 2 pairs of
	Linux hosts as sender and receiver; the sender in each pair supporting a
	different congestion control behaviour. The two senders were connected to
	separate interfaces of a Linux router, with one output interface configured 
	to have various AQMs applied. Then that interface was connected to a bridge, 
	with two ports in turn connected to the two Linux receivers.
		
	\item[Software versions:] All Linux machines (hosts and router) were
	running a modified Linux kernel v5.4-rc3
	[\href{https://github.com/L4STeam/linux/blob/testing/08-04-2020/Makefile}{testing/08-04-2020}], 
	which included v2.2 of the Classic ECN AQM Fall-back algorithm.
	Default configuration parameters were used for all software.
	
	\item[Metrics:] Unless otherwise stated for particular experiments, the
	following metrics were tracked for each Prague flow (per-flow
	variables were not tracked for `background' Cubic flows):
	
	\begin{itemize}
		\item \texttt{classic\_ecn} score (per RTT), which depends on the 5
		metrics below that are not in [brackets];
		\item Measured RTT, \texttt{acc\_mrtt} (per ACK);
		\item Adaptively smoothed RTT, \texttt{fbk\_srtt} (per ACK);
		\item {[For comparison, the pre-existing fixed gain smoothed RTT, \texttt{srtt} (per ACK)]};
		\item Adaptively smoothed mean RTT deviation, \texttt{fbk\_mdev} (per ACK);
		\item Min RTT, \texttt{rtt\_min} (per change);
		\item Slow start threshold, \texttt{ssthresh} (per change);
		\item {[Congestion window, \texttt{cwnd} (per change)]};
	\end{itemize}
	\item The following metrics were measured at the AQM for all traffic:
	\begin{itemize}
		\item The throughput per flow, which was sampled per packet each time
		taking a rolling window of the last 1\,s of transmitted data. Throughput
		was only measured individually for long-running flows; in experiments
		with short flows, the throughput of all short flows in a class was measured as if they
		were one aggregate flow;
		\bob{Change to rolling sum over 100\,ms for 20\,s}
		\item The average throughput per flow in each class of flows---scalable
		or Classic (1\,s rolling window, sampled per packet). Short flows were
		excluded from the average throughput per flow.
		\item Link utilization (1\,s rolling window, sampled per packet);
		\item Queue delay, measured per packet;
		\item Drop and marking probability, sampled every 16 base RTTs.
	\end{itemize}
	\item[Traffic scenarios:] Simple traffic scenarios with equal RTT
	long-running flows are used to measure steady-state rate for `fairness'
	metrics, irrespective of whether such scenarios are typical on the
	Internet. Scenarios with short flows and a mix of short and long flows
	are introduced later, including staggered or simultaneous start-up of
	long-running flows.
\end{description}

\subsection{`Fairness' with Long-Running Flows}\label{ecn-fallbacktr_eval_fairness}

\begin{figure}[h]
  \centering
  \includegraphics[width=0.45\linewidth]{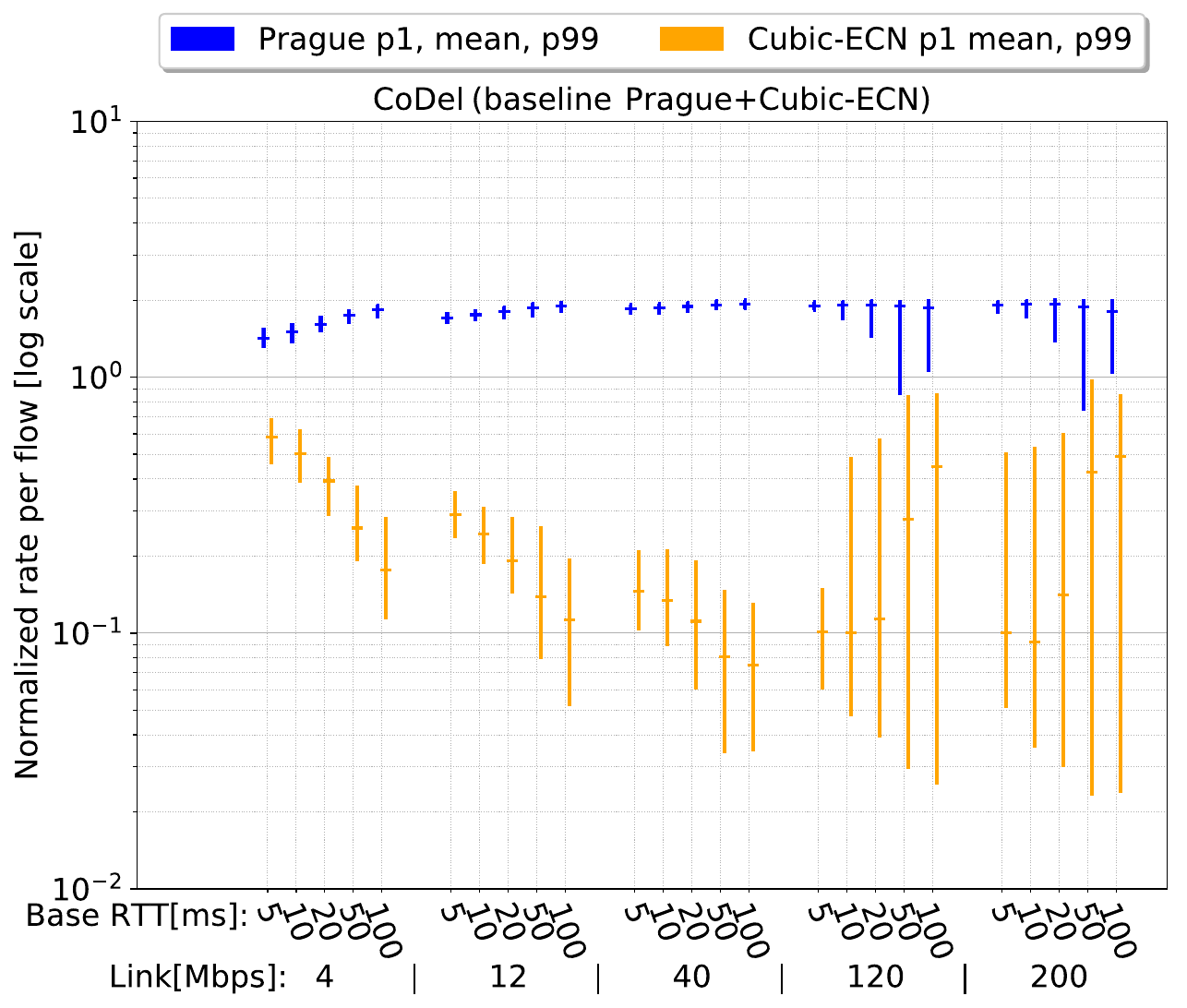}
  \includegraphics[width=0.45\linewidth]{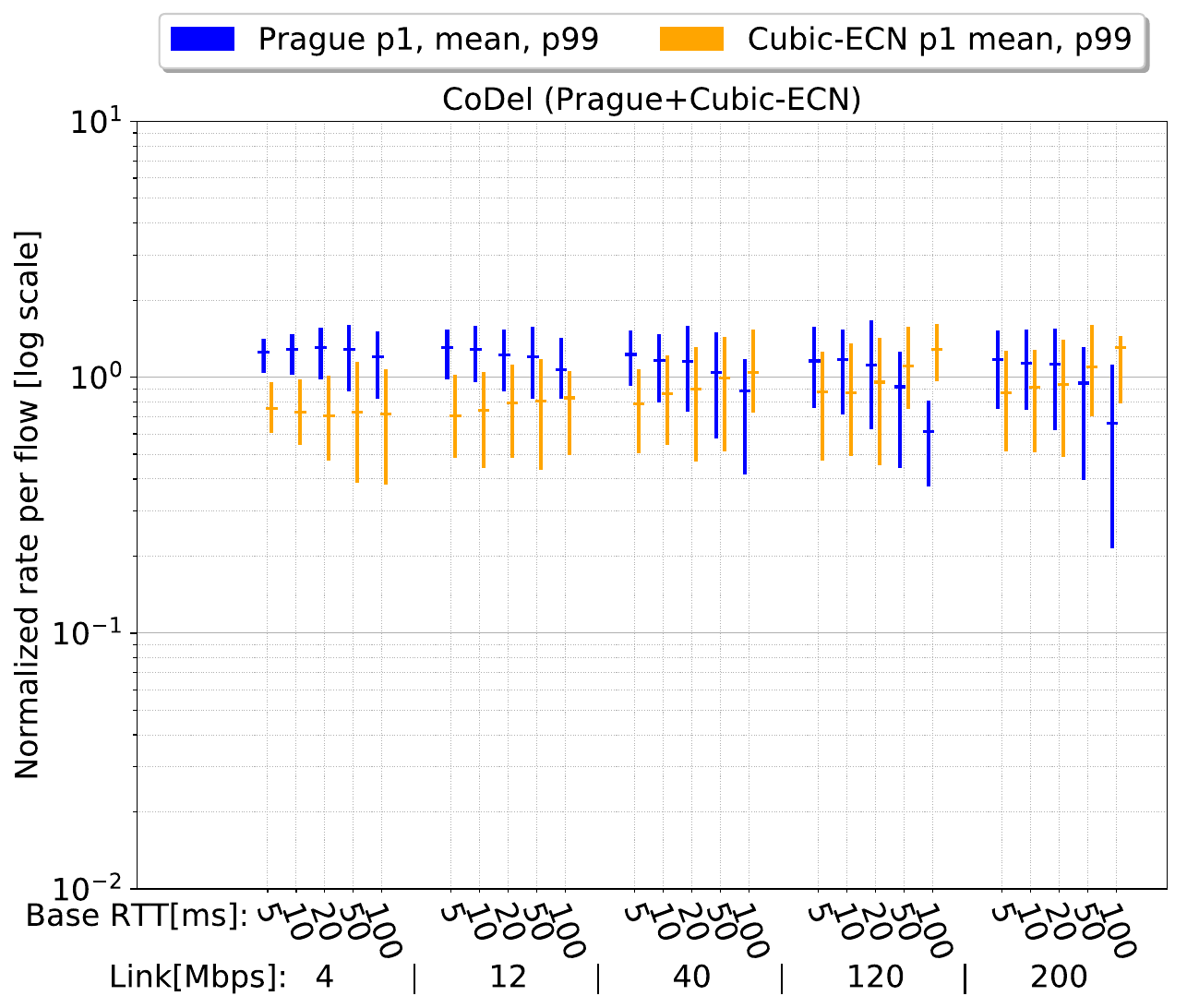}
\caption{Comparison between Classic ECN AQM Fall-back algorithm when
disabled (left) and enabled (right) in TCP Prague over 25 combinations of
link-rates and base RTTs. In both cases, the long-running TCP Prague flow
competes with a long-running TCP Cubic ECN flow over a CoDel AQM.
Normalized rate per flow is defined in the
text.}\label{fig:ecn-fallbacktr_eval_before_after}
\end{figure}

The two charts in \autoref{fig:ecn-fallbacktr_eval_before_after} give a
high level picture of the improvement in flow rate balance ('fairness')
that the passive Classic ECN AQM detection and fall-back algorithm gives. 
They show TCP Prague competing with TCP Cubic-ECN over a shared queue CoDel
AQM using a \(5\times5\) matrix of different link rates and base RTTs.
Each plot (each tall thin cross) shows the \(1^{\mathrm{st}}\)
percentile, mean and \(99^{\mathrm{th}}\) percentile of the normalized
rate per flow (see definition below). A value of 1 (\(10^0\)) is the
ideal. The left and right-hand charts show the outcome with the algorithm
disabled and enabled respectively. In all cases, it can be seen that
enabling the algorithm causes TCP Prague to pretty closely balance its
flow rate with TCP Cubic-ECN.

Normalized rate per flow is defined as the ratio of the flow rate, \(x\)
relative to an `ideal' rate. The ideal rate is defined as the link
capacity divided by \(n\), where \(n\) is the number of flows (\(n=2\) in
all cases in \autoref{fig:ecn-fallbacktr_eval_before_after}). Thus,
normalized rate per flow \(= x/(C/n)\). In this `fairness' experiment,
rate measurements were taken after allowing flows to stabilize for
\(T\)\,s, with \(T=5+x*R/100,000\), where \(x\) is the flow rate in [b/s]
and \(R\) is the RTT in [s].

It can be seen that the left-hand chart is similar to
\autoref{fig:ecn-fallbacktr_rate_flow} in
\S\,\ref{ecn-fallbacktr_problem}, which we originally used to illustrate
the coexistence problem. The link-rates and base RTTs are the same, but
in \autoref{fig:ecn-fallbacktr_rate_flow} PI2 was used as the Classic AQM
and DCTCP was used as the L4S congestion control. In
\autoref{fig:ecn-fallbacktr_eval_before_after} CoDel (not FQ\_CoDel) is
used as the Classic AQM, because it is likely to be the most difficult
single-queue Classic AQM for the algorithm to distinguish from an L4S AQM
(see \S\,\ref{ecn-fallbacktr_detection_params} for reasoning).

\begin{figure}
  \centering
  \includegraphics[width=0.49\linewidth]{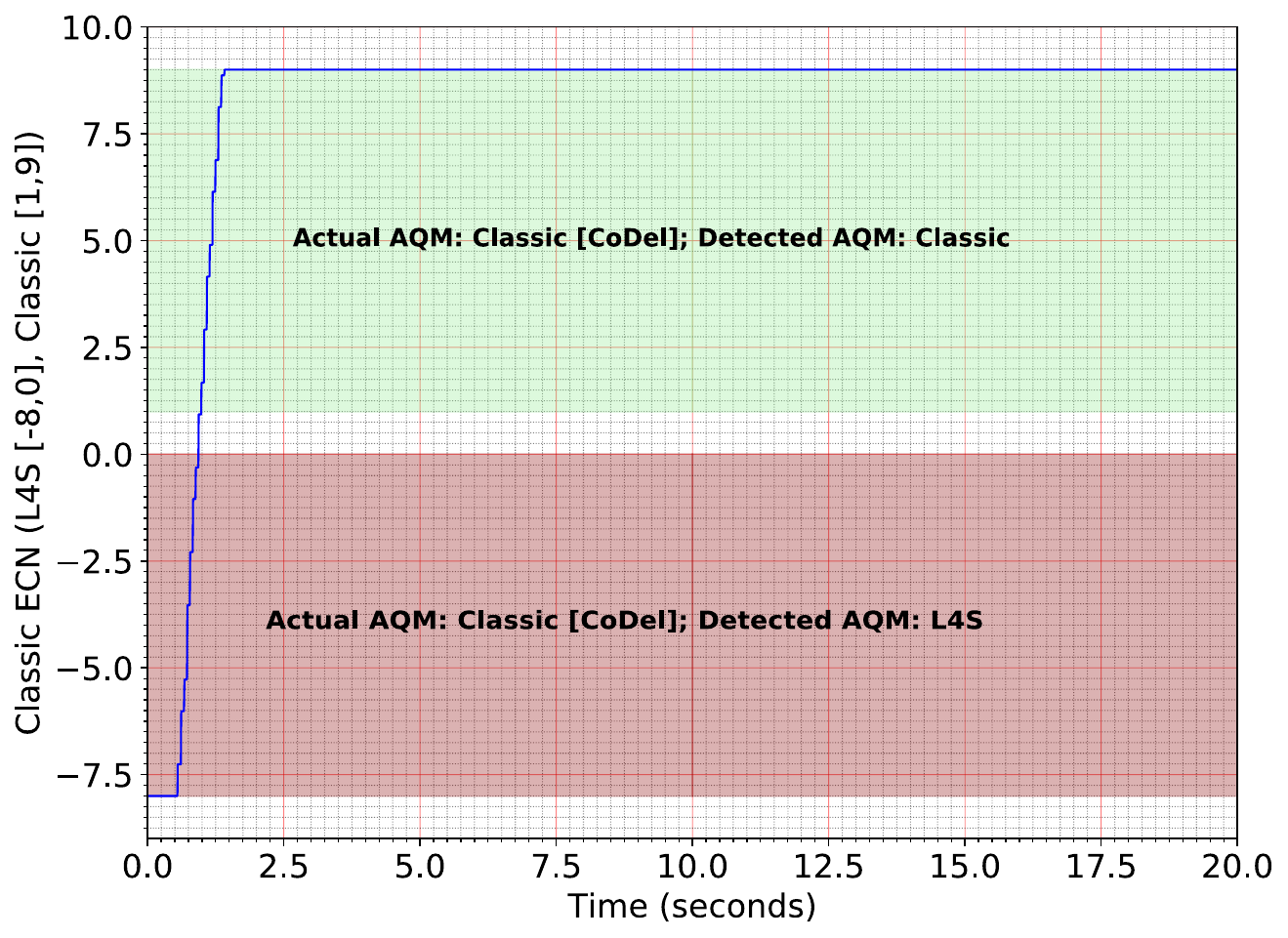}
  \includegraphics[width=0.49\linewidth]{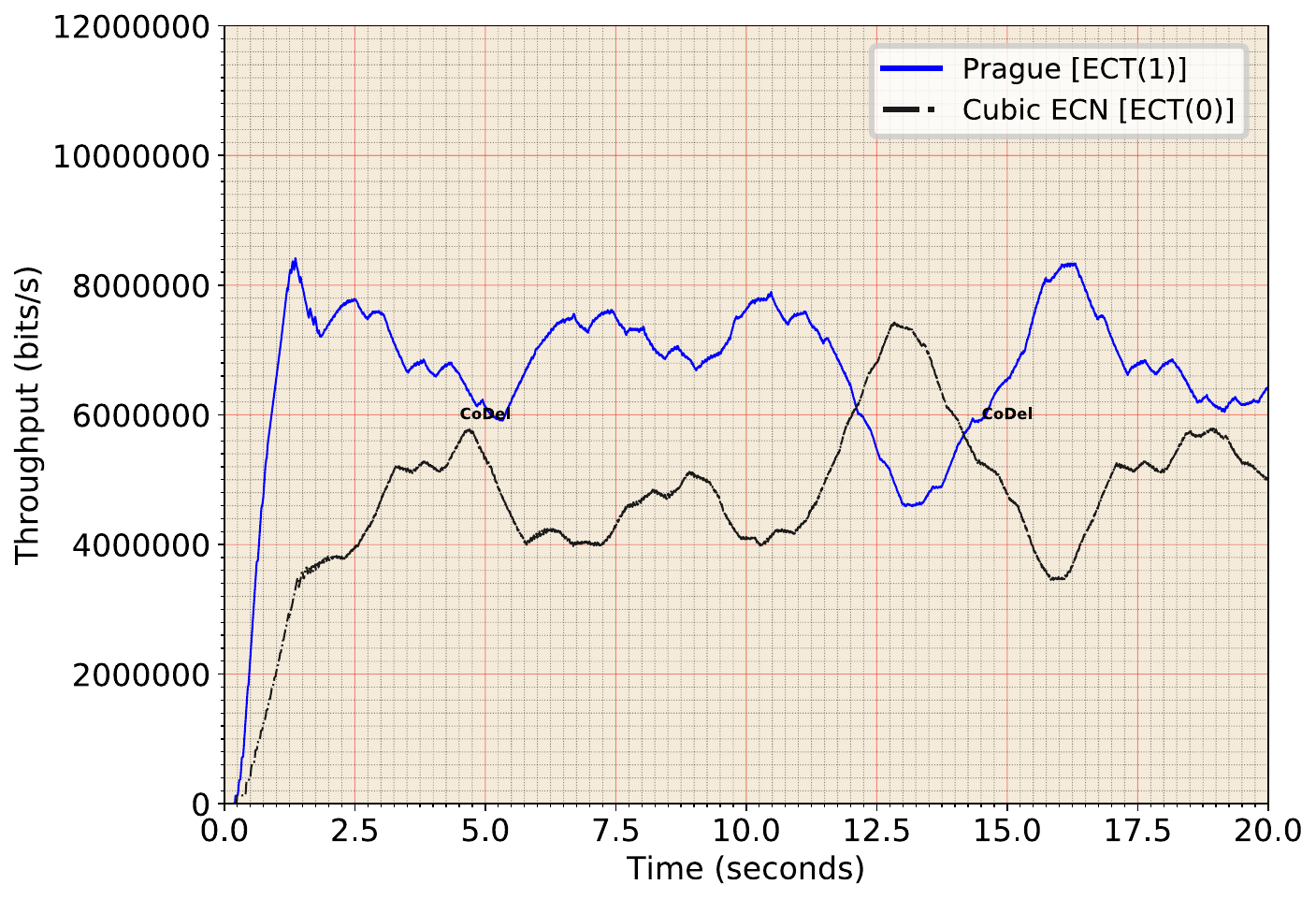}
\caption{A typical example of the evolution of the \texttt{classic\_ecn}
score (left) in a long-running TCP Prague flow while competing with 1
long-running Cubic flow in a CoDel AQM bottleneck. The algorithm detects a Classic AQM
and switches to classic ABE-Reno behaviour
(\(\mathtt{classic\_ecn}\ge1\)) after about 1\,s. The resulting
throughput of each flow is shown on the right [12\,Mb/s link; 50\,ms base
RTT].}\label{fig:ecn-fallbacktr_codel_12m_50ms_1v1_classic_ecn}
\end{figure}

In every case, the algorithm rapidly detects a Classic ECN AQM and
switches TCP Prague over to its ABE-Reno response to ECN marks, which
then competes roughly equally with Cubic.
\autoref{fig:ecn-fallbacktr_codel_12m_50ms_1v1_classic_ecn} is a typical
example time series at 12\,Mb/s and 50\,ms base RTT. In the left-hand plot, after 0.5\,s
the \texttt{classic\_ecn} variable can be seen starting to move off its
\texttt{-L\_STICKY} floor (-8) towards classic. It has already fully
transitioning to classic (at \(\mathtt{classic\_ecn} = 1\)) after 1\,s  (about
50 base round trips), then continues onward to \(\mathtt{classic\_ecn} = 9\) 
by 1.5\,s after the flow started.

The flow throughputs are shown on the right-hand side of
\autoref{fig:ecn-fallbacktr_codel_12m_50ms_1v1_classic_ecn}. The Prague
flow (in blue) fills the available capacity after about 1.5\,s; at about
the same time as its \texttt{classic\_ecn} score hits its maximum
`stickiness' of 9. Therefore, for the rest of the time, it behaves as
ABE-Reno, and competes on roughly equal terms with the Cubic-ECN flow (in
black).

\subsection{Fairness with Staggered Long-Running
Flows}\label{ecn-fallbacktr_eval_fairness_staggered}

It is necessary to test for cases in which the fall-back algorithm might
never see the true minimum RTT. This can occur if one flow starts while a
long-running flow is already maintaining a standing queue.

\begin{figure}
  \centering
  a) \includegraphics[width=0.43\linewidth]{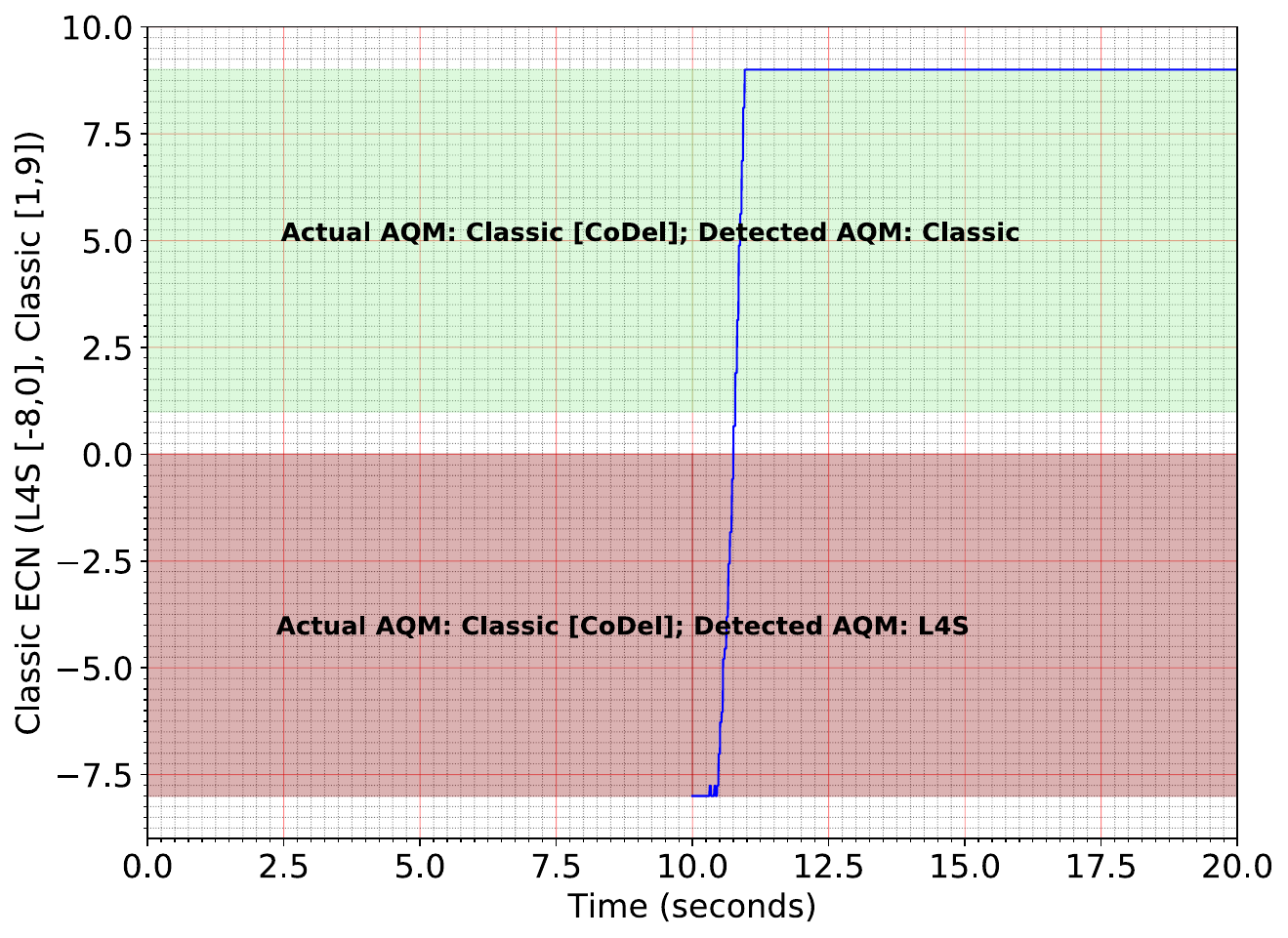}
  c) \includegraphics[width=0.43\linewidth]{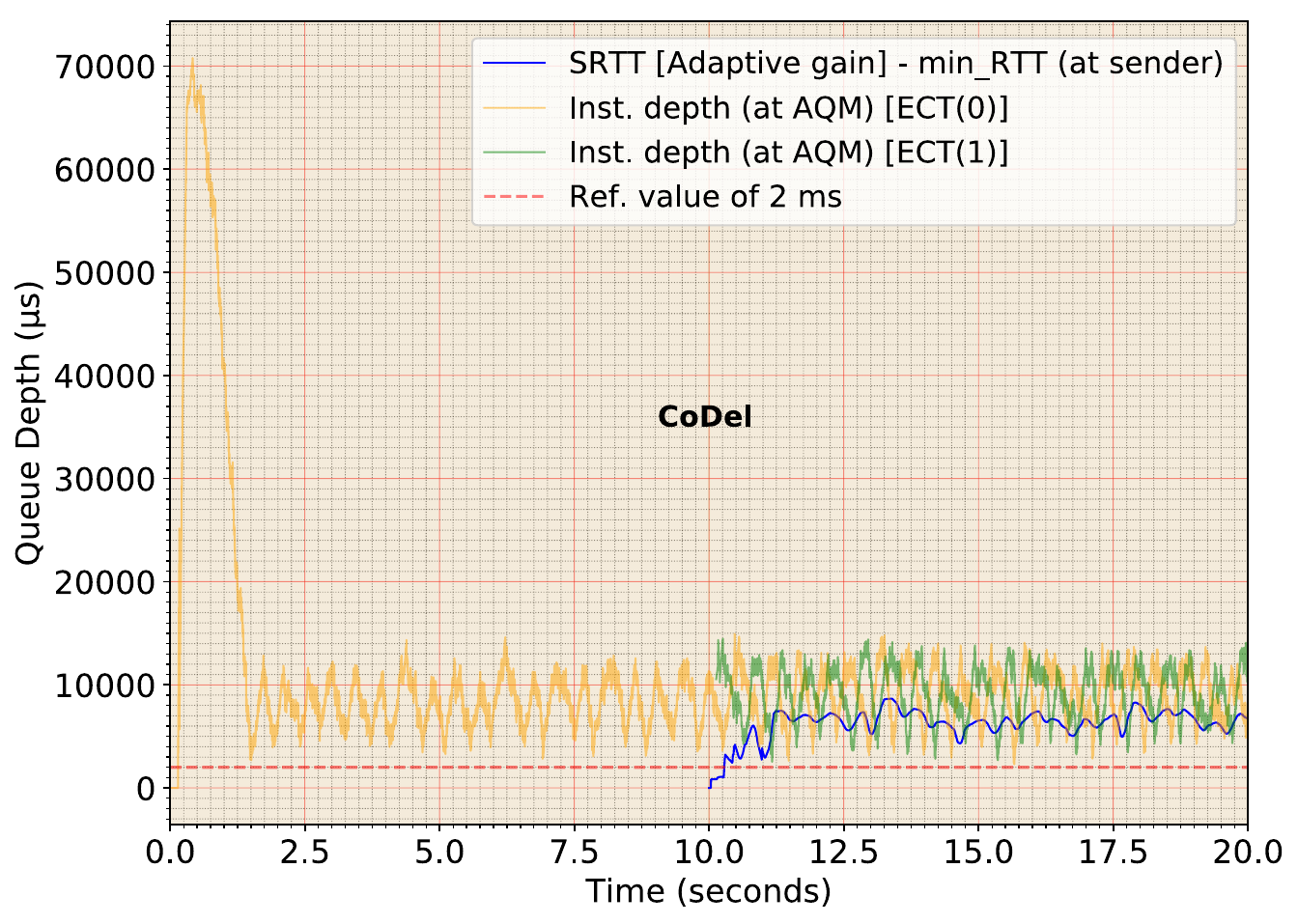}\\
  b) \includegraphics[width=0.43\linewidth]{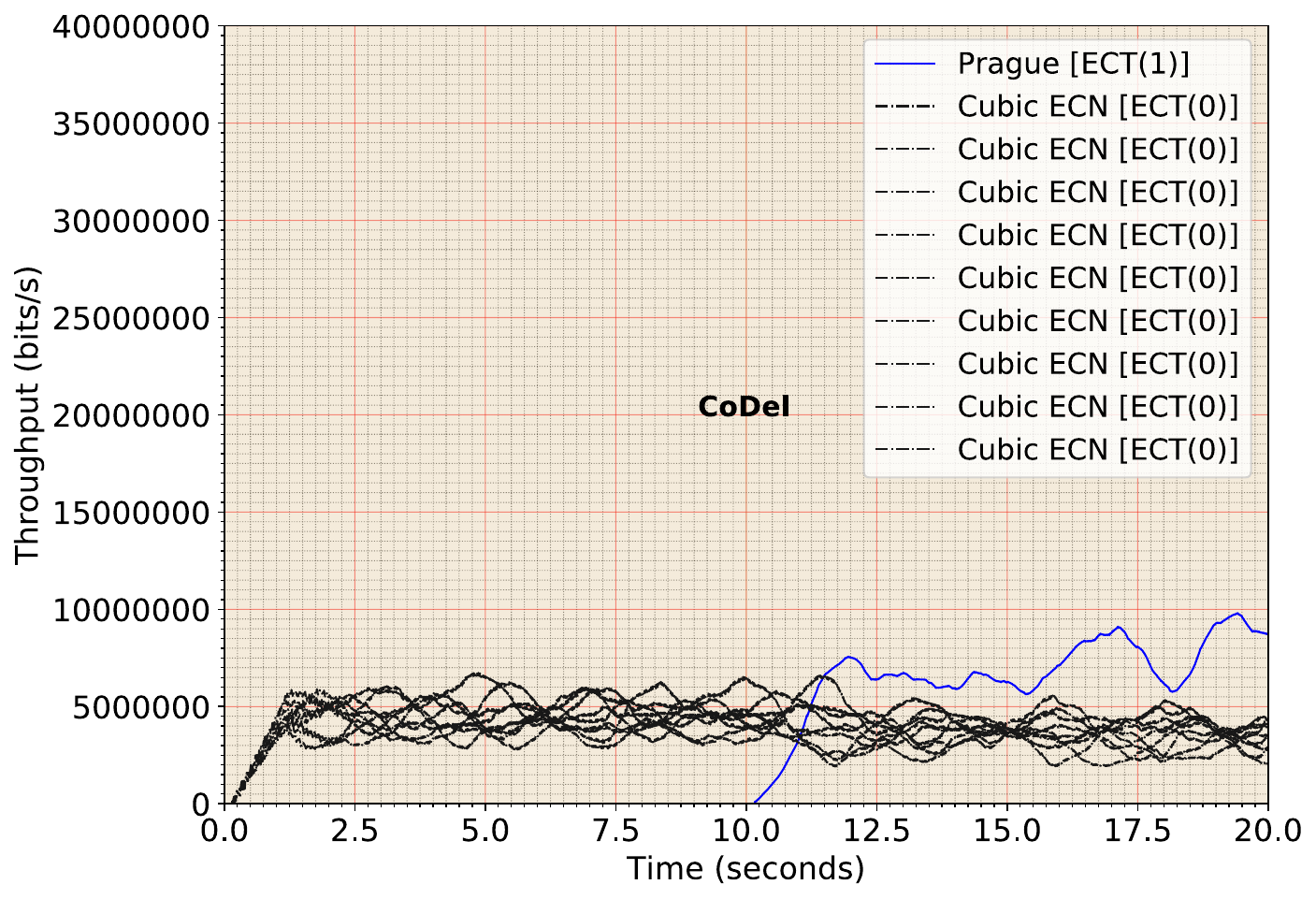}
  d) \includegraphics[width=0.43\linewidth]{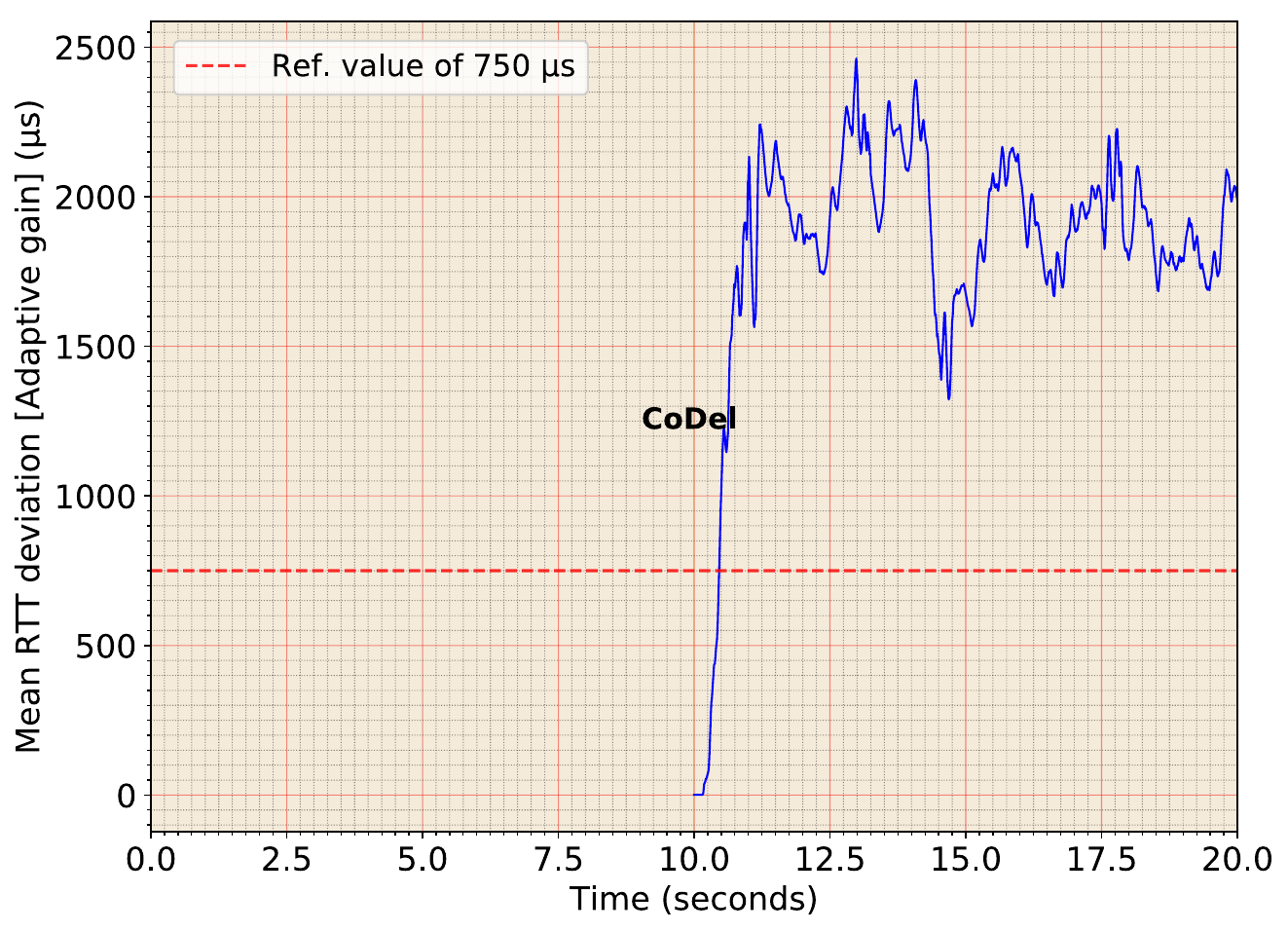}
\caption{At 10\,s a long-running Prague flow joins a standing queue
built by 9 Cubic flows over a CoDel AQM. Four variables are tracked: a)
\texttt{classic\_ecn} score; b) per-flow throughput; c) queue delay
(actual (yellow/green) and sender's estimate (blue)); d) Mean RTT
deviation. [40\,Mb/s link; 20\,ms base
RTT]}\label{fig:ecn-fallbacktr_codel_40m_20ms_stag_1v9}
\end{figure}

In the previous section, both flows started simultaneously, whereas
\autoref{fig:ecn-fallbacktr_codel_40m_20ms_stag_1v9} shows a case where a
single Prague flow joins 9 Cubic-ECN flows after 10\,s. It can be seen
from \autoref{fig:ecn-fallbacktr_codel_40m_20ms_stag_1v9}a) that the
Prague flow switches to Classic ABE-Reno behaviour about 0.75\,s after it
starts. And \autoref{fig:ecn-fallbacktr_codel_40m_20ms_stag_1v9}b) shows
that the throughput of the Prague flow (blue) competes reasonably `fairly'
with the Cubic flows (rate ratio of roughly 1.75).

Thus the fall-back algorithm works despite not correctly measuring the
queue delay, as we shall now explain with the aid of
\autoref{fig:ecn-fallbacktr_codel_40m_20ms_stag_1v9}c), which overlays
the queue delay as measured at the AQM (yellow/green)\footnote{Yellow and
green separate out the queue as measured on arrival of ECT(0) and ECT(1)
packets respectively. Being a single queue, they both give the same
reading in this case (although they are time-shifted due to slippage
between the system clocks used for the two measurements).} with the
smoothed queue delay estimated by the Prague sender's fall-back algorithm
(blue), which uses only end-to-end measurements. It can be seen that Prague's
e2e measurements persistently under-estimate the queue delay by the same
amount, which represents its over-estimate of the min RTT, due to only
ever having seen a standing queue. Nonetheless, this does not cause the
algorithm to switch back to scalable L4S behaviour, for three reasons: 
\begin{enumerate}[nosep]
	\item The queue delay estimate is not sufficiently incorrect to drop below the
	queue delay threshold (\(\mathtt{D0}=2\,\)ms) shown as a a red dashed horizontal);
	
	\item Even if the estimate did drop below the threshold, the 
	algorithm only takes note of queue delay above the threshold, not below 
	(specifically because the min RTT is unreliable);
	
	\item The algorithm also uses RTT variability, and the sender's
	instantaneous measurements still lead to a value of mean deviation that
	is higher than the \(750\,\mu\)s threshold, as illustrated by the red
	dashed horizontal in
	\autoref{fig:ecn-fallbacktr_codel_40m_20ms_stag_1v9}d).
\end{enumerate}

\subsection{Switching AQM Mid-Flow}\label{ecn-fallbacktr_eval_aqm_switch}

Experiments were also conducted to determine the effect of a change in
AQM mid-flow. Though unlikely, an AQM change can happen, e.g.\ if the
path of a flow re-routes, or if a different buffer becomes the most
bottlenecked on the path. Nonetheless, the purpose of these experiments
was more to double-check that the fall-back algorithm does not exhibit any
unexpected behaviour during such an event. Nonetheless, it was also
interesting to see just how quickly the algorithm could detect such a
change.

\begin{figure}
  \centering
  a) \includegraphics[width=0.43\linewidth]{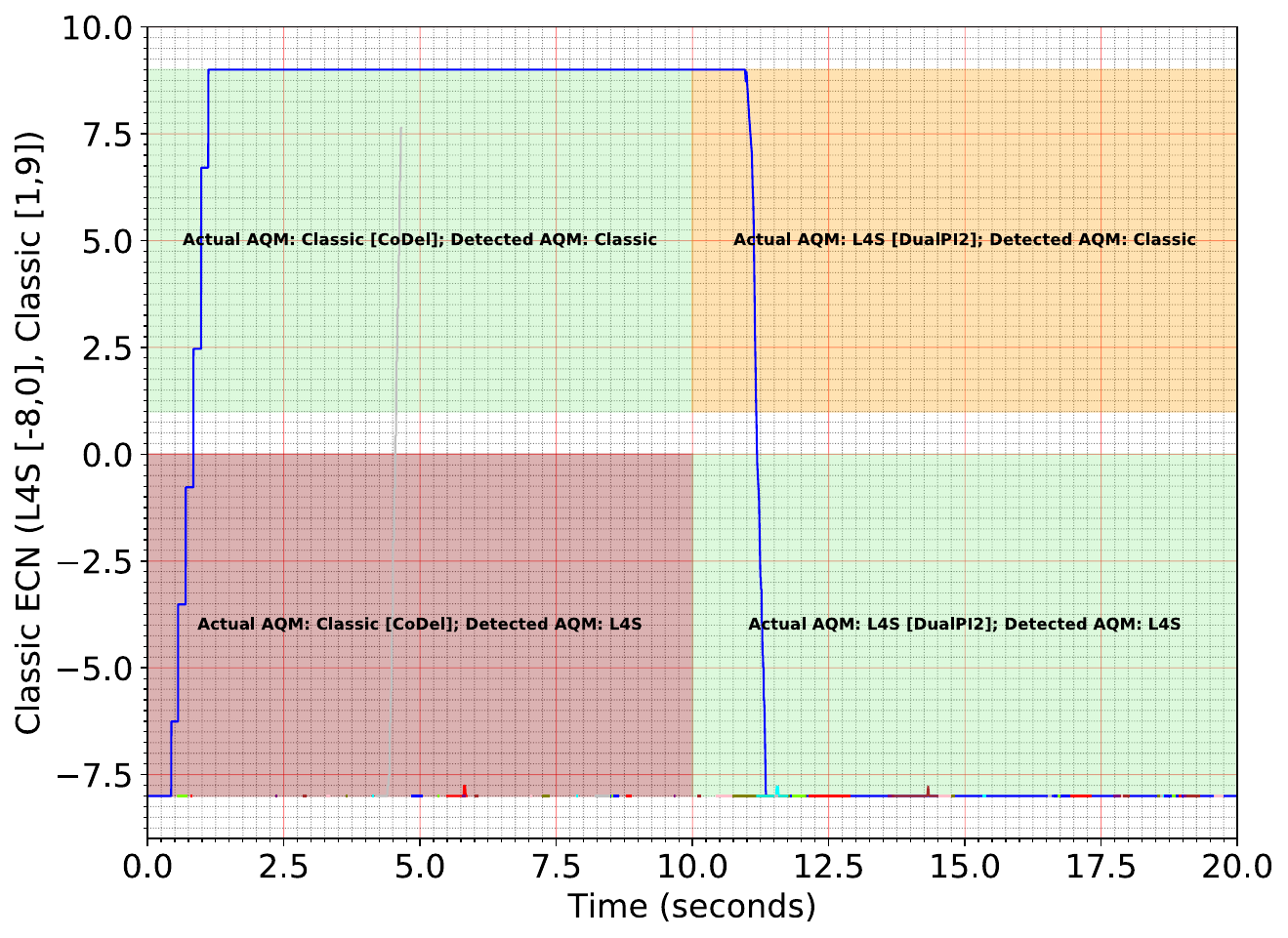}
  c) \includegraphics[width=0.43\linewidth]{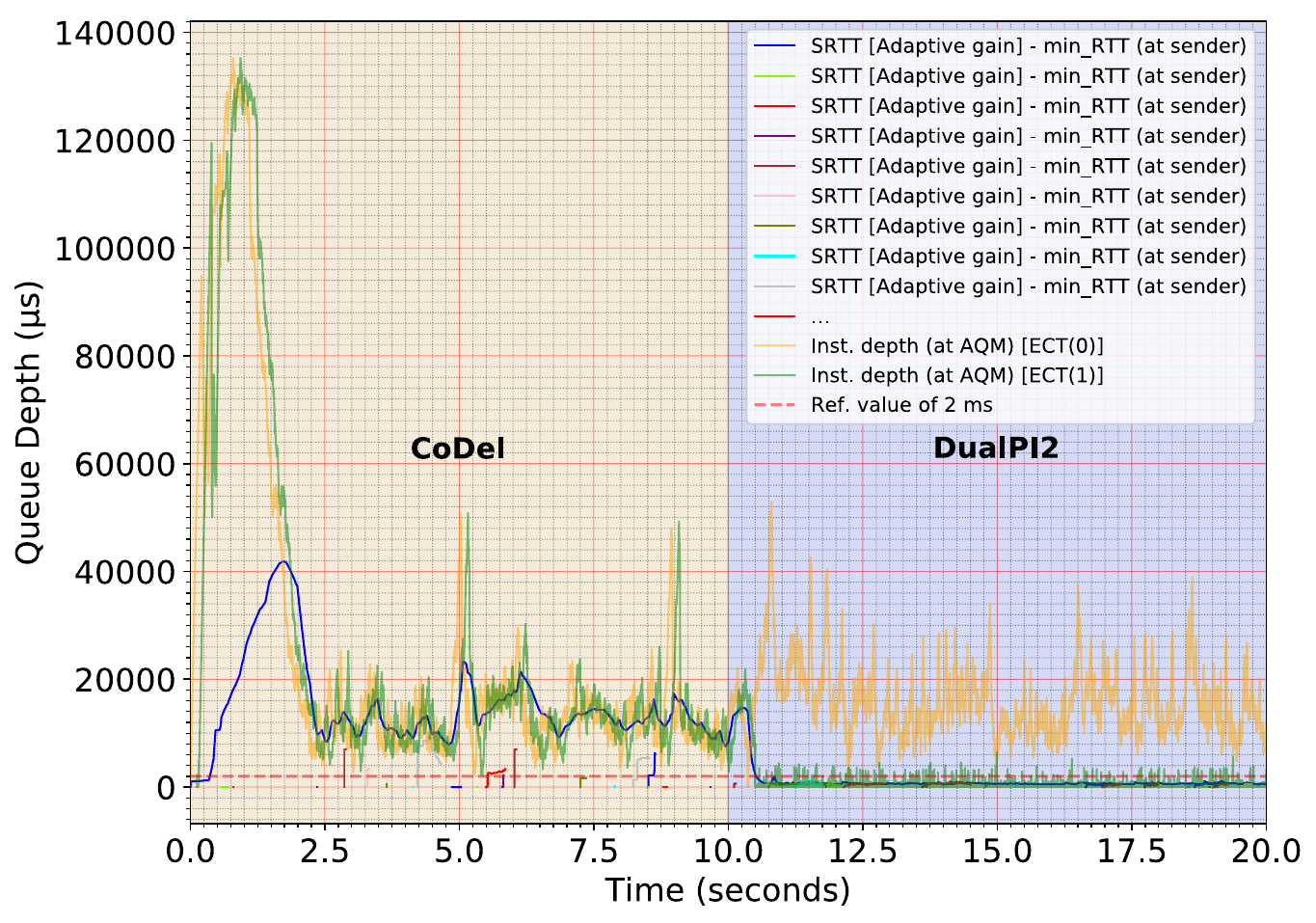}\\
  b) \includegraphics[width=0.43\linewidth]{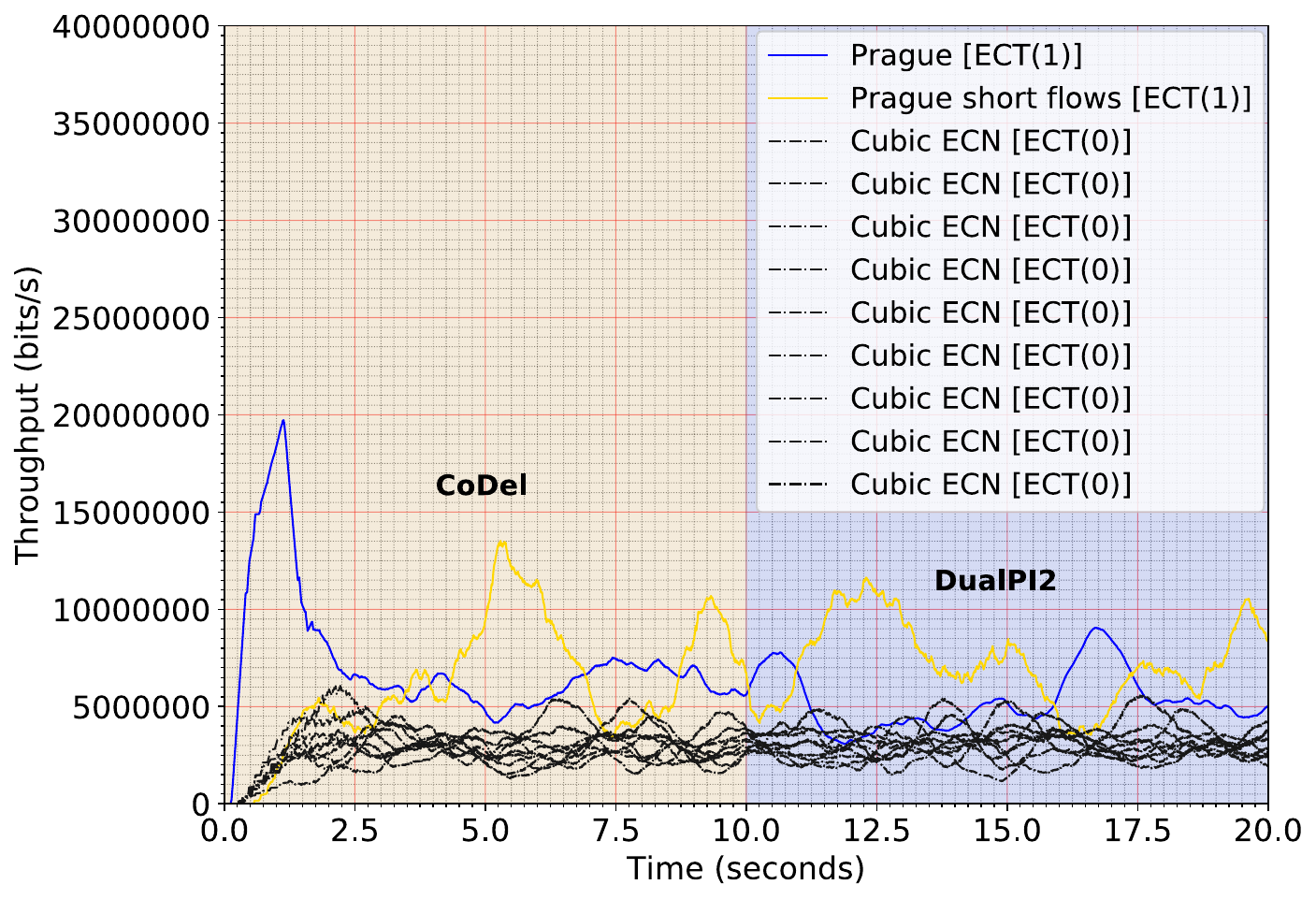}
  d) \includegraphics[width=0.43\linewidth]{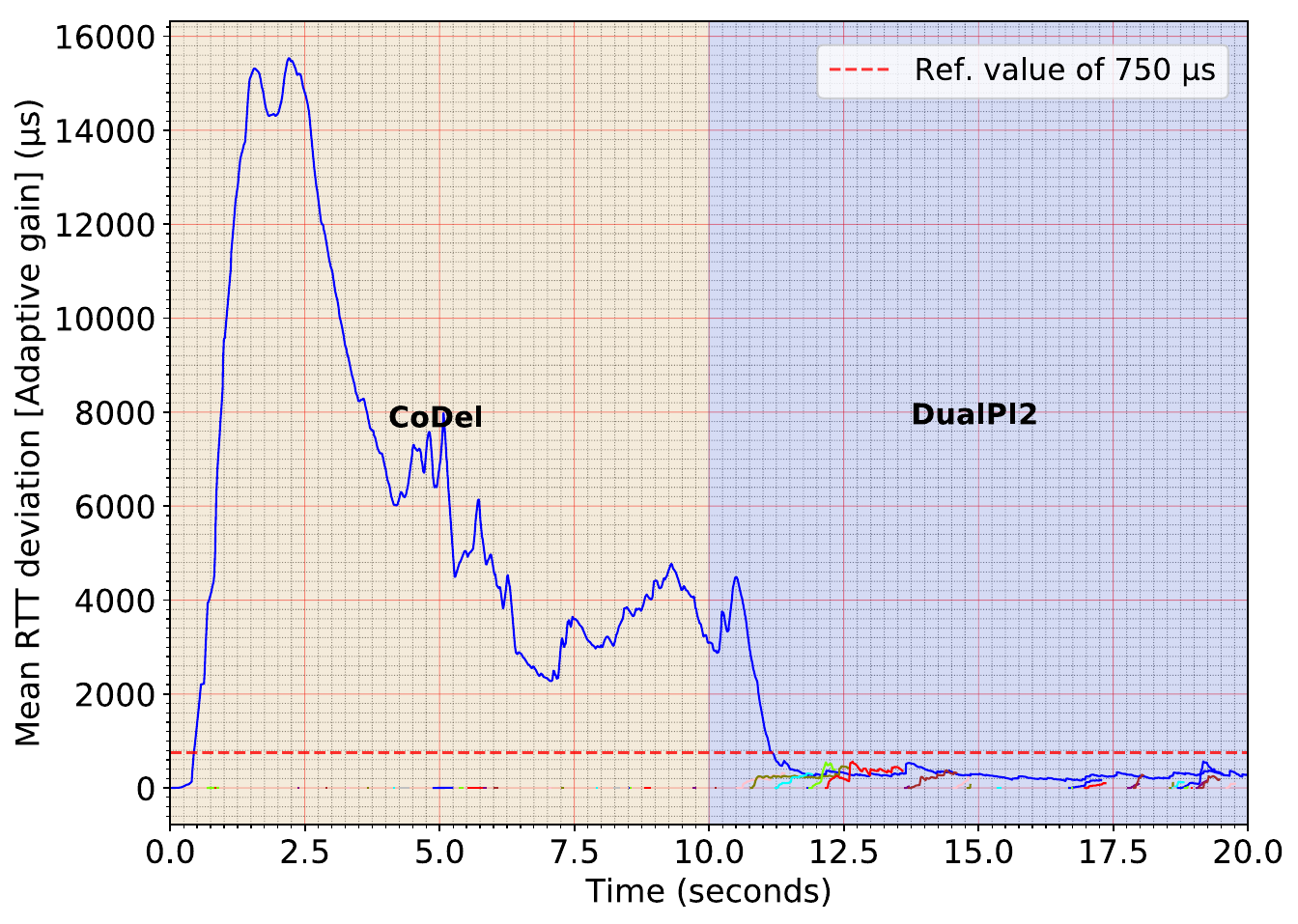}
\caption{At 10\,s the AQM switches from CoDel to DualPI2, while 1 Prague \& 9 Cubic-ECN long-running flows proceed. Also a low background load of web-like (type L) Prague flows arrive continually. Four variables are tracked: a)
\texttt{classic\_ecn} score; b) per-flow throughput; c) queue delay
(actual (yellow/green) and sender's estimate (blue)); d) Mean RTT
deviation. [40\,Mb/s link; 10\,ms base
RTT]}\label{fig:ecn-fallbacktr_codel_dualpi2_40m_10ms_1Lv9}
\end{figure}

\begin{figure}
  \centering
  a) \includegraphics[width=0.43\linewidth]{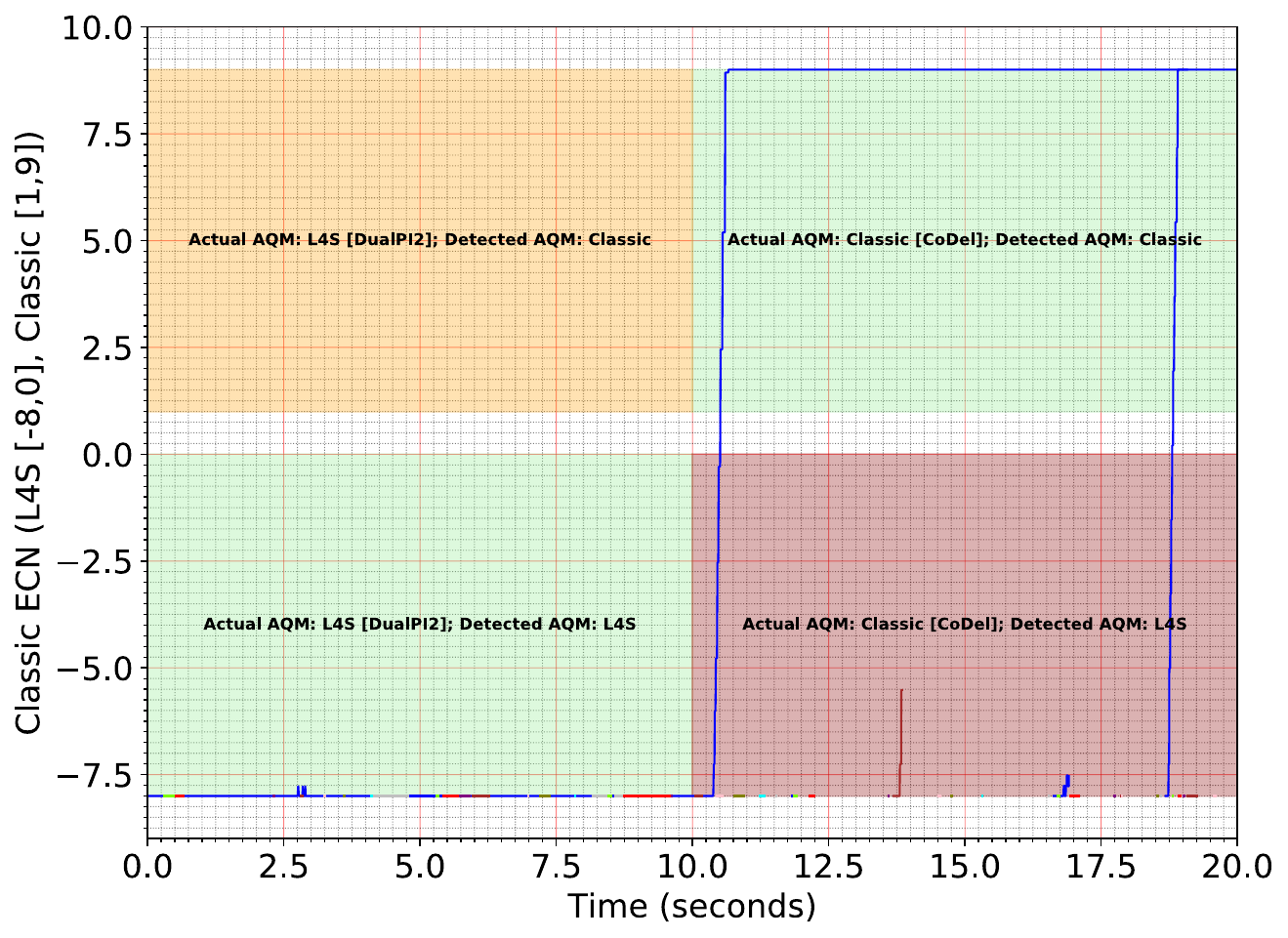}
  c) \includegraphics[width=0.43\linewidth]{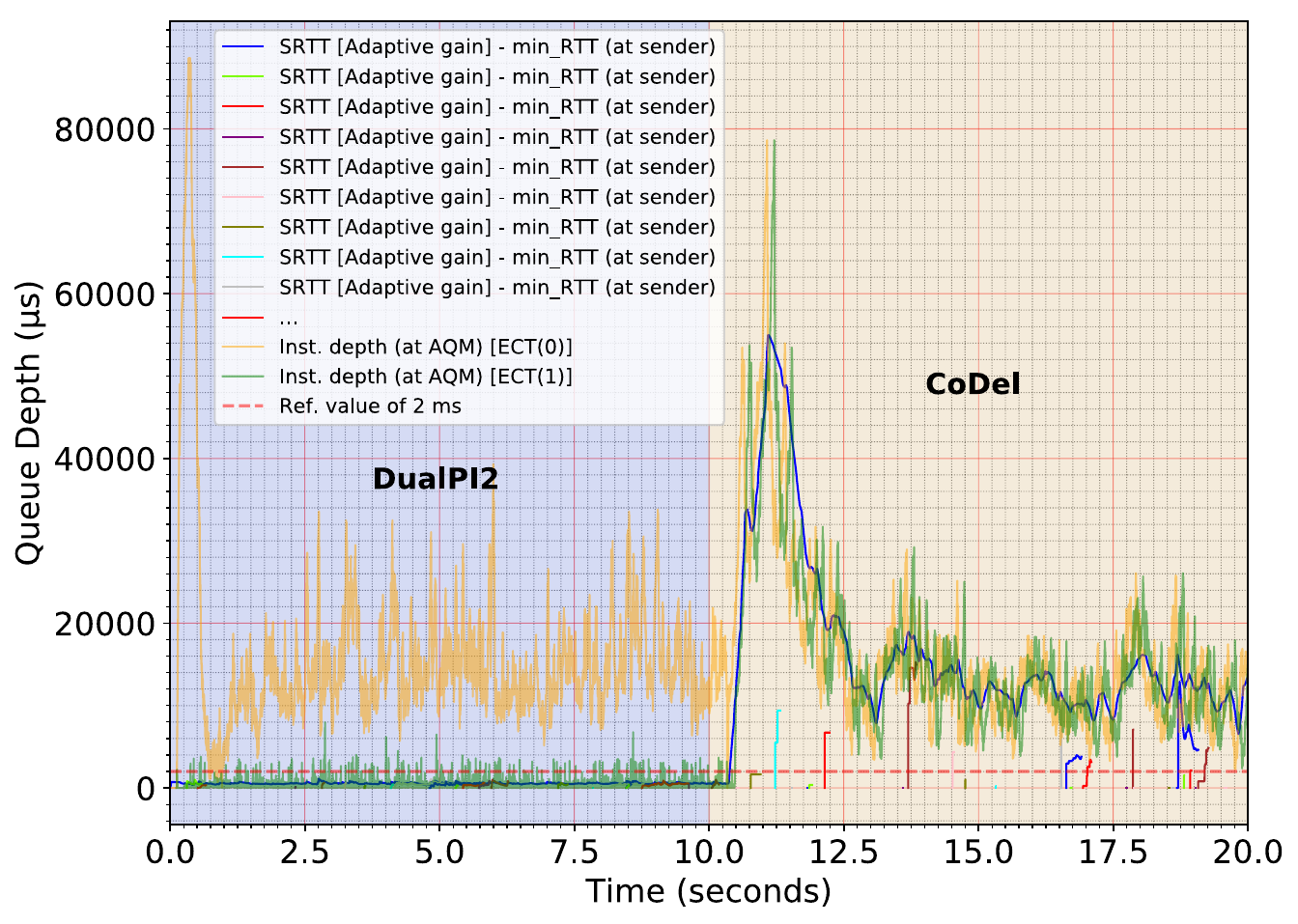}\\
  b) \includegraphics[width=0.43\linewidth]{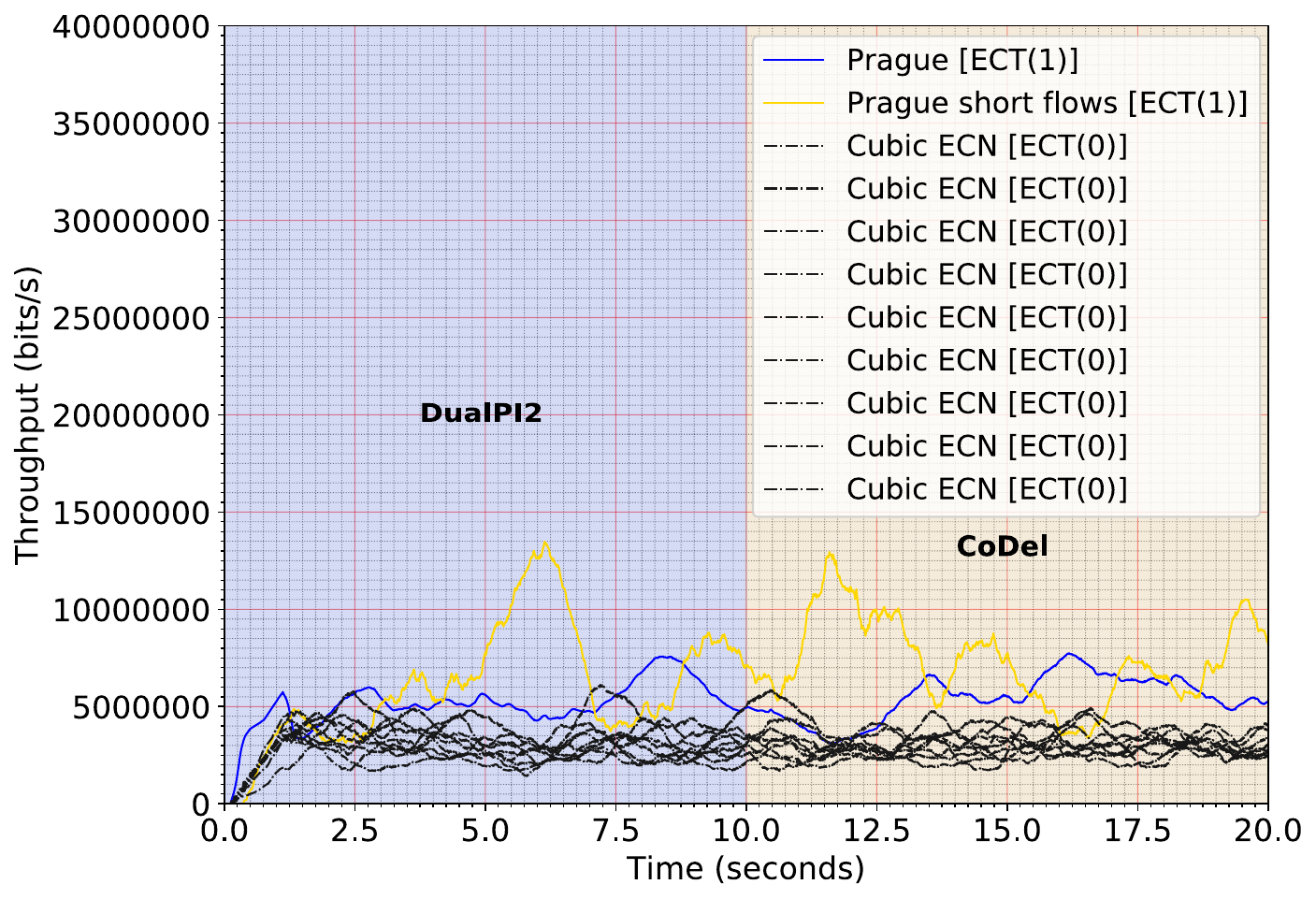}
  d) \includegraphics[width=0.43\linewidth]{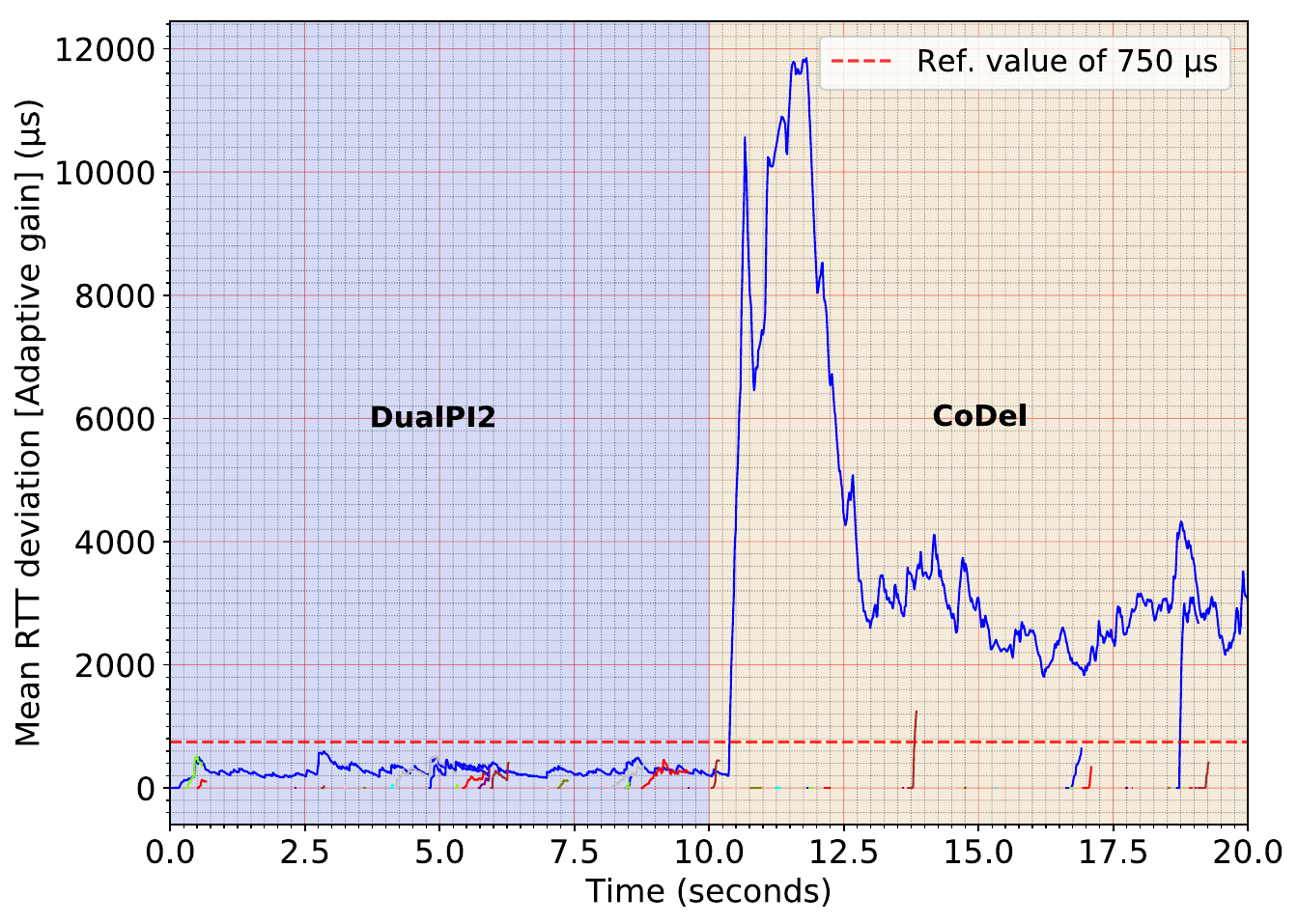}
\caption{As \autoref{fig:ecn-fallbacktr_codel_dualpi2_40m_10ms_1Lv9}, except the order of the AQM is reversed}\label{fig:ecn-fallbacktr_dualpi2_codel_40m_10ms_1Lv9}
\end{figure}

Figures \ref{fig:ecn-fallbacktr_codel_dualpi2_40m_10ms_1Lv9} \&
\ref{fig:ecn-fallbacktr_dualpi2_codel_40m_10ms_1Lv9} show the outcome in
one example scenario (40\,Mb/s link rate, 10\,ms base RTT). The scenarios
shown differ only in which AQM was applied first.

The backgrounds of the plots of the \texttt{classic\_ecn} scores (Figures
\ref{fig:ecn-fallbacktr_codel_dualpi2_40m_10ms_1Lv9}a) \&
\ref{fig:ecn-fallbacktr_dualpi2_codel_40m_10ms_1Lv9}a) have been coloured
to show the regions where the score should transition to in green (if the
algorithm correctly detects the type of AQM), and where it should
transition away from in amber or red. Red represents the unsafe situation
where a Prague flow incorrectly detects that the AQM is L4S when it is Classic.
Amber represents the safe but incorrect situation where a flow
incorrectly detects a Classic AQM when it is L4S. It can be sesen that in
both cases, the Prague flow rapidly detects and switches to the correct
AQM soon after startup, and again, soon after the switch over.

The evolution of the \texttt{classic\_ecn} score in a few of the short
Prague flows can be seen starting at their minimum value and rarely
moving before the flow finishes (by design). In
\autoref{fig:ecn-fallbacktr_dualpi2_codel_40m_10ms_1Lv9}a) a couple of
slightly longer short flows can be seen starting to move towards the
classic end of the scoring range, but only one (a blue one starting
around19\,s) lasts long enough to get there.

In both Figures \ref{fig:ecn-fallbacktr_codel_dualpi2_40m_10ms_1Lv9}b) \&
\ref{fig:ecn-fallbacktr_dualpi2_codel_40m_10ms_1Lv9}b) it can be seen
that the Prague flow's throughput seems unperturbed by the switch of AQM.
This is because it rapidly detects the type of AQM and its fall-back
algorithm quickly adapts between scalable and classic behaviour (or
\emph{vice versa}) as appropriate for the detected AQM.

Incidentally, the yellow plot tracks the total throughput of all short
flows (all Prague in this scenario). There is no expectation that their
aggregate throughput will relate in any way to the rates of the
long-running flows, because short flows act as a semi-unresponsive
aggregate.

Figures \ref{fig:ecn-fallbacktr_codel_dualpi2_40m_10ms_1Lv9}c) \&
\ref{fig:ecn-fallbacktr_dualpi2_codel_40m_10ms_1Lv9}c) give a vivid
illustration of the difference in the queuing delay that each AQM can
achieve. Over the Codel \(\rightarrow\) DualPI2 switch, the classic queue
induced by the Cubic-ECN traffic stays roughly the same, growing only
slightly, as PI2 takes over from CoDel. In contrast, once the Prague flow
detects it is in the L4S queue within the coupled DualPI2 AQM structure,
its queue rapidly reduces. The chart shows the 2\,ms threshold chosen for
the algorithm to distinguish between Classic and L4S queues, and it can
clearly be seen how the queueing delay of ECT(1) packets (green) and the
Prague flow's estimate of the queueing delay (blue) sit above or below
this threshold depending on whether the AQM is respectively Classic or
L4S.

Similarly, Figures \ref{fig:ecn-fallbacktr_codel_dualpi2_40m_10ms_1Lv9}d)
\& \ref{fig:ecn-fallbacktr_dualpi2_codel_40m_10ms_1Lv9}d) plot the
algorithm's estimate of the mean deviation of the queue, again clearly
showing that the variability of the Classic and L4S AQMs sit respectively
above and below the \(750\,\mu\)s threshold chosen to tell them apart.
Incidentally, the small coloured `scribbles' rising from the x-axis are
the estimates of mean RTT deviation starting to form in each short flow,
up to the point that each one completes.

In summary, the fall-back algorithm clearly detects a changed environment
rapidly, and switches fast so that the size of the congestion control's sawteeth 
conform to the 'rules' it expects other flows to be
following in each environment. Consequently, there is no discernible
change in per-flow throughput across the switch-over, despite the Classic
AQM not having any logic for coexistence between Classic and and scalable
flows.

\subsection{Classic ECN AQM Detection under a Range of Traffic Mixes}\label{ecn-fallbacktr_eval_detection_mix}

The online evaluations already
mentioned\textsuperscript{\ref{note:eval_url}} were used to test the
fall-back algorithm in a range of traffic scenarios. Each traffic scenario tests
one of the following flow patterns for Prague against another of this same 
list of flow patterns for Cubic-ECN:
\begin{description}
	\item[0:] denotes zero flows;
	\item[1:] denotes 1 long-running flow;
	\item[9:] denotes 9 long-running flow;
	\item[L:] denotes a `low load' of synthetic web-like short flows following an
	exponential arrival process. `Low load' equates to 1 request per second
	for the 4\,Mb/s link rate, scaled up proportionately for high rate links
	(e.g.\ 50 req/s for the 200\,Mb/s link). A lower load is used as the
	worst case, because the congestion controls and AQMs cannot settle at any
	one operating point before the load changes again. Every request opens a
	new TCP connection, closed by the server after sending data with a size
	according to a Pareto distribution with \(\alpha= 0.9\) and minimum and
	maximum sizes of 1\,KB and 1\,MB.
	\item[1L:] denotes 1 long-running flow combined with load 'L'.
\end{description}

These traffic descriptors are placed either side of a colon to denote
each traffic scenario, with the Prague flows always given first. For
example, 1:0 denotes 1 long-running Prague flow alone; while 9:L means 9
long-running Prague flows plus a low load of short, web-like Cubic-ECN
flows. 

This results in a \(4\times5\) matrix of traffic scenarios because, although there 
are 5 descriptors because there is little point in testing zero Prague flows. Each
traffic scenario was then tested in each of the combinations of 5 link rates
and 5 base RTTs already seen in \autoref{fig:ecn-fallbacktr_eval_before_after}
Thus each AQM scenario
was tested over a matrix of \((4\times5)\times(5\times5)=500\) different
traffic and environment scenarios. These are clearly far too numerous to include
exhaustively in this paper, but the general trend of the results will be described. 

Online, there is a `traffic light' visualization of all these
experiments.\textsuperscript{\ref{note:eval_url}} Each experiment was run
for 20\,s and given a colour in a large matrix of coloured squares
depending on the success of Classic ECN AQM detection within that time:
 \begin{description}
	\item[Green] denotes a true positive or true negative, (i.e.\ correctly detecting the AQM, whether Classic or L4S;
	\item[Amber] denotes a false positive, i.e.\ detecting a Classic AQM when the AQM is actually L4S;
	\item[Red] denotes a false negative, i.e.\ detecting an L4S AQM when the AQM is actually Classic;
\end{description}
False negatives are coloured red because it could be unsafe for TCP
Prague to behave as a scalable flow in a Classic AQM, if it is in a
shared queue. In such a case, TCP Prague would then outcompete any
Classic flows sharing the same bottleneck. In contrast, with false
positives, TCP Prague responds to ECN marking with a large classic
reduction, potentially under-utilizing the link, but not harming other
flows.

\begin{quote}
Note that, failure of the detection algorithm only leads to the potential
for harm not necessarily actual harm. For instance, detection is deemed
to have failed if it does not reach the correct state within the 20\,s
duration of each experiment. And in many experiments the algorithm has to
respond to a change after 10\,s, leaving it only a further 10\,s to
succeed. The choice of 20\,s and 10\,s as deadlines 
was a subjective trade-off between how long an effect is likely to be noticeable
and the impact on the run-time of all the experiments.

Often, given a little longer, the algorithm correctly detects the AQM.
Also, detection is deemed to have failed if at least a `significant
minority' of short flows switch to Classic behaviour by the time they
complete, even though they are over an L4S AQM. All these transient
effects are rarely detectable on plots of the resulting flow throughput,
let alone detectable to an end-user. 
\end{quote}

A high level summary of the results so far follows:
\begin{itemize}
	\item The fall-back algorithm correctly detected a CoDel AQM as Classic in
	497 of the 500 scenarios. All three false negatives (0.6\% of scenarios)
	involved zero Cubic-ECN flows, so there were no cases where any Cubic-ECN
	flow was at risk of being outcompeted by TCP Prague.
	
	\item The fall-back algorithm correctly detected that a DualPI2 AQM was
	L4S in 361 of the 500 scenarios. There were no unsafe false negatives,
	but the balance of the scenarios (139 or 28\%) resulted in a false
	positive. Of these:
	\begin{itemize}
		\item 75 (15\%) occurred at the 4\,Mb/s link rate, because the
		serialization delay of one packet at that rate is 3\,ms, which is greater
		than the algorithm's queue depth threshold.%
		\footnote{\label{note:eval_heuristic_plan}It is planned to address this 
		in a future revision of the algorithm by adding a heuristic that increases the 
		threshold slightly for these low flow rates.}
		
		\item 52 (10\%) occurred at the 12\,Mb/s link rate, for a similar reason,
		i.e.\ the threshold queue depth of 2\,ms is only two packet serialization
		delays at this link rate. Detection was correct in a large majority of
		the single flow tests, but only in a minority of those with multiple
		simultaneous flows (whether long-running or a mix of long and short).%
		\textsuperscript{\ref{note:eval_heuristic_plan}}
		
		\item 11 (2\%) occurred at the 40Mb/s link rate; more than half (6)
		occurring with the highest base RTT (100\,ms).\footnote{\label{note:eval_smoothing_plan}%
		Initial experiments have
		been undertaken smoothing the mean deviation of the RTT twice as fast
		(\(\mathtt{g\_mdev} = \mathtt{g\_srtt}\) rather than \( = \mathtt{g\_srtt} * 2\)). This seems to solve this
		problem without making detection of Classic worse, but a full run of
		experiments is needed to verify this.}

		\item 1 false positive occurred in the 50 scenarios at link rates above 40\,Mb/s.
	\end{itemize}
	\item Results when the AQM was switched from CoDel to DualPI2 half way
	through the run (after 10\,s) produced similar results to those over
	DualPI2 alone. However there were more false positives at the largest base
	RTT of 100\,ms, because only introducing the DualPI2 AQM after 10\,s left
	only half as long to detect it and switch to it before the 20\,s deadline. 
	Also 2 false negatives crept in. No long-running Cubic
	flows were present in either of these false-negative cases, so again
	there was no safety issue.
	
	\item When the AQM was switched from DualPI2 to CoDel half way through
	the run (after 10\,s), again the outcome was very similar to that with DualPI2
	alone, with false positives in similar types of 
	scenarios. However, this time, more false negatives appeared as well.
	Again, no long-running Cubic flows were present in any of these
	false-negative cases with one exception (1:1 at the maximum link rate and
	base RTT tested\footnote{Once the two planned alterations mentioned earlier are tested, it
	is expected that many or hopefully all of these false results will
	disappear.}).
\end{itemize}

Failures to detect the correct AQM were typically transient and, as noted
earlier, they only indicate the possibility of harm, not actual harm. No
discernible effect could be noticed on throughput balance in any of the
cases with a false result, whether false positive (amber) or false
negative (red). This is evidenced by the throughput results in
\autoref{fig:ecn-fallbacktr_eval_before_after}.

In particular, the failures at low link rate typically involved
congestion windows that were already close to TCP's minimum of 2
segments. Therefore it made little difference whether an attempt to
reduce \texttt{cwnd} was large or small, because in practice only small
reductions were possible.

\bob{ToDo: Related Work}

\section{Conclusions}\label{ecn-fallbacktr_conc}

This paper has explained and justified the design of an algorithm that
can be used by scalable congestion controls to fall back to Classic
(Reno-friendly) behaviour on detection of a Classic ECN AQM.
This is intended to ensure that no Classic
flow is beaten down by a scalable congestion control such as TCP Prague
in any Classic ECN bottlenecks that might be operational on the Internet.

Since the v1 algorithm design, the passive part of the algorithm has been implemented and extensively
evaluated over a wide range of link-rates, RTTs and traffic scenarios.
This paper now describes v2 of the algorithm. The only change 
has been to add adaptive smoothing of the RTT EWMAs, and to
set or slightly alter the setting of a number of the parameters.

The algorithm has proven itself able to make the Prague congestion
control fall back to Classic ABE-Reno behaviour whenever it proves
necessary due to a Classic AQM at the bottleneck. The resulting ratio of
the throughputs of Prague and Cubic-ECN flows easily falls within the
range of about 0.5 to 2. In contrast, without the algorithm, the
ratio can be as high as 16. 

In no test scenario has the throughput of a Prague flow been persistently
greater than twice that of a Cubic-ECN flow. Thus the algorithm satisfies 
the `Prague L4S' requirement to
fall back to Reno-friendly behaviour when operating over a Classic ECN
AQM (at least a shared one, and ideally any such AQM).
It is therefore a
candidate for use as a reference ECN fall-back design in other scalable
congestion controls.

Although so far evaluation has only been carried out with CoDel as the
Classic AQM, CoDel was chosen as the likely worst-case; being least
distinguishable from an L4S AQM. This gives some confidence that planned
evaluation with other Classic AQMs will be successful.

\section{Future Work}\label{ecn-fallbacktr_future}

The evaluation in \S\,\ref{ecn-fallbacktr_evaluation} focuses on the v2 passive
detection algorithm based on delay variation in
\S\,\ref{ecn-fallbacktr_passive_detection}. It shows promise. But it gives many
false positives at link rates below
12\,Mb/s. Also at higher link rates some false positives appear as base RTT
exceeds about 80\,ms. Initial testing of ideas for further parameter
tuning and heuristic adaptation of the thresholds were more promising
(see footnotes \textsuperscript{\ref{note:eval_heuristic_plan} \&
\ref{note:eval_smoothing_plan}}). So it was originally planned to fully explore these
avenues.

Then a large number of tests over a wider range of scenarios was planned, 
for instance:
\begin{itemize}
	\item with mixed RTTs, and emulating applications that sometimes
	self-limit the window; \item over different Classic AQMs that might be
	deployed in shared queues on the Internet (COBALT, DOCSIS PIE, RED);
	
	\item over different L4S AQMs, e.g.\ FQ\_CoDel with a shallow ECN
	threshold, DualPI2 with a ramp instead of a step;
	
	\item with different combinations of TCP Prague features enabled or
	disabled, e.g.\ RTT independence, paced chirping;
\end{itemize}

After all that, it would still be necessary to address the interaction of delay-variation measurements
with radio links, and links that add burstiness by aggregating frames.

Therefore, the focus shifted to using discrete metrics rather than delay
variation. Design ideas and a feasibility experiment on using the spacing
between marks have been written up in \S\,\ref{ecn-fallbacktr_inter-mark}. And
an idea called `exclusive' marking for altering the specification of L4S to make
it more distinct from Classic ECN marking is written up in
\S\,\ref{ecn-fallbacktr_active_solution2}. Although these approaches still
require empirical validation, the range of experiments needed should be far narrower
and less onerous.

Since the v2 algorithm was evaluated, there has also been a shift of focus
towards out-of-band testing; to validate paths prior to L4S deployment with
occasional reviews afterwards. A new section on out-of-band testing has been
added at \S\,\ref{ecn-fallbacktr_OOB_active}, where a very simple but adequate
test is defined. If out-of-band testing is deemed sufficient, nothing more would
be necessary than to verify that simple test.

\section{Acknowledgements}\label{ecn-fallbacktr_acks}

Transitioning gradually from scalable to classic behaviour, using ECT0 packets
for active detection and using the spacing between marks were based on initial
ideas suggested by Koen De Schepper. The ideas based on altering the
standardization of L4S AQMs (\S\,\ref{ecn-fallbacktr_active_solution2}) were
suggested by Alex Burr\footnote{In an email posting to the IETF's tsvwg mailing
	list:
	\url{https://mailarchive.ietf.org/arch/msg/tsvwg/mIqNOa5vmIjFk_LNdOLJHQfm5s8/}}.
Thanks are due to Olivier Tilmans and Koen De Schepper for reviewing the design
approach and the code, and to Tom Henderson for advice on presentation of the
results.

Bob Briscoe was part-funded by the Comcast Innovation Fund. The views expressed
here are solely those of the author.

\clearpage
\addcontentsline{toc}{section}{References}

{\footnotesize%
\bibliography{ecn-fallback}}

\clearpage
\appendix
\section{Adapting the RTT Smoothing Timescale}\label{ecn-fallbacktr_srtt_adapt}

Given the EWMA of the RTT is meant to be smoothed over a sawtooth, and it is updated per ACK, na\"{\i}vely, one would think the smoothing gain should be the reciprocal of the average number of ACKs in a TCP sawtooth, \(n_a\).  However, there are three reasons not to take this na\"{\i}ve approach:
\begin{itemize}
	\item The whole reason for trying to move away from Classic congestion
	controls is that the average duration of a sawtooth grows linearly with
	the congestion window. So, if the smoothing gain tracked \(n_a\), the
	classic ECN AQM detection algorithm would become linearly more sluggish
	as the congestion window scaled up.
	
	\item In practice, the duration of a Classic sawtooth does not actually
	grow linearly as link rate scales, because the longer it takes, the more
	likely it will mistake some other disturbance for congestion, such as a
	competing short flow or a bit error.\footnote{For the same reason, as link capacity
	scales, it is generally accepted that there is no need to try to maintain
	`fairness' with a congestion control that increasingly underutilizes the
	available capacity all by itself, meaning that its sawteeth do not
	continue to grow to full utilization, instead collapsing and restarting
	long before they have reached their theoretically `fair' amplitude.}
	
	\item Classic ECN AQM detection is only needed for the transition from
	Classic to scalable (L4S) congestion controls. If the L4S experiment is
	successful, there will be no need to scale the transition mechanism 
	beyond the flow rates Classic congestion controls can attain today.
\end{itemize}

Therefore, a formula for the average number of ACKs in a TCP sawtooth
\(n_a\) will be derived, but then the degree by which it scales with
congestion window will be reduced using empirical experience, to
produce a heuristic that works in practice.

\begin{align}
	n_a	&\approx \mathrm{avg\_packets\_per\_sawtooth} / d\notag\\
			&\approx  1 / (p*d),\label{eqn:ecn-fallbacktr_n_a_wrt_p}
\end{align}
where \(d\) is the delayed ACK ratio (typically 2) and \(p\) is the
probability of an ECN mark, because a mark ends a sawtooth and on average
there will be one mark per \(1/p\) packets.

It would be efficient to only recalculate the gains to use for
\texttt{srtt} and \texttt{mdev} at the end of every sawtooth, when the
slow-start threshold (\(S\)) is recalculated. We relate \(p\) to the
average congestion window \(W\) using the steady state response function
of a Classic congestion control, taking Reno as the worst
case~\cite{Mathis97:TCP_Macro}. Then it will be straightforward to relate
\(W\) and therefore \(p\) to \(S\).
\begin{align}
	W_{\mathrm{Reno}} &\approx \sqrt{\frac{3}{2p}}\notag\\
\intertext{Substituting for \(p\) in \autoref{eqn:ecn-fallbacktr_n_a_wrt_p}}
	n_a &\approx 2W_{\mathrm{Reno}}^2/3d\label{eqn:ecn-fallbacktr_n_a_wrt_W}\\
\intertext{For Reno, the average congestion window lies midway between 
\(S\) and \(S/\beta\), where \(\beta\) is the reduction factor of the congestion control:}
	W_{\mathrm{Reno}} &\approx (1+1/\beta) / 2 * S_{\mathrm{Reno}}\notag\\
\intertext{Substituting for \(W_{\mathrm{Reno}}\) in 
\autoref{eqn:ecn-fallbacktr_n_a_wrt_W}}
	n_a &\approx (1+1/\beta)^2 / (6d) * S^2\notag\\
		   &\approx 3 / 4 * S^2,\notag
\end{align}
assuming a typical delayed ACK ratio, \(d = 2\), and Reno as the
worst-case congestion control where \(\beta_{\mathrm{Reno}}=0.5\).

For efficiency, the smoothing gain should be an integer power of 2, so
that the integer power can be used as a bit-shift. A fast integer base 2
log (such as ilog2()) rounds by truncation. Therefore, for unbiased
rounding, \(n_a\) would need to be scaled up by 3/2 before taking the
log. Thus, the integer log of \(3/2*3/4*S^2 \approx S^2\) would be taken.

As explained above, in practice an empirical formula for the smoothing
gain is used in place of this theoretical formula. Nonetheless, this
theoretical number of ACKs per sawtooth helped set an upper bound to the
range of formulae to be tried in experiments. The formula for smoothing
gain, \(g_{\mathrm{fbk_srtt}}\), used in experiments was of the form:
\[g_{\mathrm{srtt}} = U2 * S^{U1}.\]

In experiments with a range of link rates between 4\,Mb/s and 200\,Mb/s
and RTTs between 5\,ms and 100\,ms, the parameters for this formula that
resulted in a good compromise between precision and speed of response
were: \[g_{\mathrm{srtt}} = 2 * S^{3/2}.\]

Currently the mean deviation is smoothed twice as slowly as this, i.e.:
\[g_{\mathrm{mdev}} = g_{\mathrm{srtt}} * 2.\]

\bob{Currently, the Linux code and the pseudocode uses g\_mdev = g\_srtt * 2}
\section{Implementation in Integer Arithmetic}\label{ecn-fallbacktr_impl}

\subsection{EWMA Precision and Upscaling}\label{ecn-fallbacktr_precision}

The pseudocode below repeats the EWMA pseudocode given in
\autoref{ecn-fallbacktr_pseudocode_srtt}, but in integer arithmetic including recommended
types for the variables and the formulae used to check for overflow.
Rationale for the recommended types is given below the pseudocode.

All the variables that are specific to Classic ECN AQM fall-back are
prefixed with \texttt{fbk\_}, so the EWMAs of srtt and mdev are called
\texttt{fbk\_srtt} and \texttt{fbk\_mdev}.\footnote{To save space in the
TCP control block, it may be preferred to solely store g\_srtt\_shift and
recalculate g\_mdev\_shift as needed.}

\begin{verbatim}
#define ACC_MRTT_MAX 0xFFFFFFUL    // < 2^24 [us] (=16.7s)
#define FBK_SSTHRESH_MAX 0x0FFF    // < 2^((20 - 2) * 2/3) - 1 = 2^(12)  (see text)
#define FBK_G_DIFF 1               // fbk_g_mdev = fbk_g_srtt * 2^(FBK_G_DIFF)

// Stored variables
u64 fbk_srtt;                // Upscaled smoothed RTT
u64 fbk_mdev;                // Upscaled mean deviation
int fbk_g_srtt_shift;        // srtt gain bit-shift (initialized dependent on ssthresh)
int fbk_g_mdev_shift;        // mdev gain bit-shift (initialized dependent on ssthresh)
u32 fbk_mdev_carry;          // Geometric carry
u32 fbk_depth_carry;         // Geometric carry

// Temporary variables
u32 acc_mrtt;                // Measured RTT from newest unack'd packet
s64 error_;
int delta_shift_;

{   // At start of connection, either after first RTT measurement or from dst cache
    fbk_srtt = acc_mrtt<<fbk_g_srtt_shift;
    fbk_mdev = 1ULL<<fbk_g_mdev_shift; // No need for conservative init, unlike for RTO
}

{   // Per ACK
    acc_mrtt = min(acc_mrtt, ACC_MRTT_MAX);
    // Update EWMAs
    error_ = (u64)acc_mrtt - fbk_srtt>>fbk_g_srtt_shift;
    fbk_srtt += error_;
    fbk_mdev += llabs(error_) - fbk_mdev>>fbk_g_mdev_shift;    // See text re llabs()
}

{   // Per ssthresh change, including when initialized
    delta_shift_ = -fbk_g_srtt_shift;               // Store old shift
    fbk_g_srtt_shift = ilog2(min(ssthresh, FBK_SSTHRESH_SHIFT_MAX));
    fbk_g_srtt_shift += fbk_g_srtt_shift>>1 + 1;    // fbk_g_srtt = 2 * ssthresh^(3/2)
    fbk_g_mdev_shift = fbk_g_srtt_shift + FBK_G_DIFF;   // fbk_g_mdev = fbk_g_srtt * 2

    // Adjust all upscaled variables
    delta_shift_ += fbk_g_srtt_shift;     // Difference between old and new shift
    if (delta_shift_ > 0) {
        fbk_srtt <<= delta_shift_;
        fbk_srtt_carry <<= delta_shift_;
        fbk_mdev <<= delta_shift_;        // Same shift for mdev
        fbk_mdev_carry <<= delta_shift_;
    } else if (delta_shift != 0) {
        delta_shift_ = -delta_shift_;
        fbk_srtt >>= delta_shift_;
        fbk_srtt_carry >>= delta_shift_;
        fbk_mdev >>= delta_shift_;        // Same shift for mdev
        fbk_mdev_carry >>= delta_shift_;
    }
}
\end{verbatim}

The question of the size of the EWMA variables can be broken down into
the number range needed prior to upscaling, and the maximum upscaling to
be supported.

\paragraph{EWMA Range Prior to Upscaling:} Both the EWMAs
(\texttt{fbk\_srtt} and \texttt{fbk\_mdev}) are fed by the measured RTT
\texttt{acc\_mrtt} so (prior to upscaling) they never need to hold a
value greater than the max usable value of \texttt{acc\_mrtt}.

RFC793 defines the maximum segment lifetime (MSL) as 2 minutes (implying
a maximum RTT of 4 minutes = \(2^{28}\,\mu\)s), but it would do no harm
if the fall-back algorithm clamped any RTT exceeding, say 15\,s to a
maximum. A bloated buffer might well lead to an average RTT of a few
seconds, but outliers beyond 15\,s would unnecessarily pollute the
average. If most RTT measurements were this high, there would be little
benefit in making classic ECN detection so sluggish. Therefore it will be
reasonable to clamp \texttt{acc\_mrtt} below \(2^{24}\,\mu\)s (16.7\,s)
and use a 32-bit unsigned long integer for it.

\paragraph{Upscaling Limit:} In order to implement an EWMA in integer
arithmetic, upscaling the EWMA by the reciprocal \(g\) of its gain
optimizes the most frequently run parts of the code---those run per ACK.
Upscaling by \(g\) is straightforward when constant gain is used. But
when the gain is adapted (see \autoref{ecn-fallbacktr_srtt_adapt}) it is
necessary to change the upscaling of the EWMAs, and one or two other
variables that depend on this upscaling.

We tried the alternative of always upscaling by a constant maximum
factor.\footnote{Another alternative might be to use the same upscaling
factor for both \texttt{fbk\_srtt} and \texttt{fbk\_mdev} even if their
gains are different.} This does indeed simplify the calculations needed
per sawtooth,
but it adds 3 or 4 bit-shifts to the per-ACK calculations. On balance, it
was decided to prioritize optimization of per-ACK processing, given the
extra per-sawtooth cost is only about half-a-dozen simple operations (a
couple of adds; an if; and a few bit-shifts).\footnote{Incidentally,
while we are talking about the `Per ACK' block, it is particularly
important not to use inefficient code here. The
\texttt{llabs()} function is used to take the absolute value of
\texttt{error\_}. The implementation of \texttt{llabs()} is not shown,
but it is intended to be similar to the assembler used in the function of
the same name in stdlib.h, which returns the absolute value of a long
long signed integer.} This involves
upscaling by a variable amount that depends on the gain. However, it
still leaves the question of what upper bound to fix for upscaling.

A workable upper bound for the number of ACKs over which smoothing is
carried out would be \(2^{20}\) (a little over 2 minutes at 200\,Mb/s
assuming 1500\,B packets and a delayed ACK every 2 packets). As flow rate
scaled beyond this, the adaptation mechanism would stop growing the
number of ACKs over which smoothing would occur and the ACK rate would
increase, so the maximum smoothing time would reduce, although this might
be mitigated somewhat by an increase in the ACK ratio. However,
\(2^{20}\) is believed to cater for sufficient scale, for the three
reasons given in \autoref{ecn-fallbacktr_srtt_adapt} (sufficient for
transition out of an unscalable regime; the likelihood of disturbance
within this time; and ensuring that detection remains responsive).

In the `Per ssthresh change' block of the above pseudocode, when
\texttt{fbk\_g\_srtt\_shift} is derived from \texttt{ssthresh}, instead
of calculating the unbounded value then clamping it, the intermediate
value is clamped to an adjusted down limit before it is raised to the
power of 3/2 and doubled. This limit is held in the constant macro
\texttt{FBK\_SSTHRESH\_SHIFT\_MAX}, which is reverse engineered as
\((20-2)*2/3 = 12\). This does not limit ssthresh itself. It is just the
value of ssthresh at which to cap the number of ACKs over which smoothing
is carried out.

\paragraph{Precision and Upscaling Summary:} \texttt{acc\_mrtt} uses 23
significant bits, and \texttt{fbk\_g\_mdev\_shift} upscales by a maximum
of 19 bits, making 42 bits for the largest EWMA. Therefore the EWMAs will
easily fit in a u64, but it will not fit in a u32.

\subsection{Iterative Fast Log Calculations with Geometric Carrying}\label{ecn-fallbacktr_ilog}

The Classic ECN detection algorithm takes the ratios between the
\texttt{fbk\_srtt} or \texttt{fbk\_mdev} metrics and their reference
values and transforms them into linear step changes in the
\texttt{classic\_ecn} score. This requires a log function but, rather
than using floating point arithmetic, it uses the base 2 integer log
function \texttt{ilog2()}, which has very low processing cost, but also
very little precision. Essentially \texttt{ilog2()} returns the binary
order of magnitude of the operand, i.e.\ the position of the most
significant bit that is set to 1 (typically using an assembler
instruction such as clz, which stands for count leading zeros).

The order-of-magnitude precision of \texttt{ilog2()} is sometimes good
enough, but not always. It is sensitive to whether the configured
threshold value is close to a step change in the integer log, which can
cause behaviour to stick then suddenly toggle. For instance if the
threshold were set to \(\mathtt{V0}=500\,\mu\)s, in floating point
arithmetic \(\mathrm{lg}(500)=8.965784285\), but in integer arithmetic
\(\mathrm{ilog2}(500)=8\). So RTT values lower than the threshold, down
to \(257\,\mu\)s would appear to be at the threshold.

Rather than resort to floating point arithmetic, we implemented an
iterative carry algorithm, \texttt{carry\_ilog2()}, to accumulate the
error and feed it back into the log function, which then outputs the
integer values either side of the precise answer, in proportion to the
error term at any one time. The cost is 3 adds, 2 bit-shifts and 1
integer multiply. In following pseudocode It is defined at the end and
used twice in the `Per RTT' block.

\begin{verbatim}
#define PRAGUE_ALPHA_BITS	20U  // Upscaling of classic_ecn
#define V_LG 1                 // Weight of queue *V*ariability metric, default 2^(-1)
#define D_LG 1                 // Weight of mean queue *D*epth metric, default 2^(-1)
#define V0_LG_US (10014684UL >> V) // Ref queue *V*ariability [us], default lg(750)<<20
#define D0 2000                    // Ref queue *D*epth [us]
#define D0_LG_US (11498458UL >> D) // Ref queue *D*epth [us], default lg(2000)<<20

// Stored variables
int classic_ecn;               // Classic ECN AQM detection score
u32 rtt_min;                   // Min RTT (uses Kathleen Nichols's windowed min tracker)

{   // At start of connection, either after first RTT measurement or from dst cache
    fbk_mdev_carry = 1 << fbk_g_mdev_shift;
    fbk_mdev_carry += 1 << (fbk_g_mdev_shift - 1);       // Initialize to 3/2 upscaled
    fbk_depth_carry = 1 << fbk_g_srtt_shift;
    fbk_depth_carry += 1 << (fbk_g_srtt_shift - 1);      // Initialize to 3/2 upscaled
}

{   // Per RTT
    // Temporary variables for readability
    u64 fbk_mdev_;
    u64 fbk_depth_;
    int fbk_mdev_lg_;
    int fbk_depth_lg_;

    fbk_mdev_ = fbk_mdev>>fbk_g_mdev_shift;               // remove upscaling
    fbk_mdev_lg_ = carry_ilog2(fbk_mdev_, fbk_g_mdev_shift, *fbk_mdev_carry)
    // V*lg(v/V0)
    classic_ecn += (u64)fbk_mdev_lg_<<(PRAGUE_ALPHA_BITS - V);
    classic_ecn -=  V0_LG_US;

    fbk_depth_ = fbk_srtt>>fbk_g_srtt_shift - rtt_min;    // Smoothed q depth
    fbk_depth_lg_ = carry_ilog2(fbk_depth_, fbk_g_srtt_shift, *fbk_srtt_carry)
    // D*lg(max(d/D0,1))
    fbk_depth_lg_ <<= PRAGUE_ALPHA_BITS - D
    if (fbk_depth_lg_ > D0_LG_US) {
        classic_ecn += (u64)fbk_depth_lg_ - D0_LG_US;
    }
}

int carry_ilog2(u64 arg, int shift,  u32 *carry) {
    // returns integer base 2 log of arg factored up by carry upscaled by shift
    int arg_lg_;

    // Multiply non-upscaled arg by upscaled geometric carry from previous round
    arg *= *carry;
    arg += 1 << (shift - 1);  // Add upscaled 1/2 to unbias truncation
    arg_lg_ = ilog2(arg) - shift;         // Non-upscaled integer log
    *carry = arg >> arg_lg_;              // Upscaled geometric carry
    return arg_lg_;
}
\end{verbatim}

Being a log function, the carry algorithm needs to be multiplicative
(geometric), not additive. The comment `Upscaled geometric carry' tags
the line that calculates the carry factor, pointed to by \texttt{carry},
by right bit-shifting the upscaled value of \texttt{arg} by its own
integer log.

For instance, if the value of \texttt{fbk\_mdev} passed to the algorithm
is \(500\,\mu\)s, \(\mathrm{ilog2}(500)=8\), so bit-shifting 500 by 8 is
equivalent to \(500/2^{8}=1.953125\). Obviously, in integer arithmetic
that would always truncate to 1, but by starting with an upscaled value
of *carry, an upscaled value of the new carry factor
(\(1.953125*2^{\mathtt{shift}}\)) is produced.

In the next iteration, the next value of \texttt{fbk\_mdev} passed to the
function as \texttt{arg} is multiplied by the new upscaled carry factor
in the line \texttt{arg *= *carry}, which produces an upscaled result.
Two lines later, the integer log of this upscaled and factored up value
is taken. Continuing the example case with \(\mathtt{fbk\_mdev} =
500\,\mu\)s, the \texttt{ilog2()} function is applied to
\(500*(1.953125<<\mathtt{shift})\) before shift is subtracted, which
produces 9 not 8. As the process iterates, if a steady state of
\(\mathtt{mdev} = 500\,\mu\)s were to remain, the output of the integer
log would be 8 only 3.422\% of the time, while it would be 9 the other
96.578\% of the time (thus averaging to \(\mathrm{lg}(500)\), which is
8.96578...) .

Even with geometric carry implemented, the average of all the integers
logs still understates the value relative to the true floating point log.
This is because every right shift to remove the extra degree of upscaling
truncates the output. To counteract this truncation bias, 0.5 is added
before the downscaling, because 0.5 is the average of all the possible
values of the truncated bits. Of course, 0.5 is upscaled by
\texttt{shift} before being added, to match the upscaling of the
\texttt{arg} that it is added to.

\paragraph{Precision:} When the EWMAs are calculated, \texttt{fbk\_srtt}
and \texttt{fbk\_mdev} are upscaled by their respective \(g\) shifts.
However, when their logs are taken, copies of these stored upscaled
values are taken and reverted to their non-upscaled values. Instead, the
carry factor is upscaled by the \(g\) shift. Thus, the same number space
is needed for upscaling, but the values of the EWMAs need two additional
bits on top of their usual maximum size (\(<2^24\) as outlined earlier).
These two bits consist of: \begin{itemize} \item The additive adjustment
of 0.5 that unbiases the integer log truncation, which consumes one bit
in the worst-case.
	
\item The unscaled value of the geometric carry factor, which takes a
value between \((1\le \mathtt{carry}<2)\), which also consumes 1 bit.
\end{itemize} Thus, \(42+2=44\) bits are needed for the local variable
holding the factored up and adjusted EWMA (\texttt{arg}) within the
\texttt{carry\_ilog2()} function.

\subsection{Combining logs}\label{ecn-fallbacktr_log_combine}

In the passive detection algorithm, both the depth (d) and variability
(v) terms involve a log function. So in practice, with
judicious choice of the parameters V and D, they could be combined
efficiently. For instance, if \texttt{V=D}, then\\
\-\qquad\texttt{V*lg(v/V0) + D*lg(max(d/D0,1))}\\
is equivalent to\\
\-\qquad\texttt{( d>D0 ? V*lg(v*d/(V0*D0)) : V*lg(v/V0) )}

Then the `Per RTT' block of pseudocode above would be replaced with that below, which aggregates two lots of steps into one.

\begin{verbatim}
// Stored variables
    fbk_ewmas_carry;        // Replaces fbk_srtt_carry

{   // Per RTT
    // Temporary variables for readability
    u64 fbk_mdev_;
    u64 fbk_depth_;
    u64 fbk_ewmas_;
    int fbk_ewmas_lg_;

    fbk_mdev_ = fbk_mdev>>fbk_g_mdev_shift;             // Remove upscaling
    fbk_depth_ = fbk_srtt>>fbk_g_srtt_shift - rtt_min;  // Smoothed q depth w/o upscaling

    if (fbk_depth_ > D0) {
        fbk_ewmas_ = fbk_mdev_ * fbk_depth_;    // Product of EWMAs
        fbk_ewmas_lg_ = carry_ilog2(fbk_ewmas_, fbk_g_mdev_shift, *fbk_ewmas_carry)
        // V*lg(v*d/(V0*D0)), assuming V = D
        classic_ecn += (u64)fbk_ewmas_lg_<<(PRAGUE_ALPHA_BITS - V);
        classic_ecn -=  V0_LG_US + D0_LG_US;
    } else {
        fbk_mdev_lg_ = carry_ilog2(fbk_mdev_, fbk_g_mdev_shift, *fbk_mdev_carry)
        // V*lg(v/V0)
        classic_ecn += (u64)fbk_mdev_lg_<<(PRAGUE_ALPHA_BITS - V);
        classic_ecn -=  V0_LG_US;
    }
}
\end{verbatim}
\section{Algorithm for Filtering Reroutes out of RTT Metrics}\label{ecn-fallbacktr_alt-srtt}

\begin{figure}
  \centering
  \includegraphics[width=0.7\linewidth]{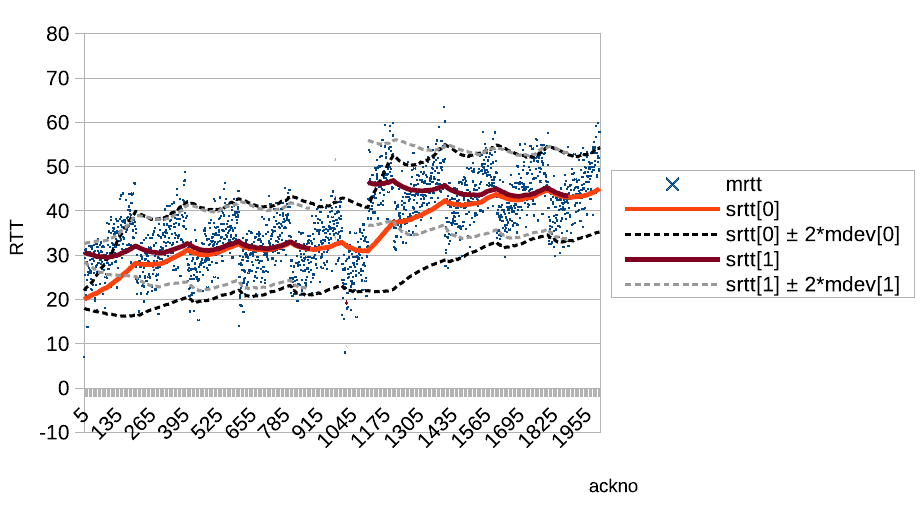}
\caption{Example plot of `incorrect' (light red) and `correct' (dark red)
smoothed RTT and their mean deviations after a reroute at ackno
1100}\label{fig:ecn-fallbacktr_reroute}
\end{figure}

If the base RTT suddenly changes, e.g.\ due to a re-route, the smoothed
RTT (\texttt{srtt}) could take a long time to converge on the new value.
In the meantime, the mean deviation (\texttt{mdev}) will expand, not
because there is more queue variability, but because \texttt{srtt} is
`incorrect'. The resultant increase in \texttt{mdev} could cause the
fall-back detection algorithm to incorrectly detect a Classic AQM, even when
it is not and there is very littel queue variability.

The pseudocode below is designed to rapidly lock-on to a better smoothed
RTT following a significant step change in the base RTT. Consequently,
the mean deviation based of this `corrected' smoothed RTT should not
increase when there is a step-change in base RTT.

\autoref{fig:ecn-fallbacktr_reroute} shows an example with both the
uncorrected smoothed RTT in bright red (labelled \texttt{srtt[0]}) and
the corrected smoothed RTT in dark red (labelled \texttt{srtt[1]}).

Numerous data points for the actual RTT measurements are shown as tiny
blue crosses. On closer inspection, a characteristic sawtoothing pattern
can be seen in data points as TCP cycles between capacity-seeking and
relaxing. 
At ackno=1100, a step increase in the sawtoothing pattern can be picked
out, despite the noisy data. For comparison, the bright red plot labelled
\texttt{srtt[0]} is the uncorrected smoothed RTT, which can be seen
slowly converging on the new smoothed average after ackno 1100. In
contrast, the dark red smoothed RTT labelled \texttt{srtt[1]} rapidly
corrects itself by jumping to a `better' value soon after the re-route.

Once the original EWMA converges with the alternative one, the 
algorithm disables calculation of the alternative to save processing,
as can be seen where the dark red plot stops. The algorithm is also useful
at the start of a flow, where it can quickly correct an unfortunate initial
value of the EWMA, as seen on the left of \autoref{fig:ecn-fallbacktr_reroute}.

The black dashed plots labelled \(\mathtt{srtt[0]} \pm
2*\mathtt{mdev[0]}\) illustrate two mean deviations either side of the
uncorrected \texttt{srtt[0]} plot. After the re-route, this mean
deviation can be seen to expand. In contrast, the grey dashed lines
illustrate two mean deviations either side of the corrected
\texttt{srtt[1]} plot (labelled \(\mathtt{srtt[1]} \pm
2*\mathtt{mdev[1]}\) ). It can be seen that \texttt{mdev[1]} hardly
increases at all during the re-route episode.

This is actually a clue to how the algorithm works; whenever there is an
outlier, it starts a new EWMA (\texttt{srtt[1]}) from that outlier and
initializes the new mean deviation (\texttt{mdev[1]}) with a slightly
inflated copy of the original mean deviation (\texttt{mdev[0]}). It only
maintains the new EWMA as long as it results in a tighter mean deviation
than that based off the original EWMA. The inflation factor for the new
mean deviation (\texttt{K1}) is contrived in such a way that the new EWMA
will quickly be rejected if it is not `better'. In particular, even if
the next data point after the outlier is as close as possible to the
first outlier without actually being an outlier (with respect to the
original EWMA), the alternative EWMA will immediately be disabled (and it
will never have been used, because the initial inflated mean deviation
was `worse' than the original). The derivation of \texttt{K1} is left as
an exercise for the reader. \bob{Strictly, a similar exercise using the
lower outlier threshold will need to be done, because the outcome of the
mean deviation approach is asymmetric. Then the greater value of K1
should be used.}

In this way, the algorithm is (usually) able to ignore TCP's sawtooth
jumps, given the mean deviation established over time includes them. By
default, outliers are defined as outside \(\texttt{K2}=2\) mean
deviations.

The pseudocode is given below, followed by a walk-through. It is written
as a general-purpose algorithm that could be used to `correct' any EWMAs
of RTT and its deviation, whether for Classic ECN AQM detection,
retransmit timer calculation or moving averages that are nothing to do
with RTT for that matter. Nonetheless, in the case of Classic ECN AQM
detection, the variables \texttt{srtt[]} and \texttt{mdev[]} are intended
to be equivalent to \texttt{fbk\_srtt} and \texttt{fbk\_mdev} as defined
in pseudocode throughout the rest of this paper (not the \texttt{srtt}
and \texttt{mdev} variables already maintained by Linux for RTO
calculation). Similarly, for Classic ECN AQM detection, the gain
parameters (\texttt{g\_srtt} and \texttt{g\_mdev}) are intended to be
equivalent to \texttt{fbk\_g\_srtt} and \texttt{fbk\_g\_mdev} in the rest
of this paper.

\begin{verbatim}
/* Macros */
#define G1 (1/g_srtt)     // The gain already used to maintain srtt
#define G2 (1/g_mdev)     // The gain already used to maintain mdev
#define K2 2              // Outlier threshold, as multiple of mdev
#define K1 1+G2*(K2*(1-G1)+(K2-1)*(1-G2)-1) // Hysteresis factor (see text later)
// SRTT or MDEV can be used wherever TCP uses srtt or mdev
#define SRTT (srtt[1] && mdev[1] < mdev[0] ? srtt[1] : srtt[0]) 
#define MDEV (srtt[1] && mdev[1] < mdev[0] ? mdev[1] : mdev[0])

/* Definitions of variables and functions
 *	srtt[1] and mdev[1] are candidate alternatives to srtt[0] and mdev[0], 
 *  which are the original uncorrected variables
 */
srtt[2];                 // Array for smoothed RTT (primary and alt)
mdev[2];                 // Array for mean deviation of RTT (primary and alt)
acc_mrtt;                // Latest measured accurate RTT (excl. delayed ACK)
error_;
sign_ = 0;

/* Initialize EWMAs */
srtt[1] = FALSE;        // holds the alt smoothed RTT, but if FALSE disables alt's.
mrtt[1] = 0;
// srtt[0] and mdev[0] will have been initialized elsewhere (resp. to acc_mrtt and 0)

{ /* Per ACK */
    error_ = acc_mrtt - srtt[0];
    // Check for inlier or outlier on opposite side to the alt srtt
    if (  (abs(error_) <= K2 * mdev[0])                                // Inlier
            ||  (srtt[1] && (sign_ * error_ <= K2 * mdev[0])) ) {      // Opp. outlier
        if (srtt[1]) {
            mdev[1] += G2 * (abs(acc_mrtt - srtt[1]) - mdev[1])); // Update alt mdev
            if (mdev[1] > mdev[0])) {    // suppress alt's if worse mean deviation
                srtt[1] = FALSE;
            } else {                     // Continue to maintain alt srtt
                srtt[1] += G1 * (acc_mrtt - srtt[1]);
            }
        }
    } else {                             // Outlier (on same side if alt srtt enabled)
        if (srtt[1]) {                   // Continue to maintain alt srtt
            mdev[1] += G2 * (abs(acc_mrtt - srtt[1]) - mdev[1]));
            srtt[1] += G1 * (acc_mrtt - srtt[1]);
        } else {                         // Initialize alt's
            mdev[1] = mdev[0] * K1;      // Inflate by K1 for hysteresis
            srtt[1] = acc_mrtt;
            sign_ = sgn(error_);         // Record which side the outlier is on
        }
    }
    
    // Regular update of non-alt srtt and mdev  (order-significant)
    mdev[0] += G2 * (abs(acc_mrtt - srtt[0]) - mdev[0]));
    srtt[0] += G1 * (acc_mrtt - srtt[0]);
}
\end{verbatim}
\paragraph{Pseudocode Walk-Through} 

The purpose of each variable is assumed to be self-explanatory from the comments in the pseudocode. So this explanation will focus on the logic, which all executes on a per-ACK basis.

The structure of the main `if' block consists of two inter-locked conditions in a bistable flip-flop arrangement:
\begin{itemize}
    \item Whether the latest RTT is an outlier with respect to \texttt{srtt[0]} (and, whether it outlies on the same side as the alternative EWMA, if one is enabled\footnote{This prevents an alternative EWMA being started from an outlier on one side of the original \texttt{srtt[0]} EWMA, but kept enabled by outliers on the other side. This would otherwise be a common occurrence given outliers at the top of a TCP sawtooth are typically followed by outliers at the bottom of the next sawtooth.});
    
    \item Whether alternative EWMAs are enabled (i.e.\ whether srtt[1] is non-zero).
\end{itemize}
Within each branch of the ``outlier?'' condition, there is an ``alt's enabled?'' condition, so the code flip-flops as follows:
\begin{itemize}
    \item If ``outlier?'' is true, and ``alt's enabled?'' is not, alt's are enabled.
    
    \item If ``outlier?'' is false, and ``alt's enabled?'' is true, then alt's are tested to see if they should be disabled (they are disabled if the alternative \texttt{mdev[1]} is worse (larger) than the original \texttt{mdev[0]}). 
\end{itemize}

In the stable state when alternative EWMAs are enabled, they are maintained as normal. And in the stable state when alternative EWMAs are disabled, they are not. This prevents the code doing unnecessary per-ACK work. In production code, a tolerance could easily be added to disable maintenance of alternative EWMAs sooner; when they are hardly any better than the originals.

Once the main `if' block completes, the original EWMAs are updated, as always.

The macros SRTT and MDEV are provided for other code to use. They compare the mean deviation of the two EWMAs and use the best (smallest).  In production code, to make these macros more efficient, when \texttt{srtt[1]} is set to FALSE to disable alternative EWMAs, \texttt{mdev[1]} could be set to a special `inifinite' value, so that these macros would not need to test \texttt{srtt[1]} as well as \texttt{mdev[1]}.

\section{Core or Peering Link with Shared-Queue Classic ECN AQM}\label{ecn-fallbacktr_common}

First, for brevity, we will use the term `common link' for either a core link or
a peering link. This appendix is about the possibility that such a common link 
could use a shared-queue Classic ECN AQM, and that it could become a bottleneck.

Initially, let us assume an ideal situation in which end systems all ensure equal flow rates at a
bottleneck and let us define the equal division of a bottleneck's capacity
among all flows bottlenecked there as the 'equitable rate' for that
bottleneck.

Usually networks are designed so that flows bottleneck in access links. 
If the number of flows converging into a common link grows, 
even though they are all bottlenecked elsewhere, there will 
come a point where the sum of all the flows feeding traffic into the common link exceeds its
capacity. If the number of flows continues to rise, the equitable rate for the
common bottleneck will continue to reduce. As the equitable rate reduces below the
highest capacity access links, the bottleneck for any lone flow in each of those
access links will move to the common link.

Let us now imagine that the equitable rate has just reduced to 50\% of the
capacity of the fastest access link feeding the common bottleneck. If it
contains one flow, that will now be running at half the access rate. If it
contains two flows, the bottleneck for them will start to move to the common
link as well.

Now let's change the scenario by replacing some classic sources with L4S. At
the common bottleneck (still assuming a Classic ECN AQM), there will be little variability in the queue because
of the high degree of multiplexing. So as the bottleneck moves there, L4S
flows will not detect a classic ECN bottleneck and they will yield less than
any classic flows. Classic flows will end up below the equitable rate and L4S
flows above it. However, as L4S flows increase, they will bottleneck in their
own access link again, which will naturally limit the inequality to \(
100\%/50\% = 2\times\).

Of course, serious anomalies might concentrate so much load at a common link
that the equitable rate reduces to less than 50\% of the fastest access, say
\(x\%\). Then the worst inequality would be \(100/x\times\). But it is
extremely rare for an anomaly to even reduce \(x\) to 50\%. In robustly
designed networks, even during an anomaly, \(x\) will only just dip below
100\%, e.g. 95\%. Then the worst inequality due to classic ECN fall-back not
detecting a classic ECN AQM in a highly multiplexed common link would be
\(100/95 = 1.05\times\).

\begin{figure}
  \centering
  \includegraphics[width=\linewidth]{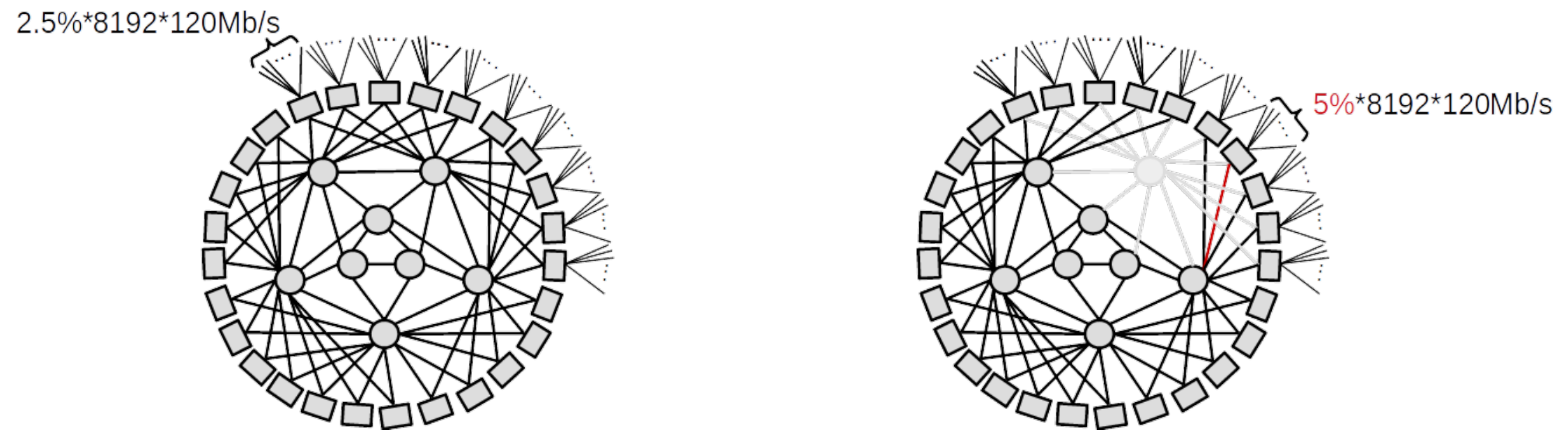}
\caption{Example before (left) and after (right) a dual anomaly that shifts 
congestion to a core link}\label{fig:ecn-fallbacktr_dual-anomaly}
\end{figure}

\paragraph{Example:} Starting with the network as designed on the left of
\autoref{fig:ecn-fallbacktr_dual-anomaly}, the 8 circles in the centre
are core routers. The rectangles connected around them are broadband
network gateways (BNGs), each dual-homed to two core routers, and each
providing Internet access to 8192 customer access links. For ease of
explanation, the access links all have the same 120\,Mb/s capacity.

The network is designed on the assumption of 2.5\% access network
utilization outwards, i.e.\ 2.5\% of access links are fully utilized at
any one time (or proportionately more than 2.5\% are partially utilized).
Thus, each BNG needs 2.5\%*8192*120\,Mb/s = 25\,Gb/s. The core links are
40\,Gb/s. Assuming reasonably equal load balancing in the core (e.g.\
randomized equal cost multipath routing), this implies each core link is
designed to be utilized at roughly \(25/(40*2) = 31\%\).

The core's design utilization is under 50\%, which is common for
dual-homed cores because, if one core router or link fails, the remaining
core capacity will still be sufficient. This keeps the BNGs as the
bottlenecks (then more costly per-customer schedulers only need to be
deployed at one node on each path).

Now let's consider two anomalies at once, as on the right hand side of
\autoref{fig:ecn-fallbacktr_dual-anomaly}. As well as a core node
failure, there happens to be unusually high utilization on one BNG,
averaging 5\%; double the design utilization. This BNG needs 50\,Gb/s of
core capacity, but it only has one 40\,Gb/s link left to connect it to
the core. This moves the bottleneck from the BNG to the interface from
the core router into the core link shown coloured red. However core
routers do not provide per-customer scheduling, so the capacity share
that each customer gets now depends on their end-system congestion
control algorithms (e.g.\ TCP), not the network's schedulers.

If every one of the 5\% of customers actively downloading at any one time
were using just a single Classic flow with perfect capacity sharing, each
access link would be utilized at \(40\,Gb/s / (5\%*8192) / 120\,Mb/s =
81\%\). Therefore, if one customer was to switch to an L4S congestion
control instead of Classic, as soon as it used more than \(1/0.81 =
1.23\times\) the Classic equitable share, its bottleneck would shift back
to its own access link, thus limiting any further advantage.

This would be no different from the similar common link bottleneck
scenario where all customers except one open a single flow. The one
opening more flows can only get \(1.23\times\) more capacity than
everyone else before their bottleneck shifts back to their own access
link.

\paragraph{In summary:} Generally, public Internet access networks do not
rely on end-system congestion controls to equitably share out capacity
between their customers. They rely on per-customer schedulers, but they
usually only deploy these at one node on the path, which they design to
be the bottleneck. A core or peering link can become the bottleneck under
anomalous conditions, so that capacity sharing does temporarily rely on
end-system congestion controls. However, if it does, any individual
congestion control only has limited scope to take advantage of the
situation before it becomes bottlenecked again within its own access
link.


\onecolumn%
\addcontentsline{toc}{part}{Document history}
\section*{Document history}

\begin{tabular}{|c|c|c|p{3.5in}|}
 \hline
Version &Date &Author &Details of change \\
 \hline\hline
00A                   &20 Oct 2019  &Bob Briscoe &First draft.\\\hline%
00B                   &21 Oct 2019  &Bob Briscoe &Moved alt-srtt algo to appendix and made optional, also corrected K1 and used MDEV\_MAX. Completed all the discussions about other factors. Just the pseudocode to pull it all together left to do.\\\hline%
00C                    &23 Oct 2019  &Bob Briscoe &First full draft. Completed pseudocode that pulls all the metrics together.\\\hline
00D                   &02 Nov 2019  &Bob Briscoe &Added active detection section. Other minor alterations.\\\hline%
01                      &02 Nov 2019  &Bob Briscoe &Issued as a complete design, but still some corners to investigate.\\\hline%
01A                   &02 Nov 2019  &Bob Briscoe &Evaluation section ToDo.\\\hline%
01B                   &06-Mar-2020  &Bob Briscoe &Corrected re-route pseudocode\\\hline%
01C                   &25-Mar-2020 &Bob Briscoe &Incorporated improvements from testing \& evaluation\\\hline%
01D                  &05-Apr-2020  &Bob Briscoe &Added appendices on Adaptive RTT Smoothing \& Implementation; Re-wrote appendix on Reroutes and added example to appendix on core links.\\\hline%
01E                   &15 Apr 2020     &Bob Briscoe &Passive Algorithm v2; Added Asad as co-author, Rewrote Problem Statement; Subdivided Passive Detection section; Added Adaptive Smoothing subsection; Rewrote Parameters subsection; Added Evaluation section; Improved throughout.\\\hline%
01F                  &18 Apr 2020     &Bob Briscoe &Improved clarity \& correctness of Eval section and added switching of AQM subsection. Added Conclusions and Further Work.\\\hline%
02                    &18 Apr 2020   &Bob Briscoe &Editorial corrections, mainly in Eval section.\\\hline
02A &19 Apr 2020     &Bob Briscoe &Adding more todo notes; editorial corrections.\\\hline
02B &16 Feb 2021     &Bob Briscoe &Updated link to online evaluation results.\\\hline
02C &17 Feb 2021     &Bob Briscoe &Restructured to add sections with new ideas. Added section on spacing between marks (\S\,\ref{ecn-fallbacktr_inter-mark}).\\\hline
\metaversion &\metadate     &Bob Briscoe &Completed new ideas sections on: exclusive marking (\S\,\ref{ecn-fallbacktr_active_solution2}) and out-of-band testing (\S\,\ref{ecn-fallbacktr_OOB_active}). Altered top and tail sections appropriately to accommodate.\\\hline
\end{tabular}

\end{document}